\documentclass[a4paper, 11pt]{book}
\usepackage[utf8]{inputenc}
\usepackage{amsmath}
\usepackage{amsfonts}
\usepackage{graphics}
\usepackage[pdftex]{graphicx}
\usepackage{multicol}
\usepackage{multirow}
\usepackage{xcolor}
\usepackage{physics}
\usepackage{mathtools}
\usepackage{subfig}
\usepackage{appendix}
\usepackage{amsmath,amsfonts,amssymb,graphicx,color, bm, xcolor}
\usepackage[colorlinks=true,citecolor=blue,linkcolor=black]{hyperref}
\usepackage[left=3.5cm , right=2.5cm]{geometry}
\usepackage[normalem]{ulem}
\usepackage[english]{babel}
\usepackage{slashed}
\usepackage{calligra}

\title{Effective-Field Theories of Analogue Gravity}
\author{Alessia Biondi}
\date{December 2022}

\newtheorem{definition}{Definition}[section]

\begin{document}

\newcommand{\lag}{\mathcal{L}}
\newcommand{\rhoo}{\rho_0}
\newcommand{\rhot}{\Tilde{\rho}}
\newcommand{\thetao}{\theta_0}
\newcommand{\thetat}{\Tilde{\theta}}
\newcommand{\xit}{\Tilde{\xi}}
\newcommand{\psit}{\Tilde{\psi}}
\newcommand{\Vt}{\Tilde{V}}
\newcommand{\lagt}{\Tilde{\lag}}
\newcommand{\mt}{\Tilde{m}}
\newcommand{\med}{\left(\frac{1}{c_s^2}-1\right)}
\newcommand{\mink}{\eta^{\mu\nu}}
\newcommand{\minks}{\eta^{\sigma\tau}}
\newcommand{\Mink}{\eta_{\mu\nu}}
\newcommand{\Minks}{\eta_{\sigma\tau}}
\newcommand{\Pmtht}{\partial^\mu\thetat}
\newcommand{\pmtht}{\partial_\mu\thetat}
\newcommand{\Pntht}{\partial^\nu\thetat}
\newcommand{\pntht}{\partial_\nu\thetat}
\newcommand{\Pstht}{\partial^\sigma\thetat}
\newcommand{\pstht}{\partial_\sigma\thetat}
\newcommand{\Pttht}{\partial^\tau\thetat}
\newcommand{\pttht}{\partial_\tau\thetat}
\newcommand{\Epiu}{E_{+,\mathrm{lab}}}
\newcommand{\Emeno}{E_{-,\mathrm{lab}}}
\newcommand{\el}{{\cal L}}
\newcommand{\de}{\partial}
\newcommand{\cs}{c_{s}}
\newcommand{\ar}{\arrowvert}
\newcommand{\ra}{\rangle}
\newcommand{\la}{\langle}
\newcommand{\da}{\dagger}
\newcommand{\xx}{{\bf x}}
\newcommand{\kk}{{\bf k}}
\newcommand{\A}{\alpha}
\newcommand{\T}{\Theta}
\newcommand{\gt}{\tilde{g}}
\newcommand{\gb}{\bar{g}}
\newcommand{\pp}{\hat{\Psi}}
\newcommand{\ff}{\varphi}
\newcommand{\ov}{\overline}
\newcommand{\cd}{\! \cdot \!}
\newcommand{\be}{\begin{equation}}
\newcommand{\ee}{\end{equation}}
\newcommand{\ba}{\begin{eqnarray}}
\newcommand{\ea}{\end{eqnarray}}
\newcommand{\ro}{\varrho}
\newcommand{\s}{\sigma}
\newcommand{\hs}{\hat{\sigma}}
\newcommand*{\myprime}{^{\prime}\mkern-1.2mu}
\newcommand*{\mydprime}{^{\prime\prime}\mkern-1.2mu}
\newcommand*{\mytrprime}{^{\prime\prime\prime}\mkern-1.2mu}
\newcommand{\rone}{\varrho}
\newcommand{\rtwo}{\tilde{\varrho}}
\newcommand{\mm}{\textcolor{red} }
\newcommand{\blue}{\textcolor{blue} }
\newcommand{\st}{\sout }
\newcommand{\oppsi}{\bm{\hat{\Psi}}}
\newcommand{\opphi}{\hat{\varphi}}
\newcommand{\opn}{\hat{n}}
\newcommand{\optheta}{\hat{\theta}}
\newcommand{\opa}{\hat{a}}
\newcommand{\opb}{\hat{b}}
\newcommand{\opc}{\hat{c}}
\newcommand{\opd}{\hat{d}}
\newcommand{\bu}{\Bar{u}}
\newcommand{\bomega}{\Bar{\omega}}
\newcommand{\wl}{\par \vspace{\baselineskip}}
\newcommand{\reale}{\mathbb{R}}
\newcommand{\f}{\mathfrak{F}}
\newcommand{\I}{\mathfrak{I}}
\newcommand{\bmx}{\bm X}
\newcommand{\bmy}{\bm Y}
\newcommand{\nablat}{\Tilde{\nabla}}
\newcommand{\Gammat}{\Tilde{\Gamma}}
\newcommand{\Wt}{\Tilde{W}}
\newcommand{\tg}{\Tilde{g}}
\newcommand{\cald}{{\calligra\footnotesize D }}
\newcommand{\cala}{{\calligra\footnotesize A }}
\newcommand{\opC}{\hat{C}}
\newcommand{\opV}{\hat{V}}
\newcommand{\opB}{\hat{B}}
\newcommand{\opQ}{\hat{Q}}
\newcommand{\phit}{\Tilde{\phi}}


\begin{titlepage}
    \begin{center}
        \vspace*{1cm}

        \includegraphics[width=0.3\textwidth]{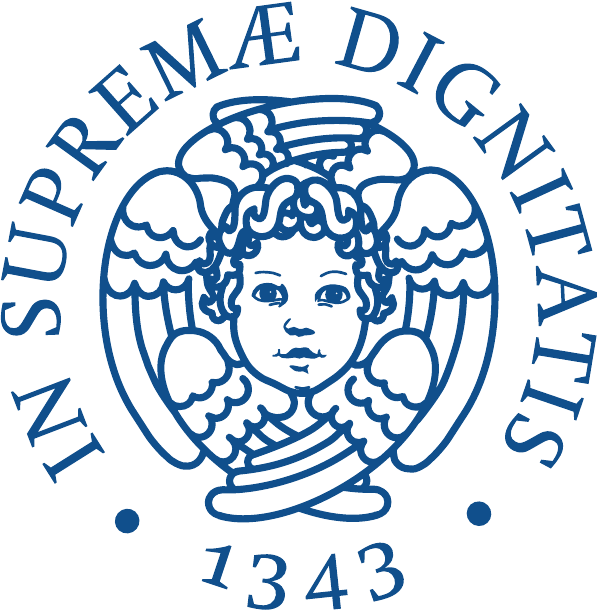}
        
        \vspace{1cm}
        
        \textsc{{\LARGE University of Pisa\\}
        \vspace{0.5cm}
        {\Large Department of Physics E. Fermi\\}
        \vspace{0.5cm}
        Master's Thesis\\}

        \vspace{4cm}
        
        {\huge{Effective-Field Theories of Analogue Gravity}}

        \vspace{2.5cm}
        
        \textbf{Candidate: \hfill Supervisors:\\}

        Alessia Biondi\hfill Maria Luisa Chiofalo\\
        \hfill Massimo Mannarelli
        \vfill
        \textsc{\LARGE Academic year 2022/2023}
        
    \end{center}
\end{titlepage}
\tableofcontents

\chapter{Introduction}

\noindent\textbf{State of the art.} In the first decades of the ’900, Einstein developed the theory of General Relativity (GR). This theory was founded on two principles: the \emph{equivalence principle} and the \emph{Mach's principle}~\cite{wald}. Thanks to the latter, it was understood that gravity is universal and that spacetime cannot be flat but it has nonvanishing curvature determined by the matter distribution. As a result, spacetime is conceived as a Lorentzian manifold and described by a metric which is affected by the presence of matter. In addition, the GR equations admit solutions that were initially rejected. One of these regards the existence of black holes, objects so dense that even light is not able to escape from their interior. All theoretical studies of black holes based on classical GR were unequivocal: a black hole cannot radiate. 
\wl
During the first half of the ’900 one other fundamental theory was developed: Quantum Field Theory (QFT). GR and QFT are very different: the first being a classical theory and the second one quantum.
In 1975, joining the two theories, Hawking~\cite{hawk1} demonstrated that, as opposed to the classical approach, any stationary black hole emits thermal radiation. Indeed, if a pair of particles is created by quantum fluctuations, one of the two particles may manage to escape from the black hole, generating a thermal distribution called Hawking radiation, instead the partner particle falls inside. The net result is a decrease of the black hole energy. According to Hawking, this loss in energy makes the black hole to shrink until it possibly evaporates. This fascinating discovery, however, opened many problems: since the emitted radiation is solely determined by the black hole mass, charge, and angular momentum, the process of black hole formation and evaporation results in the loss of information about the initial state. This problem is called Information Paradox~\cite{Mathur:2009hf}. Experimental verification of these phenomena and their full unveiling is hard due to limitations in directly detecting the Hawking radiation and investigating it in a controlled manner. Indeed, if one could access, for example, the inner part of a black hole, one could measure the correlations between the particles' pairs: this could demonstrate that the information is not lost, but is contained in the radiation.
\wl
Analogue gravity theories~\cite{analogue} represent a useful approach to investigate these problems, in which, essentially, connections between condensed matter physics and curved space QFT are established. In doing so, these theories open the possibility of engineering experimental verifications in analogue systems that are amenable to controllable conditions by means of quantum technologies. 

From the theoretical perspective, exploiting the mediums' features, the aim is to identify a so called ``analogue metric," which characterizes the propagation of the low-energy fluctuations of the systems under consideration.
Numerous analogue models exist, some of which are especially interesting from an experimental perspective, while others are important because they shed new light on puzzling theoretical topics. The most well-known of these analogies is that between sound waves in a flowing fluid and light waves in a curved spacetime. A fluid flow at transonic speed gives rise to an acoustic phenomenon analogous to a ``black hole", known as a ``dumb hole": sound waves cannot escape from its interior. This analogy can be employed to mathematically prove the presence of phonon Hawking radiation originating from the acoustic horizon. 

Essentially, two advantages can be met while playing analog gravity. Generally speaking, the very basic act of connecting notions from systems apparently unrelated, can spark original thinking that yields novel insights in both systems.
Analogue systems can even represent a useful laboratory model for QFT in curved spacetime, making practical investigations possible.

Since the idea of analog Hawking radiation was introduced, there has been a significant theoretical progress in its traits and implications. However, the analogue Hawking radiation’s detection is not so trivial. Indeed, in a cold system with a low speed of sound, the predicted power loss from the Hawking emission may be calculated to be on the scale of $P \sim 10^{-48}$W, and hence, due to the system’s finite temperature, it may be too faint to be detected above the thermal phonon background. In 2008, Carusotto et al.~\cite{PhysRevA.78.021603} predicted that Hawking radiation could be discovered by examining the density-density correlation function induced by the emitted and the absorbed Hawking particles.
In a later sequence of works, Steinhauer has first showed that the density-density correlation function could be used to measure the entanglement between the Hawking and partner particles~\cite{stein1}, and then observed spontaneous Hawking radiation emanating from an analogue black hole in an atomic Bose-Einstein condensate (BEC)~\cite{stein}. Despite the successful understanding of analog Hawking radiation in connection with the experimental observations, a number of open problems remain. Among these, especially puzzling is the question on whether and how the information about the initial black-hole state is preserved. After its radiation-driven evaporation information should never be lost, this would be inconsistent with unitary evolution of the initial quantum-mechanical black-hole state, ultimately leading to a conflict between gravity and quantum mechanics.
Addressing this problem in analogue models would be quite helpful and requires the design of useful and flexible tools to implement different spacetime geometries and beyond-mean field corrections to quantum correlations.  
\wl
\noindent\textbf{What this thesis is about.} The present thesis develops a novel method for building an analog model with a BEC, in which the analogue metric is obtained using an effective field theory and a microscopic Lagrangian with a quartic interaction, rather than with the Gross-Pitaevskii equation. Indeed, the latter describes through a mean field approximation the ground state of a dilute Bose gas of particles interacting with repulsive forces. The microscopic Lagrangian that we introduce is instead obtained adopting two essential traits, that then lead to a number of benefits: we assume that our system is described by a complex massive scalar field, and require that there must be a spontaneously broken global $U(1)$ symmetry. The benefits of our strategy are multiple indeed. First, it includes the possibility of integrating out the degrees of freedom that are not relevant to the problem at hand, while still working with an effective theory. Second, the new method is more easily applicable to any high density system where the microscopic Lagrangian exhibits a $U(1)$ and Lorentz symmetry breaking. The third advantage is to ease the incorporation of 1-loop corrections beyond mean field approximation, and include local field approximation.

While bearing these advantages, the method that we have developed provides results that are consistent with those from other approaches existing in literature. For example, we obtain the same low-energy Lagrangian determined by Greiter, Wilczek and Witten~\cite{doi:10.1142/S0217984989001400}. In addition, we compare with the next-to-leading order Lagrangian corrections worked out by Son and Wingate~\cite{Son2}, who designed an approach based on the symmetries of the system. However, this introduces not attainable parameters, yet connected to the microscopic Lagrangian. Our straightforward strategy instead, does not suffer from this problem.\\\\

\noindent\textbf{Original results.} After performing these benchmarks illustrated above, we use the developed method to obtain two main original results, that I highlight below.  
\wl
The first original result is the calculation of the next-to-leading order Lagrangian in terms of the microscopic Lagrangian’s parameters. Thus, starting from the total effective Lagrangian with only tree-level terms, we obtain a dispersion law for phonons in the presence of an acoustic horizon generated by the BEC’s flow. We observe that the phonon dispersion relation may exhibit a nontrivial minimum determined by the superfluid's velocity and the next-to-leading order low-energy constants (LECs). This minimum indicates the system's reliance on a characteristic length and consequently implies the breaking of translation symmetry. Under these conditions, a transition phase arises, leading to the hypothesis that the system is akin to a supersolid. Then, we determine the constraints on the LECs to make this happen.
\wl
The second relevant result is the design of an original procedure to calculate the density-density correlation function through the field theory tools. This object is a fundamental quantity to experimentally detect and describe the analogue Hawking radiation.
The density-density correlation function is determined for a homogeneous BEC and then applied to the microscopic and the more general effective Lagrangian which exhibits a low-energy dispersion law with a non-monotonic behavior. Our method is especially transparent because it allows to directly compute all the correlation functions in a systematic manner, besides the density-density correlation function. 
In a different method developed by Haldane~\cite{PhysRevLett.47.1840} instead, the density-density correlator is indirectly determined in a homogeneous BEC from the phonon-phonon correlation function. Needless to say, our result is consistent with that in~\cite{PhysRevLett.47.1840}. 
\wl
The tools created in the present thesis work have numerous applications, selected examples being illustrated below. The method developed for calculating the density-density correlation function for a homogeneous BEC could be used to compute the same function in a non-homogeneous BEC with a velocity profile: we plan to accomplish this task after resorting to e.g. the local- density approximation first, and then to the current-density approximation that is best suited to account for time-dependent phenomena. In addition, we can apply our method to calculate the correlation function between condensate particles and phonons: this would lead to a novel understanding of the information paradox~\cite{Tricella:2020rzl}, since we expect that this correlation function is where the information remains encoded while explaining how unitarity is preserved in the system.
Finally, the very same procedures can be adopted while considering a system where the low-energy phonons follow a non-monotonic dispersion law, like with the non-trivial minimum found for specific LECs above: this case opens up to interesting scenarios, where the horizon behavior can be related to signatures of quantum phase transitions.
\wl
\noindent\textbf{Personal contributions.} The starting question for this thesis work was to design an original approach to the information paradox problem, which I contributed to while identifying the needed steps and their conceptualization. This has led to the methodology developed in this thesis for the determination of the correlation functions in a most general and transparent manner. My personal contribution to obtain the corresponding original results outline above, has been to carry out all the analytical calculations, in most cases from scratch. I have then actively contributed to elaborate the physical interpretations of the resulting outcomes, enhancing their understanding and significance. Finally, I contributed to identify future advancements and prospects for this research, and conceptualize the steps for their development.
\wl
\noindent\textbf{Thesis contents.} In the following we briefly introduce the contents discussed in each chapter.
\wl
Chapter~\ref{chap:2} delves into a comprehensive examination of the fundamentals of differential geometry and Einstein's theory of gravity. These foundational concepts are then applied to the study of the Schwarzschild solution, which is used to represent a static, uncharged black hole.
Next, we briefly introduce QFT on curved spacetime. In particular, we investigate the thermal radiation in a static spacetime with spherical symmetry, which, applied to a black hole, is known as Hawking radiation. The presence of Hawking radiation raises intriguing issues, such as the information paradox.
\wl
Chapter~\ref{chap:3} presents an introduction to analogue gravity, which offers a powerful approach to address difficult problems in various fields using the analogy with GR. We explore selected models of analogue gravity that shed light on the underlying principles and reasonings. 
A significant part of the chapter is then devoted to discuss the analogue Hawking radiation. We examine this phenomenon from both theoretical and experimental perspectives, investigating the theoretical foundations and the most significant experimental advancements.
\wl
Chapter~\ref{chap:4} presents the original findings of this thesis. 
We construct the microscopic Lagrangian to describe the system, namely a flowing BEC, and we analyze it through an effective Lagrangian which describes the analogue Hawking radiation's behavior. In particular, we study the phononic dispersion law, discovering that, under certain constraints, it shows a non trivial minimum. Finally, we explore the density-density correlation function by employing field theory methods, resulting in a novel expression. This expression presents a fresh approach for examining fluctuations in flowing BECs.
\wl
Finally, chapter~\ref{chap:5} focuses on consolidating the key insights gained from the present thesis and outlines the potential directions for future research.

\chapter{Black Holes and Quantum Field Theories\label{chap:2}}

This chapter aims at introducing the basic tools that are preparatory to discuss how the analogy between gravity and condensed-matter systems works, that is the subject of the next Chapter~\ref{chap:3}. In particular, our final goal is to introduce the Hawking radiation~\cite{hawk1, hawk3, hawk2} and one of the connected problems, especially relevant to this thesis work: \emph{Information paradox}~\cite{Mathur:2009hf,liberatilect}.

The Chapter begins with a brief introduction to differential geometry and related needed tools that we need, like the concepts of manifold and of Killing vector~\cite{wald, carroll, hawk4, grav}, and how gravity can be interpreted in terms of geometry. We then we discuss essential concepts of General Relativity (GR), like the \emph{equivalence principle}~\cite{wald, liberatilect, PhysRevD.89.084053}, the ways it can be enunciated, and Einstein's equation. Next, we introduce the black holes~\cite{liberatilect,hawk4,dowker}, in particular examining the case of Schwarzschild black holes, the Carter-Penrose diagrams, and a brief account of trapped surfaces and apparent horizons~\cite{liberatilect}. In the end, we introduce the subject of quantum field theory on curved spacetime~\cite{liberatilect,viat,CBO9780511622632}, which id quiete useful to treat the Unruh effect~\cite{liberatilect, CBO9780511622632, unruh} and the Hawking radiation. In particular, we discuss an interesting related open problem: the \emph{information paradox}~\cite{Mathur:2009hf,liberatilect}.

\section{Differential geometry}

In the present section we summarize from~\cite{hawk4, grav, martelli} a brief introduction of essential tools for the mathematical formulation of GR.

\subsection{Manifolds}

We know that spacetime is a continuum with four dimensions, meaning that an event must be described by four numbers. Thus, spacetime can be described as a manifold~\cite{wald}. What is a manifold? We are all used to the flat Euclidean space, $\reale^n$, and its properties. For $n=1$ we are considering a curve, for $n=2$ a surface, and so on. There are also spaces which are curved, e.g. a sphere is a clearly non-flat $2$-dimensional surface. Therefore, a manifold can be thought to be a space that can be curved, but in local regions looks like $\reale^n$. Formally, manifolds can be defined in different equivalent ways, for instance:
\begin{definition}[Manifold]
   An $n$-dimensional, $C^\infty$, real \textbf{manifold} $M$ is a set together with a collection of subsets $\{O_\alpha\}$ satisfying the following properties:
    \begin{enumerate}
        \item Each $p\in M$ lies in at least one $O_\alpha$.
        \item For each $\alpha$, there is a one-to-one, onto, map $\psi_\alpha:O_\alpha\to U_\alpha$, where $U_\alpha$ is an open subset of $\reale^n$.
        \item If any two sets $O_\alpha$ and $O_\beta$ overleap, we can consider the map $\psi_\beta \circ \psi_\alpha^{-1}$ (where $\circ$ denotes composition), which relates points in $\psi_\alpha[O_\alpha\cap O_\beta]\subset U_\alpha\subset\reale^n$ to points in $\psi_\beta[O_\alpha\cap O_\beta]\subset U_\beta\subset\reale^n$. The subset of $\reale^n$ must be open and this map must be $C^\infty$.
    \end{enumerate}
\end{definition}

We can now define a topology on the manifold $M$, by demanding that all $\psi_\alpha$ are \emph{homeomorfisms}, continuous and bijective applications with continuous inverse (for more information see Ref.~\cite{manetti}). In the following we are going to consider only manifolds which are \emph{Hausdorff} \footnote{A topological space is said to be Hausdorff if distinct points admit disjoint neighborhoods. From~\cite{manetti}, pag.57.} and \emph{paracompact} \footnote{A topological space is said to be paracompact if every open cover has a locally finite open refinement. From~\cite{manetti}, pag. 134.}. 

It is now useful to define the concept of \emph{differentiability} and \emph{smoothness} of maps between the manifolds. 
\begin{definition}
    Let us consider two manifolds $M$ and $M'$ with, respectively, $\{\psi_\alpha\}$ and $\{\psi'_\beta\}$ being their chart maps. They are \textbf{diffeomorphic} if, taken a map $f:M\to M'$, $f$ is $C^\infty$ one-to-one, onto, has $C^\infty$ inverse and it is called a \textbf{diffeomorphism}. Diffeomorphic manifolds have identical manifold structure.
\end{definition}
\wl
One other fundamental concept is the notion of vector in curved geometry. Here, as in GR, the vector space structure found in flat spaces is lost. We can recover it in the limit of \emph{infinitesimal displacements} about a point. It is important to define a tangent vector in a way that refers only to the intrinsic structure of the manifold. Such a definition in provided by the notion of a \emph{tangent vector as a directional derivative}~\cite{wald}.
\begin{definition}[Tangent vector]
    Consider a manifold $M$ with $\f$ the collection of $C^\infty$ functions from $M\to\reale$. We define a tangent vector $v$ at a point $p\in M$ to be a map $v:\f\to\reale$ with the following properties:
    \begin{enumerate}
        \item it is linear
        \begin{equation}
            v(af+bg)=av(f)+bv(g),\quad\mathrm{for\ all\ }f,g\in\f;a,b\in\reale\,;
        \end{equation}
        \item it obeys the Leibniz rule
        \begin{equation}
            v(fg)=f(p)v(g)+g(p)v(f)\,.
        \end{equation}        
    \end{enumerate}
\end{definition}
It is easy to see that the collection of tangent vectors at $p$, $T_p(M)$, has the structure of a vector space under the the following laws:
\begin{align}
    &(v_1+v_2)(f)=v_1(f)+v_2(f)\\
    &(av)(f)=av(f),
\end{align}
for all $f\in\f$. An important feature of $T_p(M)$ is given by a theorem~\cite{wald}, stating that given $p\in M$, with dim $M=n$, then dim $T_p(M)=n$. At this point, we can introduce the basis $\{X_\mu\}$ of $T_p(M)$, which is often denoted as $X_\mu=\partial/\partial x^\mu$ (with $\mu=1,..,n$), and it is identified by the chart $\psi:O\to U\subset\reale$, with $p\in O\subset M$.

We can now define a \emph{smooth curve} $C$ on a manifold $M$, as a $C^\infty$ map from $\reale$ into $M$. Thus, in any coordinate basis, the components $T^\mu$ of the tangent vector to the curve are given by
\begin{equation}
    T^\mu=\frac{dx^\mu}{dt},
\end{equation}
where $x^\mu(t)$ is the mapping in $\reale^n$ of the curve $C$ on $M$.

The assignment of a tangent vector $v|_p\in T_p(M)$ is a tangent field $v$ on a manifold $M$.

Let us now consider $V$ any finite-dimensional vector space over the real numbers. The collection $V^*$ of linear maps $f:V\to \reale$, with the addition and scalar multiplication of such linear maps, is a natural vector space structure and it is called the \emph{dual vector space} to $V$. The elements belonging to $V^*$, $v^{1*},..,v^{n*}$, are called \emph{dual vectors}, and are defined by
\begin{equation}
    v^{\mu*}(v_{\nu})=\delta^\mu_\nu,
\end{equation}
where $\delta^\mu_\nu$ is 1 for $\mu=\nu$ and 0 otherwise. Thus, $\{v^{\mu*}\}$ is a basis of $V^*$, called \emph{dual basis}, and dim $V^*=$dim $V$. The vector spaces $V$ and $V^*$ are isomorphic because of the correspondence $v_\mu\longleftrightarrow v^{\mu*}$, even if this isomorphism depends on the choice of $\{v_\mu\}$. 

We can now define the notion of tensor~\cite{wald}. Taken $V$ a finite-dimensional vector space and its dual vector space $V^*$, a \emph{tensor} $T$ of type $(k,l)$ over $V$ is the multilinear map:
\begin{equation}
    T:\underbrace{ V^*\times...\times V^* }_{k}\times{ V\times...\times V }_{l}\to \reale.
\end{equation}
Thus, a tensor of type $(1,0)$ is a vector and everything that we have said before for vectors can be extended to tensors. The collection $\I(k,l)$ of all the tensors of type $(k,l)$ has indeed the structure of a vector space with dimension $n^{k+l}$, where $n=$ dim $V$.
Tensors enjoy two important operations:
\begin{itemize}
    \item The \emph{contraction} with respect to the $i$th and $j$th indices: it is a map $C:\I(k,l)\to\I(k-1,l-1)$ such that, if $T$ is a tensor of type $(k,l)$ then
    \begin{equation}
        CT=\sum_{\sigma=1}^n T(..,v^{\sigma*},..;..,v_\sigma,..),
    \end{equation}
    where $\{v_\sigma\}$ is a basis for $V$, $\{v^{\sigma*}\}$ its dual basis, and $v^{\sigma*}$ and $v_\sigma$ are in the $i$th and $j$th positions, respectively.
    \item The \emph{outer product}: given two tensors $T$ and $T'$ of type $(k,l)$ and $(k',l')$, respectively, we can construct a new tensor of type $(k+k',l+l')$ and denoted with $T\otimes T'$, as the multilinear map acting as the $T$ tensor on the $(k,l)$ indices and as the $T'$ tensor on the $(k',l')$ indices. 
\end{itemize}

Thus, if $\{v_\mu\}$ is a basis of $V$, and $\{v^{\mu*}\}$ is its dual basis, every tensor $T$ of type $(k,l)$ can be expressed as 
\begin{equation}
    T=\sum_{\mu_1,..,\nu_l}^nT^{\mu_1,..,\mu_k}\, _{\nu_1,..,\nu_l}v_{\mu_1}\otimes..\otimes v^{\nu_l*},
\end{equation}
where $T^{\mu_1,..,\mu_k}\, _{\nu_1,..,\nu_l}$ are the components of the tensor $T$ with respect to $\{v_\mu\}$.

We define $T^*_p(M)$ as the \emph{cotangent space} at $p$, and vectors in this space are called \emph{cotangent vectors}. Thus, given a coordinate system, we can construct a coordinate basis $\{\partial/\partial x^\mu\}$ of $T_p(M)$ and its dual basis $\{dx^\mu\}$. If we want to change coordinate systems, the vector components $v^\mu$ and the cotangent vector components $\omega_\mu$ transform as
\begin{equation}
    v'^\nu=\sum_{\mu=1}v^\mu\frac{
    \partial x'^\nu}{\partial x^\mu},\qquad \omega'_\nu=\sum_{\mu=1}^n\omega_\mu\frac{\partial x^\mu}{\partial x'^\nu}.
\end{equation}
In general, a tensor of type $(k,l)$ transforms as
\begin{equation}
    T'^{\mu_1,..\mu_k}\, _{\nu_1,..,\nu_l}=\sum_{\alpha_1,..,\beta_l}^n T'^{\alpha_1,..\alpha_k}\, _{\beta_1,..,\beta_l}\frac{\partial x'^{\mu_1}}{\partial x^{\alpha_1}}..\frac{\partial x^{\beta_1}}{\partial x'^{\nu_l}}.
    \label{eq:tensors}
\end{equation}

It is useful to extend the notion of continuity to vector fields. A tangent field $v$ is \emph{smooth} if for each smooth function ($C^\infty$) $f$, the function $v(f)$ \footnote{If $f$ is a smooth function then at each $p\in M$, $v(f)$ is a function on $M$.} is also smooth. A covariant vector field $\omega$ is smooth if for each smooth vector field $v$, the function $\omega(v)$ is smooth.
A tensor $T$ of type $(k,l)$ is smooth if for all smooth covariant vector fields $\omega^1,..,\omega^k$ and smooth vector fields $v_1,..,v_l$ the function $T(\omega^1,..,\omega^k;v_1,..,v_l)$ is smooth.

Considering the reasoning that we have made for a vector field, we can extend this notion even for a covariant vector field and, more in general, for any tensor of type $(k,l)$
\wl
We want to give now the notion of metric~\cite{wald}. 
\begin{definition}[Metric]
    A metric $g$ on a manifold $M$ is a tensor field of type $(0,2)$ such that, given $p\in M$,     for all $v_1,v_2\in T_p(M)$, it is
    \begin{enumerate}
        \item symmetric, that is $g(v_1,v_2)=g(v_2,v_1)$;
        \item nondegenerate, that is $g(v_1,v_2)\ne0$ with $v_1$ and $v_2\ne0$\,.
    \end{enumerate}

\end{definition}
The metric $g$ can be expanded in terms of its components $g_{\mu\nu}$ as
\begin{equation}
    g=\sum_{\mu,\nu}g_{\mu\nu}dx^{\mu}\otimes dx^{\nu}.
\end{equation}
In GR it is customary to omit the outer product symbol and to use instead of the metric the infinitesimal interval squared:
\begin{equation}
    ds^2=\sum_{\mu,\nu}g_{\mu\nu}dx^\mu dx^{\nu}.
\end{equation}
Given a metric $g$, we can always find an orthonormal basis $v_1,..,v_n$ of the tangent space at each point $p$, such that $g(v_\mu,v_\nu)=\pm\delta_{\mu\nu}$. The number of "+" and "-" signs is called \emph{signature} of the metric. We can distinguish two types of metric:
\begin{itemize}
    \item \emph{positive definite}, or \emph{Riemannian}: these metrics have signature $(+,..,+)$, i.e. all the signs are "+";
    \item\emph{Lorentzian}: these metrics have signature $(-,+,..,+)$, i.e. one "-" and the remaining "+". These are the spacetimes we focus on.
\end{itemize}

\subsection{Curvature}

Here we want to define the notion of \emph{intrinsic curvature} of a Manifold $M$. To do this, as we are considering a general manifold, we first need the concept of covariant derivative and parallel transport~\cite{hawk4}.
From now on we use the Einstein notation:
\begin{equation}
    v^\mu\omega_\mu=\sum_{\mu=1}^n v^\mu\omega_\mu.
\end{equation}
\begin{definition}[Covariant derivative]
    A \textbf{covariant derivative} $\nabla$ on a manifold $M$ is a map which takes each differentiable ($C^r$ with $r\ge1$) vector field to a differentiable vector field, such that:
    \begin{enumerate}
        \item$\nabla_{ X} Y$ is a tensor in the argument $ X$, i.e. for any $f,g\in\f$ and for all vector fields $ X,\ Y,\ Z$
        \begin{equation}
            \nabla_{fX+gY}Z=f\nabla_XZ+g\nabla_YZ;
        \end{equation}
        \item $\nabla_XY$ is linear in $Y$, i.e. for all vector fields $Y,Z,X$ and for all $\alpha,\beta\in\reale$
        \begin{equation}
            \nabla_X(\alpha Y+\beta Z)=\alpha\nabla_XY+\beta\nabla_XZ;
        \end{equation}
        \item for all $f\in\f$ and for all vector fields $Y$
        \begin{equation}
            \nabla_X(fY)=X(f)Y+f\nabla_XY.
        \end{equation}
    \end{enumerate}
\end{definition}
If we now consider the associated coordinated basis $\{e_a\}$ and $\{e^a\}$, the components of $\nabla Y$ can be written as $V^a\,_{;b}$, thus:
\begin{equation}
    \nabla Y=Y^a\,_{;b}e^b\otimes e_a.
\end{equation}
We now define the Christoffel symbols $\Gamma^a_{bc}$ as
\begin{equation}
    \nabla e_c=\Gamma^a_{bc}e^b\otimes e_a.
\end{equation}
As a consequence, the covariant derivative of a vector can be cast in the form
\begin{equation}
    \nabla Y=\nabla(Y^c e_c)=\frac{\partial Y^c}{\partial x^a}e^a\otimes e_c+Y^c\Gamma^a_{bc}e^b\otimes e_a,
\end{equation}
and the $\nabla Y$ components are
\begin{equation}
    Y^a\,_{;b}=\frac{\partial Y}{\partial x^b}+\Gamma^a_{bc}Y^c.
\end{equation}
Notice that $\Gamma^a_{bc}$ is not a tensor because it does not follow Eq.~\eqref{eq:tensors}, it is instead a set of $n^3$ functions determined by the indices. However, if we have two different connections $\nabla$ and $\nablat$, they are distinguished by two different $\Gamma^a_{bc}$ and $\Gammat^a_{bc}$. Their difference gives rise to a tensor
\begin{equation}
    \nabla Y-\nablat Y=(\Gamma^a_{bc}-\Gammat^a_{bc})Y^ce^b\otimes e_a
\end{equation}
and, as a consequence, the difference $(\Gamma^a_{bc}-\Gammat^a_{bc})$ is a tensor.
The definition of covariant derivative can be extended to any differentiable tensor~\cite{hawk4}.
\begin{definition}
    Given $T$ a $C^r$ tensor field of type $(k,l)$, then the covariant derivative $\nabla T$ is a $C^{r-1}$ tensor field of type $(k,l+1)$ such that:
    \begin{itemize}
        \item $\nabla$ is linear and commutes with the contractions;
        \item for all tensor fields $T,\ S$, Leibniz rule holds:
        \begin{equation}
            \nabla(S\otimes T)=\nabla S\otimes T +S\otimes \nabla T;
        \end{equation}
        \item $\nabla f=\frac{\partial f}{\partial x^a}e^a$ for any $f\in\f$.
    \end{itemize}
\end{definition}
The covariant derivative of a $C^r$ tensor field $T$ along a $C^r$ curve $\lambda(t)$ is defined as $DT/\partial t$. If one chooses local coordinates, so that $\lambda(t)$ has coordinates $x^a(t)$, in terms of the components of a vector field $Y$ one has:
\begin{equation}
    \frac{DY^a}{\partial t}=\frac{\partial Y^a}{\partial t}+\Gamma^a_{bc}Y^c\frac{dx^b}{dt},
    \label{eq:pt}
\end{equation}
where $X^a=\frac{dx^a}{dt}$ is the tangent vector to the curve $\lambda$. If in Eq.~\eqref{eq:pt} $DT/\partial t=0$ is satisfied, then the tensor $T$ is said to be \emph{transported in parallel} along $\lambda$.
A particular case is the covariant derivative of the tangent vector to the curve $\lambda$: the curve $\lambda$ is called a \emph{geodesic curve} if there is a function $f$ such that 
\begin{equation}
    \nabla_XX=\frac{D}{\partial t}\left(\frac{\partial}{\partial t}\right)_\lambda=fX\qquad X^a\,_{;b}X^b=fX^a  .  
\end{equation}
Here, it is possible to find a parameter, called \emph{affine parameter}, which physically corresponds to the proper time, and it is related to $t$ by $v=at+b$, with $a$ and $b$ constants such that
\begin{equation}
    \frac{D}{\partial v}\left(\frac{\partial}{\partial v}\right)_\lambda=0.
\end{equation}
Thus, the associated tangent vector $V=(\partial/\partial)_\lambda$ is parallel to $X$ but has a scale determined by $V(v)=1$:
\begin{equation}
    V^bV^a\,_{;b}=0\Longleftrightarrow\frac{d^2x}{dv^2}+\Gamma^a_{bc}\frac{dx^b}{dv}\frac{dx^c}{dv}=0.
\end{equation}

Given a $C^r$ connection $\nabla$, one can define a $C^{r-1}$ tensor field of type $(1,2)$:
\begin{equation}
    T(X,Y)=\nabla\nabla_XY-\nabla_YX-[X,Y]\qquad T^i\,_{jk}=\Gamma^i\,_{jk}-\Gamma^i\,_{kj}=\Gamma^i\,_{[jk]},
\end{equation}
where $X$ and $Y$ are arbitrary $C^r$vector fields, and $[X,Y]=XY-YX$, which is called \emph{torsion tensor}. If we consider a parallelogram generated by two infinitesimal vectors parallel transported along each other, its non-closure is given by the the torsion. We work only with torsion-free connections $T=0$. Thus, the connections that satisfy this condition are called symmetric because $\Gamma^i\,_{jk}=\Gamma^i\,_{kj}$. This object is important for the choice of the connection.

Given $C^{r+1}$ vector fields $X,Y,Z$, we can define a $C^{r-1}$ vector field by a $C^r$ connection $\nabla$
\begin{equation}
    R(X,Y)Z=\nabla_X(\nabla_YZ)-\nabla_Y(\nabla_XZ)-\nabla_{[X,Y]}Z\qquad R^a\,_{bcd}Z^b=Z^a\,_{;dc}-Z^a\,_{;cd}.
\end{equation}
This is called \emph{Riemann tensor}. It gives us the measure of the non-commutation of the covariant derivative. Physically, curvature is the rotation that a vector experiences when it is parallel carried around a closed curve. The Riemann tensor, $R(X,Y)Z$, is linear in $X,Y,Z$. Its value in $p$ only depends on the values of $X,Y,Z$ at $p$.
The expression for the coordinate components of the Riemann tensor is given by
\begin{equation}
    R^a\,_{bcd}=\frac{\partial\Gamma^a\,_{db}}{\partial x^c}-\frac{\partial \Gamma^a\,_{cb}}{\partial x^d}+\Gamma^a\,_{cf}\Gamma^f\,_{db}-\Gamma^a\,_{df}\Gamma^f\,_{cb}.
\end{equation}
The contraction of this tensor gives rise to the \emph{Ricci tensor} $R^a\,_{bad}=R_{bd}$, and to the \emph{scalar curvature} $R=R^a\,_a$.

If we now consider a metric $g$ on a manifold $M$, there is a unique connection $\nabla$ which is symmetric and with the property $g_{ab;c}=0$. This connection is called \emph{Levi-Civita connection}. We use it from now on, to build the GR formalism. From this connection it is possible to compute the Christoffel symbols
\begin{equation}
    \Gamma^a\,_{bc}=\frac12g^{ad}\left(\frac{\partial g_{bd}}{\partial x^c}+\frac{\partial g_{dc}}{\partial x^b}-\frac{\partial g_{cb}}{\partial x^d}\right).
\end{equation}
Note that there are other gravitational theories, like in~\cite{jose}, where the torsion tensor or the covariant derivative of the metric are different from zero, while $R$ vanishes. 

\subsection{Lie derivative and Killing vectors}

As a final step for this part of the Chapter, it is useful to introduce the isometries and the Killing vectors~\cite{hawk4}. In order to do so, we first need to introduce preliminary important notions.

If $f$ is a function on $M'$, the mapping $\psi:M\to M'$ defines the function $\psi^*f$ on $M$ such that, given $p\in M$,
\begin{equation}
    \psi^*f(p)=f(\psi(p)).
\end{equation}
If we consider a curve $\lambda(t):\reale\to M$ such that there is a value of $t$, $t_0$, so that $\lambda(t_0)=p\in M$, then $\psi(\lambda(t))$ is a curve in $M'$ which passes through $\psi(p)$. Also, the tangent vector to this curve at $\psi(p)$ is denoted as $\psi_*(\partial/\partial\lambda)_\lambda|_p$. Thus, $\psi_*$ is a linear map $\psi_*:T_p(M)\to T_{\psi(p)}(M')$ which maps vector from one manifold to another, and it can be characterized as follows:
\begin{equation}
    X(\psi^*)|_p=\psi_*X(f)|_{\psi(p)},
\end{equation}
where $f$ is a $C^r$ function at $\psi(p)$ and $X$ is a vector at $p$.

If we consider any $C^r$ vector field $X$ on $M$, there is one maximal curve $\lambda(t)$ through each point of $M$, such that $\lambda(0)=p\in M$ and $X|_{\lambda(t)}=(\partial/\partial t)_{\lambda(t)}$. We consider $\{x^i\}$ as the local coordinates, thus the curve $\lambda(t)$ has coordinates $x^i(t)$ and the vector $X$ has components $X^i$. This curve is called the \emph{integral curve} of $X$ with initial point $p$, and it is locally a solution of the set of differential equations:
\begin{equation}
    \frac{dx^i}{dt}=X^i(x^i(t),..,x^n(t)).
\end{equation}
For each point $q$ of $M$, there is an open neighborhood $U$ of $q$ and an $\epsilon>0$, such that $X$ defines a family of diffeomorphisms $\psi_t:U\to M$ whenever $|t|<\epsilon$, obtained by taking each point $p$ in $U$ a parameter distance $t$ along the integral curves of $X$.

There is one type of differentiation which is defined naturally by the manifold structure and which we have not treated yet: \emph{Lie differentiation}. 
The \emph{Lie derivative} $L_XT$ of a tensor field $T$ with respect to $X$ is defined as
\begin{equation}
    L_XT|_p=\lim_{t\to0}\frac1t(T|_p-\psi_{t*}T|_p).
\end{equation}
If we consider the Lie derivative $L_XY$ of a vector field $Y$ with respect to $X$, this can be written as
\begin{equation}
    L_XY=[X,Y].
\end{equation}
If it vanishes, then if we start from $p\in M$ and go a parameter distance $t$ along the integral curves of $X$, then a parameter distance $s$ along the integral curves of $Y$, we arrive at the same point as if we went a parameter distance $s$ along the integral curves of $Y$, then a parameter distance $t$ along the integral curves of $X$. 
\wl
A diffeomorphism $\psi:M\to M$ is called \emph{isometry} if it leaves the metric unchanged, and this happens if the mapped metric $\psi_*g$ is equal to $g$ in each point. This means that $\psi_*:T_pM\to T_{\psi(p)}M$ preserves scalar products
\begin{equation}
    g(X,Y)|_p=\psi_*g(\psi_*X,\psi_*Y)|_{\psi(p)}=g(\psi_*X,\psi_*Y)|_{\psi(p)}.
\end{equation}
If the diffeomorphisms $\psi_t$ generated by the vector field $K$ are isometries, then $K$ is called a \emph{Killing vector field}. In this case, the Lie derivative of the metric vanishes
\begin{equation}
    L_Kg=\lim_{t\to0}\frac1t(g-\psi_{t*}g)=0
\end{equation}
and
\begin{equation}
    K_{a;b}+K_{b;a}=0.
\end{equation}

\section{General Relativity's tools}

Before the advent of GR, the gravitational effects were described by Newton's theory. However, the main issue was that the Newton description is inconsistent with special relativity~\cite{grav}. Indeed, Newton's theory provides that interactions propagate instantaneously, while in special relativity interactions propagate at a finite velocity, and that the upper limit of this speed is the speed of light. 

One could think that an approach to solve this problem could be modifying Newton's theory to fit within the framework of special relativity. Einstein did not follow this path, he built a new theory from two key ideas:
\begin{enumerate}
    \item \emph{equivalence principle};
    \item \emph{Mach's principle}.
\end{enumerate} 
Mach's principle~\cite{wald} is based on the idea that, contrary to the notions of spacetime in special relativity and prerelativity, all matter in the universe should contribute to the structure of spacetime and, in particular, that "inertial and non-rotating motion" are influenced by matter.

The equivalence principle can be formulated in different ways~\cite{PhysRevD.89.084053, liberatilect}:
\begin{itemize}
    \item Newtonian Equivalence Principle (NEP): inertial and gravitational masses of a body are equal;
    \item Weak Equivalence Principle (WEP): the free-falling observer is universal, i.e. the motion of bodies under gravitational forces does not depend on their composition;
    \item Gravitational Weak Equivalence Principle (GWEP): the distinctive features of the test bodies themselves have no impact on the world lines of small freely-falling test bodies (this amounts to WEP plus self-gravity condition);
    \item Einstein Equivalence Principle (EEP): any physical non-gravitational test mass in any position in spacetime is locally unaffected by the gravitational field when in a free-falling frame (WEP plus Local Lorentz Invariance plus Local Position Invariance);
    \item Strong Equivalence Principle (SEP): local gravity experiments are not affected by the presence of a background gravitational field (GWEP plus Local Lorentz Invariance plus Local Position Invariance).

\end{itemize}
The last principle, SEP, restricts the possible theories of gravity to GR and Nordstrom gravity~\cite{nord}. 

The Nordstrom gravity is the first relativistic theory of gravitation described by the interactions of a massless scalar field and it was inspired by electrostatic. Indeed Newton's law for the gravitational force has the same form of Coulomb's law for the force between two opposite charges 
\begin{equation}
    F_N=-G\frac{m_am_b}{r^2}\qquad F_C=\frac{e_ae_b}{4\pi r^2},
\end{equation}
one could try to describe them with a similar language. In electrostatics the electric field $\bm E$ can be written in terms of a scalar potential $\phi$ as $\bm E = -\bm \nabla \phi$. Given the static Lagrangian
\begin{equation}
    \lag=\frac12\bm E^2-\rho\phi=\frac12(\bm \nabla \phi)^2-\rho\phi,
\end{equation}
where $\rho$ is the charge density, the equation of motion is $\nabla^2\phi=-\rho$. As a consequence, if one interprets gravitational interaction as a scalar massless field $\phi$, the correspondent Lagrangian is
\begin{equation}
    \lag=-\frac{1}{8\pi G}(\bm \nabla\phi)^2-\rho\phi,
\end{equation}
where$\rho$ is the mass density, the equation of motion of $\phi$ is 
\begin{equation}
    \nabla^2\phi=4\pi G \rho.
    \label{eq:nord}
\end{equation}
In order to make this theory relativistic and dynamic, Nordstrom generalized Eq.~\eqref{eq:nord} in
\begin{equation}
    \Box\phi=-4\pi G\rho,
\end{equation}
where $\Box=\partial_t^2-\nabla^2$ is the d'Alembertian operator. This theory had an immediate flaw in that it could not be deduced from an action principle, and, in addition, it has been emphasized that the mass density should be the trace $T$ of the system's energy-momentum tensor $T_{\mu\nu}$. However, if one assumes that $\rho\propto T$, one finds that the field equation is
\begin{equation}
    \phi\Box\phi=-4\pi G T,
\end{equation}
which is not linear. Since the electromagnetic field has $T=0$, it has no gravitational interactions, and thus, there would be no light bending around stellar masses in this hypothesis: Nordstrom theory has been ruled out.
\wl
Another pillar of General relativity is the Einstein equation. It can be obtained as the relativistic version of the Poisson equation for the Newtonian potential $\Phi$~\cite{carroll}
\begin{equation}
    \nabla^2\Phi=4\pi G\rho,
    \label{eq:poisson}
\end{equation}
where $\rho$ is the mass density and $G$ is the Newton's gravitational constant. A relativistic generalization should take the form of an equation between tensors. The tensor generalization of the mass density is the energy momentum tensor $T_{\mu\nu}$, with $\mu=0,1,2,3$. The gravitational potential is generalized by the metric tensor $g_{\mu\nu}$: gravity is universal and can be described as a fundamental feature of the background on which matter fields propagate, so that spacetime is identified as a curved manifold. Our new equation therefore have 
\begin{equation}
    G_{\mu\nu}\propto T_{\mu\nu},
    \label{eq:einstein}
\end{equation}
where $G_{\mu\nu}$ is second-order in metric tensor derivatives. Because of energy conservation, $\nabla^\mu T_{\mu\nu}=0$, thus also the left-hand-side of Eq.~\eqref{eq:einstein} must behave like $\nabla^\mu G_{\mu\nu}$. Furthermore, the Riemann tensor $R^\mu_{\nu\rho\sigma}$ is, obviously, a quantity that does not vanish. It is constructed from second derivatives of the metric, and its contractions, the Ricci tensor $R_{\mu\nu}$ and the curvature scalar $R$, obey the Bianchi identity:
\begin{equation}
    \nabla^\mu R_{\mu\nu}=\frac12\nabla_\nu R.
\end{equation}
Thus, we can assume that 
\begin{equation}
    G_{\mu\nu}\equiv R_{\mu\nu}-\frac{1}{2}g_{\mu\nu}R=kT_{\mu\nu},
\end{equation}
where now $G_{\mu\nu}$ is the Einstein tensor and $k$ is a constant of proportionality. To obtain the value of $k$, we consider that in the non-relativistic and weak field limit, which in formula is translated as
\begin{align*}
    &g_{\mu\nu}=\eta_{\mu\nu}+h_{\mu\nu}\quad \mathrm{with} \quad |h_{\mu\nu}|\ll1, \qquad h_{00}=-2\Phi\\
    &R_{00}=-\frac12\nabla^2h_{00}\,\qquad T_{00}=\rho,
\end{align*}
Eq.~\eqref{eq:poisson} must be satisfied. In this way one gets the Einstein's equation for general relativity
\begin{equation}
    R_{\mu\nu}-\frac12 g_{\mu\nu}R=8\pi G T_{\mu\nu}.
\end{equation}
This equation describes how the presence of energy momentum affects the curvature of spacetime.

The Einstein equation can be formally derived by minimizing the functional variation of the action \hfill\break$S=\frac{1}{16\pi G}S_{EH}+S_m$, where 
\begin{equation}
    S_{EH}=\int d^4x\sqrt{-g}R,
\end{equation}
is the Einstein-Hilbert action~\cite{hawk4}, with $g=\mathrm{det}(g_{\mu\nu})$. On the other hand, the matter action is
\begin{equation}
    S_m=\int d^4x\sqrt{-g}\lag_m,
\end{equation}
with $\lag_m$ the Lagrangian density containing all fields associated to particles in the model. The energy-momentum tensor is then defined by
\begin{equation}
    T_{\mu\nu}=-2\frac{1}{\sqrt{-g}}\frac{\delta S_m}{\delta g^{\mu\nu}},
\end{equation}
where $g$ is the determinant of the metric.

\section{Black holes}

The first idea of a black hole was given by Michelle in 1783 ~\cite{michelle}. In those years it was already well known that light had a speed of about 300.000 km/s~\cite{kthorne}. He considered Newton's laws of gravity and a corpuscolar description of light. Then, he computed the radius of a star, having the same density of the sun, so that a body has an escape velocity equal to the speed of light. From these results, he concluded that Universe might contain \emph{dark stars}, as he named them, invisible from Earth because the corpuscoles of light emitted from its surface are pulled back down. Today, we know that those dark stars, thought by Michelle, are black holes (BH). 
\wl
In this section, we introduce a definition of BH~\cite{liberatilect} in terms of properties of the spacetime. In particular, we analyze the simplest case of a Schwarzschild black hole, the simplest. Then, we introduce the global methods worked out by Carter and Penrose~\cite{hawk4} and, to conclude, we treat the Hawking radiation.

\subsection{Generalities}

Black holes are physical objects that possess horizons and singularities. We are going to define both.

\subsubsection{Singularity}

What is a singularity? This phenomenon is recognized as a sequence of events in which the metric shows a pathology or the curvature reaches infinity. However, we cannot assume the metric a priori, because it is a result of our field equations. Therefore, only when the pair $(M,g)$ is precisely defined at a spacetime event, then it has a physical significance. The implication of this is that a singularity cannot be a collection of events in spacetime; rather, it must be characterized as a collection of missing events in $(M,g)$, where spacetime is not defined.

To explain in a more formally way what the presence of a singularity in the spacetime implies, we now introduce two notions.
A spacetime which misses a point, i.e., more formally, which is not symmetric to a proper subset of another spacetime, is called \emph{inextendible spacetime}.
A geodesic which cannot extend beyond a finite value of the affine parameter, is called an \emph{incomplete geodesic}.
Thus, \emph{an inextendible spacetime has a singularity if there is at least one incomplete geodesic}.

\subsubsection{Event horizon \label{subsub:eh}}

The event horizon of a black hole can be identified by a light ray which does not fall into the black hole, neither escapes into infinity. Thus, it is a mathematical notion and can be defined in a more formal way.
To define a horizon we need to introduce the preliminary concepts as follows. 

A \emph{time orientable spacetime} is a spacetime $(M,g)$, in which for each of its point we can locally distinguish a future and a past light cone. Here, there must be a globally specified, timelike, continuous, and non-zero vector field $v^\mu$.

A \emph{space orientable spacetime} is a spacetime $(M,g)$, where there is a three-form \footnote{A $k$-form is a tensor field of type $(0,k)$ completely antisymmetric.} $\epsilon_{[\mu\nu\lambda]}\ne0$ everywhere and continous, and if $v^\mu\epsilon_{[\mu\nu\lambda\rho]}=0$ for any $v^\mu$ timelike.

There is a everywhere non-null, continuous, globally defined
4-form $\epsilon_{[\mu\nu\rho\lambda]}$ if and only if a spacetime is \emph{orientable}. If a spacetime is time orientable and space oriantable, then it is orientable.

The set of points $q$ of the manifold $M$ such that there is a past directed timelike curve, which starts in $p$ and ends in $q$, is known as the \emph{chronological past} of a point $p$, and it is denoted by the symbol $I^-(p)$.
In analogy, the set of points $q$ of the manifold $M$ such that there is a future directed timelike curve, which starts in $p$ and finishes in $q$, is known as the \emph{chronological future} of a point $p$, and it is denoted by the symbol $I^+(p)$.

The \emph{causal future} of a point $p$, $J^+(p)$, is the set of points $q\in M$ such that there exists a future directed causal curve $\gamma$, with $\gamma(0)=p$ and $\gamma(1)=q$. The definition is very similar in the case of \emph{causal past}, $J^-(p)$, the only difference being that the curve $\gamma$ is a past directed causal curve.

\begin{figure}[ht!]
    \centering
    \includegraphics[scale = 0.8]{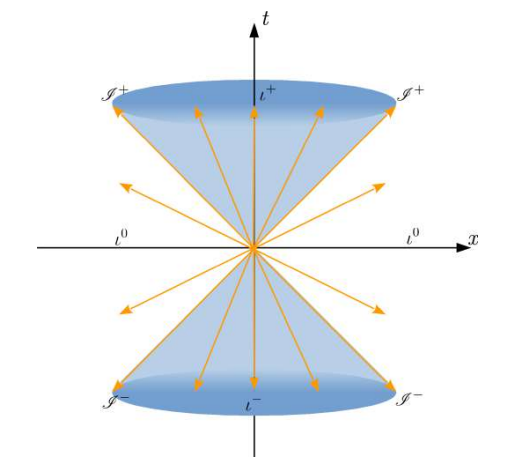}
    \caption{\emph{Representation of the light cone and of the all possible infinities. Past and future null infinity are indicated with {\calligra\footnotesize I}$\ \,^\pm$. Spatial, the past time and the future time infinity are indicated with $\iota^0,\iota^-, \iota^+$, respectively. From }\cite{liberatilect}.}
    \label{fig:lightcone}
\end{figure}

In general, we can define 5 types of infinity (see Fig.~\ref{fig:lightcone}).
One is a spatial infinity $\iota^0$. Other two are time infinities: one past-time infinity, $\iota^-$, and one future-time infinity $\iota^+$. The remaining two are null infinity, one future null infinity $\mathcal{I}^+$, and one past null infinity $\mathcal{I}^-$ (the concepts will be more clear while illustrating the Carter-Penrose diagrams).

Besides than for a point $p\in M$, we can define all the above concepts also referring to a subset of the manifold $U\subset M$ as
\begin{equation}
    I^\pm(U)=\bigcup_{p\in U}I^\pm(p)\qquad J^\pm(U)=\bigcup_{p\in U}J^\pm(p).
\end{equation}
We are finally in a position to define a \emph{black hole} region $\mathcal{B}_{bh}$, that is
\begin{equation}
    \mathcal{B}_{bh}=M-J^-(\mathcal{I}^+).
\end{equation}
From this definition, we see that there are portions of event in our spacetime, that will never be causally connected with $\mathcal{I}^+$. If we now introduce the border of the topological closure of $J^\pm$ as $\dot{\Bar{J}}^\pm(U)=\Bar{J}^\pm(U)-J^\pm(U)$, with $\Bar{J}$ the intersection of all closed sets containing $J^\pm(U)$, and $J^\pm(U)$ being the union of all the open sets containing $J^\pm(U)$, we can define the \emph{future event horizon}~\cite{town} as
\begin{equation}
    \mathcal{H}^+\equiv\dot{\Bar{J}}^-(\mathcal{I}^+).
\end{equation}
Consequently, one can define also a \emph{white hole} region and a \emph{past event horizon} as
\begin{equation}
    \mathcal{B}_{wh}=M-J^+(\mathcal{I}^+)\qquad \mathcal{H}^-\equiv \dot{\Bar{J}}^+(\mathcal{I}^-).
\end{equation}

From these definitions, as a result, in order to accurately define the event horizon, we must know in advance how our spacetime will evolve infinitely in the future and, in particular, what will be the $\mathcal{I}^+$ of the manifold $M$. This definition of the event horizon has a mathematical nature but, as we shall show, other definitions exist that render a more physical and effective idea, when we want to describe physical processes and interactions with black holes.

\subsection{Schwarzschild black holes \label{sub:schw}}

The spherically symmetry static solution of the in vacuum Einstein equation describing a non-rotating mass $m$ is the Schwarzschild metric:
\begin{equation}
    ds^2=g_{\mu\nu}dx^\mu dx^\nu=-\left(1-\frac{2m}{r}\right)dt^2+\left(1-\frac{2m}{r}\right)^{-1}dr^2+r^2d\theta^2+r^2\sin^2\theta d\phi^2,
\end{equation}
where $r$ is the radial coordinate. For the \emph{Birkhoff's theorem}~\cite{birkhoff}, this is the unique spherically symmetric solution to the vacuum equation. Since this metric is time independent, $\partial/\partial t$ is a timelike Killing vector and is invariant under the group of isometries $SO(3)$ oparating on the spacelike two-spheres with $t,r$ constants. The solution is asymptotically flat, as for large $r$ the metric has the form $g_{\mu\nu}\simeq\eta_{\mu\nu}+O(\frac1r)$.

The metric presents two singularities, $r=0$ and $r=2m$, thus we must cut these points out of the manifold defined by the coordinates $(t,r,\theta,\phi)$. Cutting out the point $r=2m$ divides the manifold into two disconnected components:
\begin{itemize}
    \item $0<r<2m$;
    \item $r>2m$.
\end{itemize}
Since we took the spacetime manifold to be connected, we must consider only one component, and the obvious choice is the one that represents the external field, $r>2m$.

\begin{figure}
    \centering
    \subfloat[]{\includegraphics[scale =0.7]{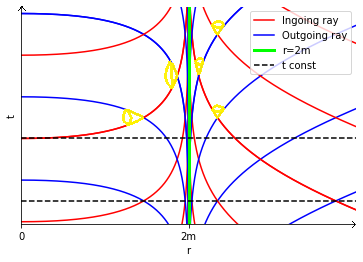}}\qquad
    \subfloat[]{\includegraphics[scale =0.7]{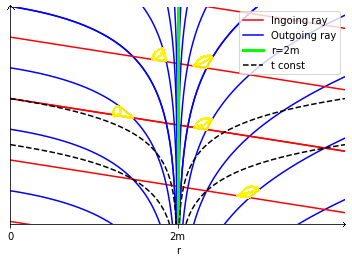}}
    \caption{\emph{Schwarzschild black hole. Constant section $(\theta,\phi)$ of the Schwarzschild solution. Each point represents a two-sphere of area $4\pi r^2$. Red lines: ingoing rays with $v=const$. Blue lines: outgoing rays with $w=const$. The solutions with $t= const$ and selected light cones (yellow) are shown. Notice the apparent singularity in $r=2m$. In (a) the coordinates $(t,r)$ are used, while in (b) the coordinates $(v,r)$.}}
    \label{fig:sch_eddfink}
\end{figure}

However, $r=2m$ is not a real physical singularity: the curvature does not diverge for $r\to 2m$, as it happens for $r\to0$. This divergence seems to be due to a bad choice of coordinates. Therefore, the manifold $M$
with the Schwarzschild metric $g$ is extendible, so that it is possible to map $M$ onto a bigger manifold $M'$, the Lorentz metric $g'$ on $M'$ is appropriate, and coincides with $g$ on the image of $M$. To confirm this, we define
\begin{equation}
    v\equiv t+r^*\qquad w\equiv t-r^*,
\end{equation}
with
\begin{equation}
    r^*\equiv\int\frac{dr}{1-2m/r}=r+2m\log\left|\frac{r}{2m}-1\right|,
    \label{eq:rindlernull}
\end{equation}
the so-called \emph{tortoise coordinate}. Thus, $v$ and $w$ are, respectively, the \emph{advanced null coordinate} and the \emph{retarded null coordinate}. In physical terms, as we can see in Fig.~\ref{fig:sch_eddfink}(a), these coordinates are the radial ingoing and outgoing trajectories of a beam of light: we can see that the curves does not cross the horizon $r=2m$. If we use the coordinates $(v, r,\theta,\phi)$, the metric takes the ingoing Eddington-Finkelstein~\cite{ed,fink} form
\begin{equation}
    ds^2=g'_{\mu\nu}dx'^\mu dx'^\nu=-\left(1-\frac{2m}{r}\right)dv^2+2dvdr+r^2d\theta^2+r^2\sin^2\theta d\phi^2.
    \label{eq:eddfinin}
\end{equation}
Thus, the metric $g'$ is no longer singular in $r=2m$: the manifold $M'$ is well defined in the region $r>0$. The region of $(M',g')$ for which $0<r<2m$ is isometric to the region of the Schwarzschild metric for which $0<r<2m$, and therefore we have extended the metric. Fig.~\ref{fig:sch_eddfink}(b) reports the Finkelstein diagram with these new coordinates. In the manifold $M'$, the surface $r=2m$ is a null surface \footnote{A null surface is a surface in which null geodesics travel along it. For more details~\cite{liberatilect, dowker}.} and here the parameter $t$ becomes infinity. The surface $r=2m$ acts like a one-way membrane: only the ingoing (from $r>2m$ to $0<r<2m$) future-directed timelike and null curves can cross this horizon.

The same reasoning can be followed for $w$ instead of $v$, thus using the coordinates $(w,r,\theta,\phi)$: one finds the extension $M''$ of $M$ with the outgoing Eddington-Finkelstein metric
\begin{equation}
    ds^2=g''_{\mu\nu}dx''^\mu dx''^\nu=-\left(1-\frac{2m}{r}\right)dw^2 -2dwdr+r^2d\theta^2+r^2\sin^2\theta d\phi^2,
\end{equation}
for $r>0$. This solution is the same as the previous one~\eqref{eq:eddfinin}, with the difference that the isometry between the region $0<r<2m$ in $M$ and the regions $0<r<2m$ in $M''$ reverses the direction of time.

\begin{figure}[ht]
    \centering
    \includegraphics[scale = 1]{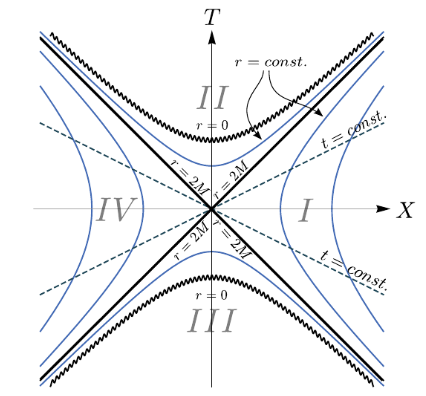}
    \caption{\emph{The Kruscal diagram. The coordinates $\theta$ and $\phi$ are suppressed; the wavy lines represent the singularity in $r=0$. The solutions with $r=const$ and $t=const$ are shown with blue and black dashed curve lines, respectively. The former are hyperbolas, while the latter are straight lines. The hyperbole's asymptotes are given by the straight lines $r=2m$, that are null curve lines and identify the horizons. From}~\cite{dowker}.}
    \label{fig:kruskal}
\end{figure}

The extentions $(M',g')$ and $(M'',g'')$ can be performed simultaneously: there is an even larger manifold $M^*$ still larger, with metric $g^*$, in which both the manifolds $M'$ and $M''$ can be embedded. This extension was given by Kruskal~\cite{PhysRev.119.1743}. He defined a new pair of coordinates $(W,V,\theta,\phi)$ in the region $r>2m$, so-called \emph{Kruskal-Szekeres coordinates}:
\begin{align}
    \label{eq:kruskalszekeres1}
    &W\equiv-\exp\left(-\frac{t-r^*}{4m}\right)=- \left|\frac{r}{2m}-1\right|^{1/2}\exp\left(-\frac{t-r}{4m}\right)\\
    &V\equiv\exp\left(\frac{t+r^*}{4m}\right)=\left|\frac{r}{2m}-1\right|^{1/2}\exp\left(\frac{t+r}{4m}\right).
    \label{eq:kruskalszekeres2}
\end{align}
For any value of $r$, $W<0$ and $V>0$. Knowing that $r$ is linked to $W,V$ through the relation
\begin{equation}
    WV=-\left(\frac{r-2m}{2m}\right)\exp\left(\frac{r}{2m}\right),
\end{equation}
one gets a new metric 
\begin{equation}
    ds^2=-\frac{32m^3}{r}\exp\left(-\frac{r}{2m}\right)dWdV+r^2d\theta^2+r^2\sin^2\theta d\phi^2,
    \label{eq:kruskal}
\end{equation}
which is well defined for all $r>0$. However, we notice that for $r<2m$ one must have $WV>0$, which is not possible. So, we forget for the moment about the definitions of the coordinates $W$ and $V$, and we study a new spacetime with the metric in Eq.~\eqref{eq:kruskal} and extended coordinates $W,V\in(-\infty,+\infty)$. For convention, we define two new coordinates $T=W+V$ and $X=W-V$. The manifold that we get is $M^*$ and the spacetime is shown in Fig.~\ref{fig:kruskal}.
The wavy lines correspond to the singularity in $r=0$. Here, the blue curve lines represent the solutions for $r=const$, that are hyperbolas; the dashed black lines are the solutions for $t=const$, that are straight lines. In this diagram, we can see that the spacetime is divided in four regions by the null curve lines $r=2m$:
\begin{itemize}
    \item Region I: it is defined for $X>|T|$, and is the region which was covered by the original Schwarzschild coordinates $(t,r,\theta,\phi)$. Physically, this region corresponds to $r>2m$ (thus, outside an event horizon).
    \item Region II: it is defined for $T>|X|$ and, along with region I, is the region covered by ingoing Eddington-Finkelstein coordinates. Physically, this region corresponds to the interior part of a black hole ($r<2m$ with the horizon which crosses only the ingoing radial null geodesics).
    \item Region III: it is defined for $T<|X|$ and, along with region I, is the region covered by outgoing Eddington-Finkelstein coordinates. Physically, this region corresponds to the interior part of a white hole ($r<2m$ with the horizon which crosses only the outgoing radial null geodesics). 
    \item Region IV: it is defined for $X<|T|$, and is covered only by Kruskal-Szekeres coordinates. It is a mirror image of region I.
\end{itemize}

The Kruskal extension $(M^*,g^*)$ is the unique analytic and locally inextendible extension of the Schwarzschild solution.

\subsubsection{Surface gravity}

Consider now the Killing vector $\xi=\frac{\partial}{\partial t}$. This can be written in terms of the exponentially null coordinates $W$ and $V$ as
\begin{equation}
    \xi=\frac{\partial}{\partial t}=\frac{\partial V}{\partial t}\frac{\partial }{\partial V}+\frac{\partial W}{\partial t}\frac{\partial}{\partial W}.
    \label{eq:killing}
\end{equation}
From Eq.~\eqref{eq:kruskalszekeres1} and~\eqref{eq:kruskalszekeres2} we have that
\begin{equation}
    W=- \left|\frac{r}{2m}-1\right|^{1/2}\exp\left(-\frac{t-r}{4m}\right)\qquad V=\left|\frac{r}{2m}-1\right|^{1/2}\exp\left(\frac{t+r}{4m}\right).
\end{equation}
We notice that the tortoise coordinates have a logarithmic divergence at the black hole horizon ($r=2m$), that is $W=0$ here. Considering $W=0$ and $\frac{\partial V}{\partial t}=\frac{V}{4m}$, Eq.~\eqref{eq:killing} becomes
\begin{equation}
    \xi=\frac{\partial}{\partial t}=\frac{\partial V}{\partial t}\frac{\partial }{\partial V}=\frac{V}{4m}\frac{\partial}{\partial V}.
\end{equation}
The acceleration per unit mass required to be still and hover just over the horizon, when measured from infinity, is called the \emph{surface gravity} $\kappa$. This is defined as
\begin{equation}
    \xi^\mu\nabla_\mu\xi^\nu\equiv\kappa\xi^\nu.
\end{equation}
In the Schwarzschild context, the surface gravity is thus $\kappa=\frac{1}{4m}$.

\subsection{Carter-Penrose diagrams}

In this section, we describe a practical technique for displaying the causal organization of an infinite spacetime on a limited amount of paper. Performing a conformal change on the metric is required~\cite{dowker}.
\begin{definition}
    A conformal transformation is a map $\psi:(M,g)\to(M,\Tilde{g})$ such that
    \begin{equation}
        \Tilde{g}_{\mu\nu}(x)=\Lambda^2(x)g_{\mu\nu}(x),
    \end{equation}
    where $\Lambda(x)$ is a smooth function of the spacetime coordinates which does not vanish for any $x$.
\end{definition}
The causal structure of a spacetime is preserved by conformal transformations.

The idea behind a Penrose diagram~\cite{PhysRevLett.10.66, Carter:1966zza} is the following. First, one has to bring the "infinity" to a finite coordinate distance, and here, the metric typically diverges. To fix this, we apply a conformal transformation on $g$ to produce a new, regular-on-the-edges metric called $\Tilde{g}$. In this manner, $(M,\Tilde{g})$ is a good representation of the original $(M,g)$: it has exactly the same causal structure. The points at infinity are often added to $M$ to create a new manifold, $\Tilde{M}$. The resutling spacetime $(\Tilde{M}, \Tilde{g})$ that results is what is referred to as the conformal compactification of $(M, g)$.
Although the conformally compactified spacetime $(\Tilde{M},\Tilde{g})$ provides a good representation of the causal structure of the physical spacetime $(M, g)$, it is not physical. Indeed, the curvature tensors are in general not preserved under conformal transformations.

\begin{figure}
    \centering
    \includegraphics[scale = 0.8]{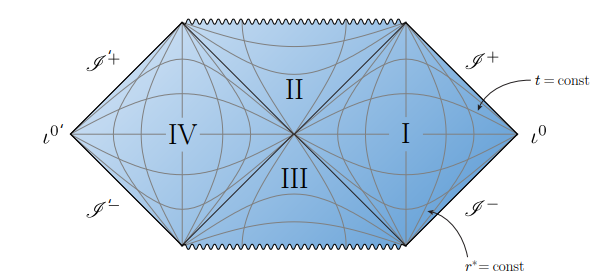}
    \caption{\emph{Carter-Penrose diagram for a Schwarzschild black hole. We can see that this diagram is still divided in four regions with the same correspondence as in the Kruskal diagram. The wavy lines represent the singularity in $r=0$. In the diagram, the solutions with $r=const$ and $t=const$ are shown as the vertical and horizontal lines, respectively. The lines indicated as {\calligra\footnotesize I}$\ \, ^\pm$ and {\calligra \footnotesize I}$\ \, '^\pm$ represent the past ($-$) and future ($+$) null infinity, thus the whole universe in displayed in a limited amount of paper. From}~\cite{liberatilect}.}
    \label{fig:pc_sc}
\end{figure}

One can apply this reasoning to the Kruskal space. The Kruskal metric is that in Eq.~\eqref{eq:kruskal}. We define a new set of null coordinates:
\begin{equation}
    \begin{cases}
        &W=\tan\Wt \\
        &V=\tan\Vt
    \end{cases}\qquad \mathrm{such\ that}\quad -\frac{\pi}{2}<\Wt,\Vt<\frac{\pi}{2}.    
\end{equation}
After applying a conformal transformation, the new metric is
\begin{equation}
    ds^2=\left[-4\frac{32m^3}{r}\exp\left(-\frac{r}{2m}\right)d\Wt d\Vt+r^2\cos^2\Wt\cos^2\Vt(d\theta^2+\sin^2\theta d\phi^2)\right],
\end{equation}
and we add the points at infinity. The obtained spacetime is displayed in Fig.~\ref{fig:pc_sc}.
The wide curve line is still the singularity in $r=0$. Note that this diagram is still divided in four region by the null curves identified by $r=2m$, each of which has still the same meaning as in Kruskal diagram in Fig.~\ref{fig:kruskal}. The lines indicated as {\calligra\footnotesize I}$\ \, ^\pm$ and {\calligra \footnotesize I}$\ \, '^\pm$ represent the past ($-$) and the future ($+$) null infinity, thus the whole universe in displayed in a limited amount of paper. The solutions with $r=const$ are represented by the vertical lines, while the solutions with $t=const$ are represented by horizontal lines. The light cones' form is unchanged and thus a light ray which comes from the past null infinity $\mathcal{I}^-$, and that collapses into the black hole up to the singularity in $r=0$, has a trajectory represented by a straight line parallel to that which separates the sectors I and II from the sectors III and IV (see Fig.~\ref{fig:pc_sc}). From this diagram, we understand what really are all these infinities that we have defined in Sec.~\ref{subsub:eh}. The future time infinity is a point where every timelike trajectory outside the black hole go, i.e. if we stay at $r>2m$ fixed, for $t\to\infty$ we reach $\iota^+$. The past time infinity $\iota^-$ is the time reversed of $\iota^+$. The space infinity is the point which collects all the spacelike trajectories with time $t$ fixed at $r\to \infty$. The future null infinity, $\mathcal{I}^+$, is the set of points in which all the outgoing rays fall; the past null infinity, $\mathcal{I}^-$, is the set of points from which all the ingoing rays start.

\subsection{Trapped surfaces and apparent horizon}

We now introduce, in a brief and intuitive manner, the concepts of trapped surfaces and apparent horizon.

\subsubsection{Trapped surfaces}

\begin{figure}
\centering

\raisebox{-\height}{%
  \begin{tabular}{@{}c@{}}
  \subfloat[]{%
    \includegraphics[width=0.35\textwidth]{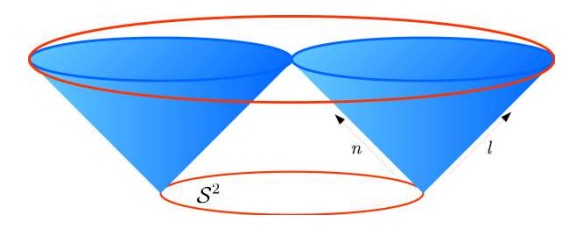}%
    \label{tr1}%
  } \\
  \subfloat[]{%
    \includegraphics[width=0.35\textwidth]{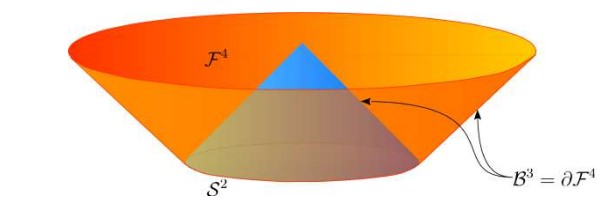}%
    \label{tr2}%
  }
  \end{tabular}%
}\qquad
\raisebox{-\height}{%
  \subfloat[]{%
     \includegraphics[height=6cm]{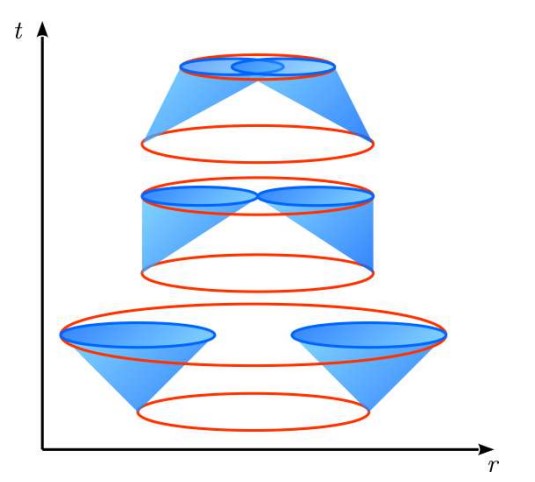}%
     \label{ts3}%
  }%
}
\caption{\emph{Trapped surfaces. Light cones in different spacetimes. For each image, the vertical axis is time, and a $t$ constant slice is the radius and one angle. (a) and (b) Light cones in a flat spacetime are shown. The vectors $n$ and $l$ are null vectors and identify, respectively, the ingoing and outgoing directions of emitted light rays. (c) Light cones in a curved spacetime, to show what trapped surfaces are. From }\cite{liberatilect}.}
\label{fig:tr}

\end{figure}

Consider the manifold shown in Fig.~\ref{fig:tr}(a).
In the vertical axis there is the time, and on a $t$-constant slice there is the radius and one angle. From each point of the initial spacelike 2-surface $\mathcal{S}^2$, the null rays can be emitted along the vector $n$ as ingoing light rays or along vector $l$ as outgoing light rays. Obviously, $l$ and $n$ are null vectors, so they are both orthogonal to $\mathcal{S}^2$. 

In a flat Lorentzian spacetime, all of the incoming light beams concentrate into a single point, creating a sort of cone with a "bowl" constructed with the departing light rays surrounding it, as shown in Fig.~\ref{fig:tr}(b).

In spacetime with gravity, as for example with Eddington-Finkelstein coordinates, wee can see in Fig.~\ref{fig:sch_eddfink}(b) that the light cones are not always straight, as they get close to the horizon they begin to tilt. 
This observation can be used to figure out the meaning of trapped region. Timelike observers must move through their own light cones, and if they are sufficiently slanted during a gravitational collapse, the region $\mathcal{S}^2$ will eventually be causally related to smaller and smaller regions, as shown in Fig.~\ref{fig:tr}(c). This corresponds to having vectors $n$ and $l$ converge: we have a \emph{trapped surface}.

Thus, trapped surfaces are closed spacelike surfaces such that their area tends to decrease locally along any possible future direction~\cite{2011IJMPD..20.2139S}.

A \emph{trapped region} on a 3-D hypersurface $\Sigma$ is the set of all the points in $\Sigma$ through which a trapped surface passes.

\subsubsection{Apparent horizon}

An \emph{apparent horizon} $\mathcal{A}$ is the boundary of a total trapped region on $\Sigma$. 

What is the difference between the apparent horizon and the event horizon? For a stationary black hole, the two horizons coincide $\mathcal{A}=\mathcal{H}$, but this is not true in general. Indeed, for example, during a collapse, $\mathcal{A}\subset\mathcal{H}$, i.e. the apparent horizon is contained within the event horizon. The main difference between the two concepts is that to define the event horizon we need to know $J^-(\mathcal{I}^+)$, i.e. it is the boundary of a region through which at any time anything cannot escape. The apparent horizon is instead a local concept, i.e. it is the boundary of a region through which at some time anything can’t escape.

\section{Quantum field theory in curved spacetimes}

The aim of this section is to introduce quantum field theory in non-trivial metric background~\cite{CBO9780511622632, viat, Jacobson:2003vx}, and to apply this result to explain the Unruh effect~\cite{unruh, unruh2} and the Hawking radiation~\cite{hawk1,hawk2, hawk3}.

\subsection{Second quantization in curved spacetime}

In this section we want to briefly introduce the second quantization in a curved spacetime. In the following, we use the particle-physics signature $(-,+,+,+)$. 

Consider a real scalar field $\phi(x)$ minimally coupled to gravity. To obtain the action we replace in the action, for a scalar field in Minkowski spacetime
\begin{equation}
    \begin{split}
        &\eta^{\mu\nu}\to g^{\mu\nu}\\
        &\partial_\mu\to\nabla_\mu\\
        &d^4x\to\sqrt{-g}d^4x,
    \end{split}
\end{equation}
so that one assures that the equation of motion reduces to the Klein-Gordon equation in flat space. As a result, the non-interacting Lagrangian density is as follows:
\begin{equation}
    \lag=\frac{\sqrt{-g}}{2}\left(g^{\mu\nu}\nabla_\mu\phi\nabla_\nu\phi-m^2\phi^2\right),
\end{equation}
with the corresponding equation of motion 
\begin{equation}
    (\Box+m^2)\phi(x)=0,
\end{equation}
where $\Box=\nabla^\mu\partial_\mu$ is the d'Alembertian operator holding for scalar fields.
The canonical quantization of $phi$, which is now an operator, is what we do next. To do that, we foliate the manifold $M$, and use the Arnowitt-Deser-Misner (ADM) formalism~\cite{Arnowitt:1962hi} to write the metric $g_{\mu\nu}$. The conjugate momentum $\pi$ is
\begin{equation}
    \pi(x)\equiv\frac{\partial \lag}{\partial(\nabla_t\phi)}=\sqrt{-g}g^{t\mu}\nabla_\mu\phi,
\end{equation}
which reduces to the standard expression in flat spacetime. The commutation relations are as follows:
\begin{align}
    &[\phi(t,\bm x),\pi(t,\bm y)]=\frac{i}{\sqrt{-g}}\delta^{(3)}(\bm x - \bm y)\\
    &[\phi(t,\bm x),\phi(t,\bm y)]=[\pi(t,\bm x),\phi(t,\bm y)]=0.
\end{align}
If we consider a foliation $\Sigma_t$ at time $t$, then the scalar product between two solutions of the Klein-Gordon equation $\phi_1,\phi_2$, which is independent of the specific $\Sigma_t$ that we choose, is defined as
\begin{equation}
    (\phi_1,\phi_2)\equiv-i\int_{\Sigma_t}\left(\phi_1\nabla_\mu\phi_2^*-\phi_2^*\nabla_\mu\phi_1\right)n^\mu\sqrt{h}d^3x=-i\int_{\Sigma_t}\left(\phi_1 \overleftrightarrow{\nabla}_\mu\phi_2^*\right)n^\mu\sqrt{h}d^3 x,
\end{equation}
where $h$ is the spatial metric which describes $\Sigma_t$, and $n^\mu$ is the unit normal to the surface. This scalar product defines the negative and positive norm modes that form a complete orthonormal basis $\{f_i\}$ of the solutions of the Klein-Gordon equation solutions:
\begin{equation}
    \begin{split}
        &(f_i,f_j)=\delta_{ij}\qquad(f_i^*,f_j^*)=-\delta_{ij}\qquad(f_i,f^*_j)=(f^*_i,f_j)=0\quad \forall i,j\\
        &\mathrm{such\ that} \qquad \partial_t f=-i\omega f\qquad \partial_tf^*=+i\omega f^*.
    \end{split}
\end{equation}
If we now define the creation and annihilation operators $\opa_i,\ \opa_i^\dagger$, such that satisfy the commutation relations
\begin{equation}
    [\opa_i,\opa_j^\dagger]=\delta_{ij}, \qquad [\opa_i,\opa_j]=[\opa_i^\dagger,\opa_j^\dagger]=0,
\end{equation}
the field $\phi$ can be written in terms of the basis $\{f_i\}$ as
\begin{equation}
    \phi(x)=\sum_i\left[\opa_if_i(x)+\opa_i^\dagger f_i^*(x)\right].
    \label{eq:scalarfield1}
\end{equation}
These operators define a vacuum state in this basis such that
\begin{equation}
    \opa_i|0\rangle_f=0 \quad\forall\ i\,.
\end{equation}
We can thus build the Fock space with $n$-particles states for the mode $i$, and also the \emph{number operator} $\hat{N}_{i|f}$ as:
\begin{equation}
    |n_i\rangle_f\equiv\frac{1}{\sqrt{n!}}\left(\opa_i^\dagger\right)^n|0\rangle_f\qquad \hat{N}_{i|f}\equiv\opa_i^\dagger\opa_i.
\end{equation}
The choice of the basis is not unique. We can consider a different set of solutions of the Klein-Gordon equation, $\{g_A\}$ (with indices $A,B,..$), with a corresponding different set of annihilation and creation operators, $\opb_A,\opb_A^\dagger$, respectively, such that the scalar field $\phi$ can be cast in the form:
\begin{equation}
    \phi(x)=\sum_A\left[\opb_Ag_A(x)+\opb_A^\dagger g_A^*(x)\right].
    \label{eq:scalarfield2}
\end{equation}
The corresponding vacuum $|0\rangle_g$ is different from the one defined before, $|0\rangle_f$, since it is defined as the state annihilated by the set of operators $\opb_A$ for any $A$. However, the field $\phi$ is the same, only the way we expand it is different: thus, there must be a relation between the two bases. These relations are called \emph{Bogoliubov transformations}~\cite{CBO9780511622632}, and are linear relations between $g_A$ and $f_i$. Upon comparing Eq.~\eqref{eq:scalarfield1} with Eq.~\eqref{eq:scalarfield2}, we obtain that
\begin{align}
    \label{eq:bogol1}
    &g_A=\sum_i\left[\alpha_{Ai}f_i+\beta_{Ai}f^*_i\right]\\
    \label{eq:bogol2}
    &f_i=\sum_A\left[\alpha^*_{Ai}g_A-\beta_{Ai}g^*_A\right],
\end{align}
where $\alpha_{Ai}$ and $\beta_{Ai}$ are called Bogoliubov coefficients. They are explicitly given by
\begin{equation}
    \alpha_{Ai}=(g_A,f_i),\qquad \beta_{A_i}=-(g_A,f_i^*),
    \label{eq:coefficients}
\end{equation}
and satisfy the following normalization conditions
\begin{align}
    \label{eq:coeffrelations1}
    \sum_k\left(\alpha_{Ak}\alpha_{ik}^*-\beta_{Ak}\beta_{ik}^*\right)=\delta_{Ai},\\
    \label{eq:coeffrelations2}
    \sum_k\left(\alpha_{Ak}\beta_{ik}-\beta_{Ak}\alpha_{ik}\right)=0.
\end{align}
As a consequence, the creation and annihilation operators can be expressed as
\begin{align}
    \opa_i=\sum_A\left[\opb_A\alpha_{Ai}+ \opb^\dagger_A\beta_{Ai}^*\right]\\
    \opb_A=\sum_i\left[\opa_i\alpha^*_{Ai}-\opa^\dagger_i\beta^*_{Ai}\right].
\end{align}
Since the two vacua are not equivalent, it could be that the vacuum relative to the basis $\{f_i\}$, which is not annihilated by the $b_A$ operators. This means that the observer in the basis $\{g_a\}$ sees a number of particles in the vacuum of the observer in the basis $\{f_i\}$ which is different from zero. Indeed,
\begin{equation}
    _f\langle0|\hat{N}_{A|g}|0\rangle_f=\ _f\langle0|\opb_A^\dagger\opb_a|0\rangle_f=\sum_{ij}\beta_{Ai}\beta_{Aj}^* \ _f\langle0|\opa_i \opa_j^\dagger|0\rangle_f=\sum_i|\beta_{Ai}|^2,
\end{equation}
where the $\beta_{Ai}$ coefficients are defined in Eq.~\eqref{eq:coefficients}. From Eq.~\eqref{eq:coeffrelations1} and~\eqref{eq:coeffrelations2}, it is clear thet $\beta_{Ai}$ "mixes" the annihilation operator of one basis with a creation operator of the other basis. Notice that although the concept of "particle" is Lorentz invariant, it is not covariant in general: if $\beta_{Ai}\ne0$, the mixing of positive and negative energy modes in the basis change suggests that the observer associated with the quantization in the $\{g_A\}$ basis perceives the vacuum described by one observer using the $\{f_i\}$ basis, as if it were filled with particles. 

What we have just exposed generates two very important phenomena: the Unruh effect and Hawking radiation. In the following, we analyze them in detail.

\subsection{Unruh effect \label{sub:unruh}}

The Unruh effect~\cite{liberatilect, unruh2} represents the fact that Rindler observers, i.e. observers moving with uniform acceleration $a_p$ in Minkowski spacetime, associate a thermal bath of Rindler particles at a temperature
\begin{equation}
    k_BT_U=\frac{\hbar a_p}{2\pi c},
\end{equation}
 to the Minkowski vacuum, i.e. the no-particle state of inertial observers.
To understand the mechanism, we focus on the $1+1$-dimensional case. If we consider a flat spacetime with metric $g_{\mu\nu}=\mathrm{diag}(1, -1)$, then
\begin{equation}
    ds^2=dt^2-dx^2.
\end{equation}
An observer moving with velocity $u^\mu$, and having constant proper acceleration $a_p$, such that
\begin{equation}
    u^\mu\nabla_\mu u^\nu=a_p^\nu,
\end{equation}
lies along the hyperbola $x^2-t^2=a_p^{-2}$. We can perform the change of coordinates $(x,t)\to(\xi,\eta)$, such that the $\xi=\mathrm{const}$ lines are the hyperbola
\begin{equation}
    t=\frac{e^{a\xi}}{a}\sinh(a\eta)\qquad x=\frac{e^{a\xi}}{a}\cosh(a\eta),
    \label{eq:coordinates1}
\end{equation}
with $a$ a parameter. Indeed:
\begin{equation}
    x^2-t^2=\left[\frac{e^{a\xi}}{a}\right]^2=\left[\frac{1}{a_p}\right]^2.
\end{equation}
\begin{figure}[ht]
    \centering
    \includegraphics[scale=0.8]{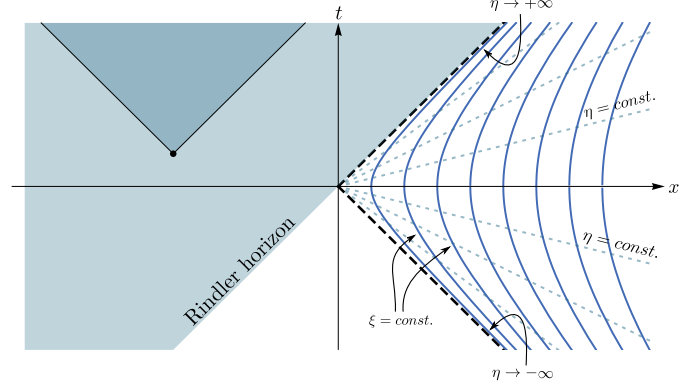}
    \caption{\emph{Unruh effect. The Rindler space is visually represented by the dashed black null lines encircling the right wedge. Within this space, the dashed blue lines are annotated with constant values of $\eta$, while the blue lines trace the worldlines of observers who are not in an inertial state. It is important to note that certain events within the darkened area, such as the black dot and others, remain concealed from these non-inertial observers due to the effects of their acceleration. The demarcation between the shaded and un-shaded regions is referred to as the Rindler horizon, denoting a crucial boundary that separates different aspects of the spacetime experienced by these observers. This illustration provides insights into the concepts of the Rindler wedge and the associated elements, shedding light on the interplay between acceleration, worldlines, and concealed events. The information source for this depiction is}~\cite{dowker}.}
    \label{fig:rindler}
\end{figure}
In Fig.~\ref{fig:rindler} the permitted surfaces for $\eta=const$ are straight lines with different angles emerging from the origin; for this reason, the variable $\eta$ is called \emph{the hyperbolic opening angle of the Rindler wedge} $x>|t|$.
With these new coordinates, the metric becomes
\begin{equation}
    ds^2=e^{2\xi a}(d\eta^2-d\xi^2).
\end{equation}
As we can see in Fig.~\ref{fig:rindler}, the coordinates in Eq.~\eqref{eq:coordinates1} does not cover the full Minkowski space, only the right side where $x>|t|$. Thus, the Rindler observer sees an horizon in $t=x$ (or $\eta\to+\infty$), which is generated by a Killing vector 
\begin{equation}
    \chi\equiv\frac{\partial}{\partial\eta}=a\left(x\frac{\partial}{\partial t}+t\frac{\partial}{\partial x}\right).
\end{equation}
The surface gravity is
\begin{equation}
    \partial_\mu\chi^2=-2\kappa\chi_\mu|_{\mathcal{H}^+}.
\end{equation}
Changing again the coordinates $(\xi, \eta)\to(u,v)$, so that
\begin{equation}
    u=\eta-\xi\qquad v=\eta+\xi
\end{equation}
are the null coordinates, the metric is
\begin{equation}
    ds^2=e^{a(v-u)}dudv.
\end{equation}

We now want to quantize the fields in the Rindler wedge. In Minkowski coordinates, a free massless scalar field $\phi$ can be expanded as
\begin{equation}
    \phi(t,x)=\int_{\mathbb{R}}(\opa_k\bu_k+\opa^\dagger_k\bu_k^*)dk\qquad\mathrm{with}\quad \bu_k=\frac{e^{ikx-i\omega t}}{\sqrt{(2\pi)2\omega}},
    \label{eq:phim}
\end{equation}
where $\opa_k$ and $\opa_k^\dagger$ are the destruction and creation operators, respectively,~\cite{CBO9780511622632,peskin} associated to the basis. In~\eqref{eq:phim}, $\omega=|k|$ and $\partial_t\bu_k=-i\omega\bu$, so that these are positive modes with respect to the global timelike Killing vector $\partial_t$. As mentioned before, The Minkowski vacuum is the no-particle state with respect to the inertial observer. It is defined as
\begin{equation}
    |0_M\rangle,\qquad\mathrm{such\ that}\quad \opa_k|0_M\rangle=0
\end{equation}

\begin{figure}
    \centering
    \includegraphics[scale=0.9]{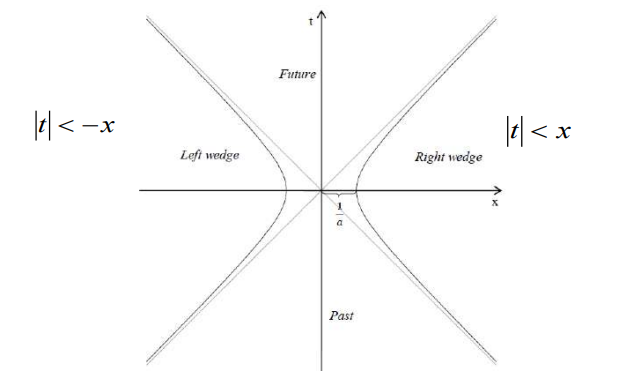}
    \caption{\emph{Minkowski spacetime. The light cone is shown, with indicated the wedges used in Rindler coordinates. From}~\cite{liberatilect}.}
    \label{fig:rindler2}
\end{figure}
Quantization is more delicate in Rindler coordinates, because the latter cover only a region in the right side (see Eq.~\eqref{eq:coordinates1} and Fig.~\ref{fig:rindler}). Thus, we need to patch the full Minkowski space.
Since the future and past regions (Fig.~\ref{fig:rindler2}) are causally connected with the Rindler wedges, in that region one can decompose the modes as combinations of the wedges one. As a result, we need to cover only the left wedge, that can be pathced considering $x\to-x$ in Eq.~\eqref{eq:coordinates1}: in this way we have a complete basis for the scalar field $\phi$ which constrains right and left modes. The scalar-field quantization can then be written as
\begin{equation}
    \phi(\eta,x)=\int_{\mathbb{R}}\left[\left( \opb_k^Lu_k^L+\opb^{L\dagger}_ku_k^{L*} \right)+\left( \opb_k^Ru_k^R+\opb^{R\dagger}_ku_k^{R*} \right)\right]dk, 
\end{equation}
with
\begin{equation}
    \begin{cases}
        u_k^L=\frac{e^{ikx+i\bomega\eta}}{\sqrt{(2\pi)2\bomega}}\ & \ x<-|t|\\
        u_k^R=\frac{e^{ikx-i\bomega\eta}}{\sqrt{(2\pi)2\bomega}}\ & \ x>|t|\\
        0 & \mathrm{otherwise},
    \end{cases}
\end{equation}
where $\opb_k^{R/L}$ and $\opb_k^{RL\dagger}$ are the destruction and creation operators, respectively, associated to the basis, and $\bomega=|k|$ in this case as well. In addition, $u_k^R$ and $u_k^L$ are positive norm-modes with respect to the Killing vectors $+\partial_\eta$ and $-\partial_\eta$, in the right wedge and in the left wedge, respectively. Thus: 
\begin{equation}
    \partial_\eta u_k^R=-i\bomega u_k^R,\qquad -\partial_\eta u_k^L=-i\bomega u_k^L.
\end{equation}
Finally, we can define the Rindler vacuum as the state of no-particle with respect to the Rindler observer, and it can be written as
\begin{equation}
    |0_R\rangle, \qquad \mathrm{such\ that}\quad \opb_k^{R/L}|0_R\rangle=0.
\end{equation}

The quantity of particles in the Minkowski vacuum as determined by an accelerating observer is what we are trying to calculate. Indeed, the vacuum states are not the same. This is evident from the fact that the functions $u_k^R$ do not smoothly go over to $u_k^L$, as one passes from the right to the left wedge, and so the modes are not analytic in the origin. Instead, the Minkowski modes are analytic, and this analytic-like property remains true of any linear superposition of these positive-frequency Minkowski modes. Thus, Rindler modes must contain negative- and positive-frequency Minkowski modes, meaning that the two observers cannot have the same vacuum.
We now have to identify the Bogoliubov transformation~\cite{CBO9780511622632} between the two sets of modes. Following the approach proposed by Unruh~\cite{unruh}, one finds an analytic combination of $u_k^L$ and $u_k^R$, which shares the positive-frequency analytic-like property of the Minkowsky modes $\bu_k$, so that they have to share also a common vacuum state $|0_M\rangle$. The combination is:
\begin{align}
    &f_k^{(1)}\equiv\frac{1}{N_k}e^{\frac{\pi\bomega}{2a}}u_k^R+e^{-\frac{\pi\bomega}{2a}}u_{-k}^{L*}\\
    &f_k^{(2)}\equiv\frac{1}{N_k}e^{-\frac{\pi\bomega}{2a}}u_{-k}^{R*}+e^{\frac{\pi\bomega}{2a}}u_{k}^{L},
\end{align}
with $N_k$ a normalization coefficient. Defining $\opd^{(j)}_k$, $j=1,2$, as the operators associated to the new basis, Eq.~\eqref{eq:phim} can be recast as
\begin{equation}
    \phi(\eta,\xi)=\int_{\mathbb{R}}\left[2\sinh\left(\frac{\pi\bomega}{a}\right)\right]^{-1/2}\Big\{\left[ \opd_k^{(1)}f_k^{(1)}+\opd_k^{(1)\dagger}f_k^{(1)*}\right]+\left[ \opd_k^{(2)}f_k^{(2)}+\opd_k^{(2)\dagger}f_k^{(2)*}\right] \Big\}dk.
\end{equation}
Exploiting the following scalar products:
\begin{equation}
    \begin{split}
        &(f_k^{(1)},u_{k'}^R)=\frac{e^{\frac{\pi\bomega}{2a}}}{N_k}\delta_{k,k'}\qquad (f_k^{(2)*},u_{k'}^R)=\frac{e^{-\frac{\pi\bomega}{2a}}}{N_k}\delta_{-k,k'}\\
        &(f_k^{(1)*},u_{k'}^R)=0\qquad\qquad\ \ (f_k^{(2)},u_{k'}^R)=0,
    \end{split}
\end{equation}
we obtain the Bogoliubov relations between the operators $\opb_k$ and $\opd_k$
\begin{align}
    &\opb_k^L=\left[2\sinh\left(\frac{\pi\bomega}{a}\right)\right]^{-1/2}\Big\{e^{\frac{\pi\bomega}{2a}}\opd_k^{(2)}+e^{-\frac{\pi\bomega}{2a}}\opd_{-k}^{(1)\dagger}\Big\} \\
    &\opb_k^R=\left[2\sinh\left(\frac{\pi\bomega}{a}\right)\right]^{-1/2}\Big\{e^{\frac{\pi\bomega}{2a}}\opd_k^{(1)}+e^{-\frac{\pi\bomega}{2a}}\opd_{-k}^{(2)\dagger}\Big\}.
\end{align}
We can thus compute the number of particles measured by an accelerating observer in the Minkowski vacuum:
\begin{equation}
    \langle0_M|\opb_k^{R\dagger}b_k^{R}|0_M\rangle=\frac{1}{e^{\frac{2\pi\bomega}{a}}-1}.
\end{equation}
This is the Planck spectrum for radiation at temperature $T_U=\frac{a}{2\pi k_B}$.

\subsection{Hawking radiation}

The first hint that balck holes can radiate was put forward by Yakov Borisovich Zeldovich in 1971~\cite{zeldovich}, even if, as Kip Thorne writes in his divulgative book 'Black Holes \& Time Warps: Einstein's Outrageous Legacy'~\cite{kthorne}, nobody paid any attention. According to Zeldovich, much like a spinning metal sphere emits electromagnetic radiation when electromagnetic vacuum fluctuation tickle it, so similarly a spinning black hole should emit gravitational waves when gravitational vacuum fluctuations graze its horizon. 
It was in september 1973 when Stephen Hawking met Zeldovich and found fascinating his conjecture. In 1975, Hawking demonstrated that any stationary black hole emits particles with a thermal spectrum~\cite{hawk1}.

In the following, we would like to expose the essential traits of this fascinating effect and one of the problems that this phenomena opens: the \emph{Information Paradox}~\cite{liberatilect, Mathur:2009hf}.

Consider a $3+1$-dimensional spacetime with a massless scalar field $\phi$ in a spherically symmetric collapsing star geometry. Let us assume that the field $\phi$ represents the radiation, and that the same result holds in presence of any type of field. Even if the formation of a black hole involves complicated dynamics, in the far asymptotic past $\mathcal{I}^-$ and future $\mathcal{I}^+$ the spacetime is approximately stationary, therefore we can perform a second quantization associated with stationary observers near $\mathcal{I}^-$, and a second quantization associated with stationary observers near $\mathcal{I}^+$.
The second quantization in the far past leads to the $in$-vacuum. This latter can be built after specifying a complete set of positive-frequency modes on $\mathcal{I}^-$. For the quantization in the far future, we have to consider $\mathcal{H}^+\bigcup\,\mathcal{I}^+$ and specify a complete set on it, since only $\mathcal{I}^+$ is not a Cauchy surface. Therefore, we define a set of positive frequencies on $\mathcal{I}^-$, $\{f_i\}$, a set of positive frequencies on $\mathcal{I}^+$ and zero on $\mathcal{H}^+$, $\{g_i\}$, and a set of positive frequencies on $\mathcal{H}^+$ and zero on $\mathcal{I}^+$, $\{h_i\}$. To link these two bases, we leverage the Bogoliubov transformations~\eqref{eq:bogol1},~\eqref{eq:bogol2}. The sets $\{f_i,f_i^*\}$ and $\{g_i,g_i^*,h_i,h_i^*\}$ are complete. Thus, defining the creation and annihilation operators from the two basis, we can expand the the field $\phi$ in two ways
\begin{equation}
    \phi(x)=\sum_i\opa_if_i(x)+a_i^\dagger f_i^*(x)=\sum_i \opb_i g_i(x)+\opc_i h_i(x)+\opb^\dagger_i g^*_i(x)+\opc_i^\dagger h_i^*(x).
\end{equation}
Through the basis $\{f_i\}$, we can define the $in$-vacuum as
\begin{equation}
    a_i|0_{in}\rangle=0 \ \forall\ i.
\end{equation}
As in Sec~\ref{sub:unruh}, we want to understand if an observer on $\mathcal{I}^+$ sees a number of particles different from zero in the vacuum of $\mathcal{I}^-$. To do this, we \emph{trace back in time} the solution of the Klein-Gordon equation in Schwarzschild spacetime from $\mathcal{I}^+$ to $\mathcal{I}^-$.

Given the coordinates $(t, r^*,\theta,\varphi)$, where $r^*$ is defined in Eq.~\eqref{eq:rindlernull}, the Schwarzschild metric becomes
\begin{equation}
    ds^2=\left(1-\frac{2m}{r}\right)\left(-dt^2+dr^{*2}\right)+r^2 d\Omega^2,
\end{equation}
with $d\Omega^2=d\theta^2+\sin^2\theta d\varphi^2$. In these coordinates and with this metric, the Klein Gordon equation reads
\begin{equation}
    \left[\partial_t^2-\partial_{r^*}^2+V_l(r^*)\right]\chi_l=0,
    \label{eq:motohawk1}
\end{equation}
where we have expanded the solution for the Klein-Gordon in spherical harmonics \\
$\phi(t,r^*,\theta,\varphi)=\chi_l(r^*,t)Y_{lm}(\theta,\varphi)$, and we have defined the potential 
\begin{equation}
    V_l(r)=\left(1-\frac{2m}{r}\right)\left[\frac{l(l+1)}{r^2}+\frac{2m}{r^3}\right].
\end{equation}
This latter represents the potential barrier that a particle has to climb, in order to escape from gravitational attraction of the collapsing star.
If we set $\chi_l(r^*,t)=e^{-i\omega t}R_{l\omega}(r^*)$, we get
\begin{equation}
    \left(\partial_{r^*}^2+\omega^2\right)R_{\omega l}=V_l R_{\omega l}.
    \label{eq:motohawk}
\end{equation}

\begin{figure}
    \centering
    \includegraphics[scale =0.8]{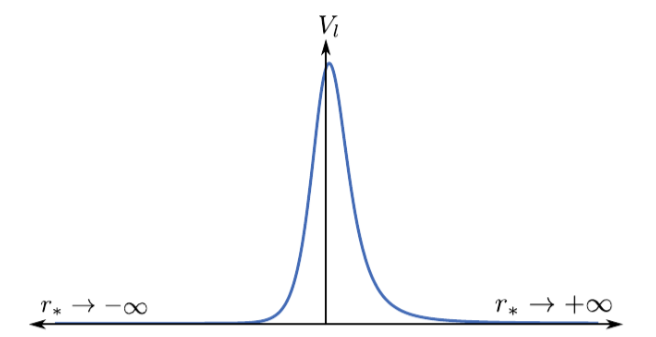}
    \caption{\emph{Hawking radiation. The potential $V_l(r^*)$ for $l=1$. We can see that for $r^*\to\pm\infty$, i.e. at the horizon and at $\mathcal{I}^\pm$, the potential vanishes and presents instead a barrier in $r^*\to0$. From}~\cite{dowker}.}
    \label{fig:vl}
\end{figure}
From Fig.~\ref{fig:vl}, we can see that the potential $V_l$ vanishes near the horizon ($r^*\to-\infty$) and near $\mathcal{I}^\pm$ ($r^*\to\infty$), thus if we consider how any solution of~\eqref{eq:motohawk} can evolve in time in presence of a potential barrier such as that shown in Fig.~\ref{fig:vl}, we imagine that the solution is partially transmitted and partially reflected.

Since for $r^*\to+\infty$ the potential $V_l(r^*)$ vanishes, the solutions to~\eqref{eq:motohawk1} near $\mathcal{I}^\pm$ are just plane waves. Indicating with "+" the outgoing modes and with "-" the ingoing modes, at $\mathcal{I}^-$ we can define the \emph{early modes} 
\begin{equation}
    f_{lm\omega+}=\frac{1}{\sqrt{2\pi\omega}}e^{-i\omega u}\frac{Y_{lm}}{r}\qquad f_{lm\omega-}=\frac{1}{\sqrt{2\pi\omega}}e^{-i\omega v}\frac{Y_{lm}}{r},
\end{equation}
and at $\mathcal{I}^+$ the \emph{late modes}
\begin{equation}
    g_{lm\omega+}=\frac{1}{\sqrt{2\pi\omega}}e^{-i\omega u}\frac{Y_{lm}}{r}\qquad g_{lm\omega-}=\frac{1}{\sqrt{2\pi\omega}}e^{-i\omega v}\frac{Y_{lm}}{r}.
\end{equation}
Since we will be interested in ingoing early and outgoing late modes, we can use the shorthand notation
\begin{equation}
    f_\omega=f_{lm\omega-}\qquad g_\omega=g_{lm\omega+}.
\end{equation}
We now want to understand which are the Bogoliubov coefficients that identify the transformation expressing $g_\omega$ in terms of $f_{\omega'}$ and $f^*_{\omega'}$ on $\mathcal{I}^-$, i.e.
\begin{equation}
    g_i=\sum_j A_{ij}f_j+B_{ij}f^*_j.
\end{equation}
The waves $g_\omega$ are plane waves, thus completely delocalised on $\mathcal{I}^+$. By using the superposition principle, we can construct a localised wave-packet peaked at some $\omega_0$. While the wave travels inwards from $\mathcal{I}^+$, $r^*$ decreases, and the wave-packet encounters the potential barrier. Part of the wave is reflected, $g_\omega^{(r)}$, and comes back to $\mathcal{I}^+$ with the same frequency $\omega$: this corresponds to consider $A_{\omega\omega'}\propto\delta(\omega-\omega')$. The part transmitted through the barrier, $g_\omega^{(t)}$, instead, enter the collapsing matter. Here, we do not know precisely the spacetime geometry, since we are interested in a packet peaked at large $u_0$ and finite $\omega_0$, because of the gravitational blue-shift, we know that the packet is peaked at a very high frequency as it enters the collapsing matter. This implies that the wave takes the form $g_\omega\sim A(x)e^{iS(x)}$ with $A(x)$ varying very slowly with respect to $S(x)$, as the packet obeys the \emph{geometric optics approximation}. The possible solutions for such a wave are $\nabla^\mu S\nabla_\mu S=0$, i.e.. the surfaces of constant phase are null. Therefore, we can trace back in time the surfaces of a given wave with constant phase, by following null geodesics.

\begin{figure}
    \centering
    \includegraphics[scale = 0.9]{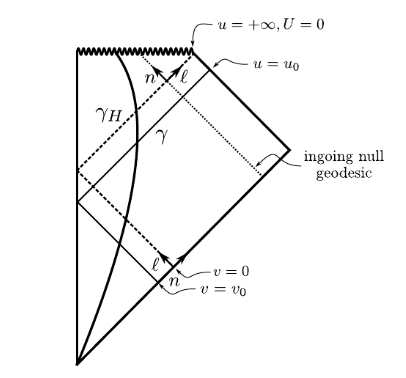}
    \caption{\emph{Hawking radiation. Carter-Penrose diagram for a collapsing star with spherical symmetry. The curve $\gamma_H$ lies on the future event horizon $\mathcal{H}^+$ and has been extended into the past until it crosses $\mathcal{I}^-$ for $v=0$. The geodesic $\gamma$ is infinitesimally close to $\gamma_H$, starts from $\mathcal{I}^-$ with advanced time $v_0$ and ends in $\mathcal{I}^+$ with retarded time $u_0$. From}~\cite{dowker}.}
    \label{fig:pc_hawk}
\end{figure}
Consider now two curves $\gamma_H$ and $\gamma$, as shown in Fig.~\ref{fig:pc_hawk}. The first is the null generator of the horizon $\mathcal{H}^+$ which has been extended into the past until it crosses $\mathcal{I}^-$ for $v=t+r^*$ fixed at $v=0$. The second is a geodesic which starts from $u_0$ in $\mathcal{I}^+$ and reaches $v_0$ in $\mathcal{I}^-$, which is infinitesimally close to $\gamma_H$ and $v_0<0$. Still considering Fig.~\ref{fig:pc_hawk}, one can take a connecting vector $n$ between the two curves and a generator vector of the Killing horizon $\ell$ such that $n\cdot\ell=-1$. To understand which is the form of $g_\omega$ at $\mathcal{I}^-$, we need to find how the affine distance along the vector $n$ changes while going into the past back to the time in which $\gamma$ reaches $\mathcal{I}^-$. Since $u$ is not well defined in correspondence of the horizon, we have to consider the Kruskal coordinate $U=-e^{-\kappa u}$, where $\kappa=1/4m$ is the surface gravity. This establishes an affine distance along $n$, that can be leveraged to compute the distance between the two curves. The coordinate $v$ is an affine parameter along the null geodesic integral curves of $n$ at $\mathcal{I}^-$. Since $\gamma_H$ and $\gamma$ are infinitesimally close, we can expand the affine distance between the curves at $\mathcal{I}^-$ in powers of $U_0$:
\begin{equation}
    v=\alpha U_0+O(U_0^2)\Longrightarrow u=-\frac{1}{\kappa}\ln\left(-U\right)=-\frac1\kappa\ln\left(-\alpha v\right),
\end{equation}
with $\alpha>0$ some constant. Therefore, if a mode takes the form $g_\omega$ on $\mathcal{I}^+$
\begin{equation}
    g_\omega\sim e^{-i\omega u}\ \Longrightarrow\ g_\omega^{(t)}\sim
    \begin{cases}
        e^{i\frac{\omega}{\kappa}\ln(-v)} & v<0\\
        0 & v>0
    \end{cases}\quad \mathrm{on}\ \mathcal{I}^-.
\end{equation}
This is analogous to the Rindler modes in the Unruh effect with $\kappa\longleftrightarrow a$. Therefore, we can define
\begin{equation}
    A_{\omega\omega'}=e^{-\pi\omega/\kappa}B_{\omega\omega'}.
\end{equation}
Following a similar reasoning than for the Unruh effect, we obtain that the number of particles seen by an observer at $\mathcal{I}^-$ in $\mathcal{I}^+$ is:
\begin{equation}
    \langle N_\omega\rangle\propto\frac{1}{e^{\hbar\omega/k_B T_H}-1},
\end{equation}
where
\begin{equation}
    T_H=\frac{\hbar\kappa}{2\pi k_B}
\end{equation}
is named Hawking temperature.
We can notice that $T_H\propto\kappa\propto 1/m$ and thus the black hole heats up as it evaporates.

Notice that there is a potential barrier and not all $g_\omega$ is transmitted. Thus, we should consider a \emph{gray-factor} $\Gamma_l(\omega)$ 
\begin{equation}
    \langle N_\omega\rangle\propto \frac{\Gamma_l(\omega)}{e^{\hbar\omega/k_B T_H}-1},
\end{equation}
to take into account the attenuation of the transmitted wave.

\subsubsection{Information paradox}

\begin{figure}
    \centering
    \includegraphics[scale = 0.9]{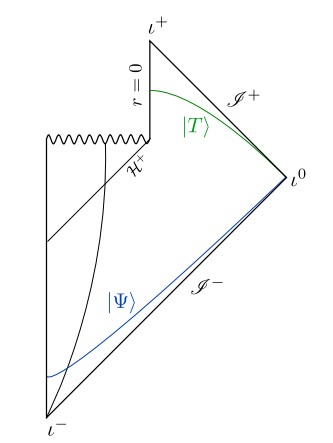}
    \caption{\emph{Information paradox. Carter-Penrose diagram for a black hole that is evaporating. At the beginning there is the state $|\Psi\rangle$ of a collapsing star which creates a black hole with event horizon $\mathcal{H}^+$ that evaporates. The process ends up in $|T\rangle$, when the black hole is completely evaporated. These two states are defined over two spacelike slices of the manifold, $\Sigma_1$ and $\Sigma_2$, respectively. We remind that the wavy line corresponds to the singularity in $r=0$. From}~\cite{liberatilect}.}
    \label{fig:ip}
\end{figure}

The discovery of Hawking radiation led to new questions. Here, we step onto the information paradox~\cite{Mathur:2009hf, liberatilect}. Hawking radiation predicts that a black hole evaporates in a finite time, thus it cannot be eternal and after a certain time the singularity disappears. The Carter-Penrose diagram of such a situation in shown in Fig.~\ref{fig:ip}.
We can consider that on an early Cauchy hypersurface $\Sigma_1$, at the quantum level we are in a state $|\Psi\rangle$ and that, after the black hole has completely evaporated, we end up on a late Cauchy hypersuface $\Sigma_2$ with a state $|T\rangle$ after the black hole is completely evaporeted. Since the quantum evolution is unitary 
\begin{equation}
    |\Psi(t)\rangle=e^{i\hat{H}t}|\Psi(0)\rangle,
\end{equation}
according to an Hamiltonian operator $\hat{H}$, if we start with a certain pure state $|\Psi\rangle$, after a certain amount of time, we could end up in $|T\rangle$, which is instead a mixed state. Indeed, $|T\rangle$ can only be described by a density matrix, much like when a piece of paper burns away and the information is encoded in correlations between radiation quanta. Too see this, we can notice that when the black hole completely evaporates away, the emitted radiation quanta have a non-zero entanglement, but since there is nothing left that they are entangled with, the final state cannot be described by any quantum wavefunction. As a result, unless a measurement has been performed, the black hole evaporation process cannot happen without violating unitarity.

Let us express this telling in a more formal way. To this end, let us consider an ensemble of particles. The state of the system can be characterized by the density matrix operator
\begin{equation}
    \hat{\rho}=\sum_nP_n|\psi_n\rangle\langle\psi_n| \qquad \mathrm{Tr}\hat{\rho}=\sum_n P_n=1,
\end{equation}
where $P_n$ is the probability of the $|\psi_n\rangle$ state of the system.
Given a general observable $\hat{A}$, its ensemble average is defined as
\begin{equation}
    \langle \hat{A}\rangle = \mathrm{Tr}\left[\hat{\rho}\hat{A}\right]=\sum_n P_n\langle \psi_n|\hat{A}|\psi_n\rangle.
\end{equation}
What we are saying is that the mixed state is created by the superposition of pure states $|\psi_n\rangle$, which can be measured with a probability $P_n$. A pure state is always written in the form
\begin{equation}
    \hat{\rho}=|\psi_{\Bar{n}}\rangle\langle \psi_{\Bar{n}}|,
\end{equation}
and thus each time we perform a measurement we obtain always the same answer. The mixed and pure states are distinguishable for one characteristic: a pure state $\hat{\rho}$ satisfies $\hat{\rho}^2=\hat{\rho}$, condition that is not satisfied instead by a mixed state. Indeed:
\begin{equation}
    \hat{\rho}^2= |\psi_{\Bar{n}}\rangle\langle\psi_{\Bar{n}}|\psi_{\Bar{n}}\rangle\langle \psi_{\Bar{n}}|=\hat{\rho}\qquad \mathrm{Tr}\hat{\rho}^2=\sum_n P_n^2<1.
\end{equation}

The \emph{information} of a system can be measured via the concept of von Neumann entropy 
\begin{equation}
    S\equiv-\mathrm{Tr}\left(\hat{\rho}\log\hat{\rho}\right).
\end{equation}
For pure states $S=0$, while for mixed states $S>0$.

Consider now the unitary operator $\hat{U}$, with $\hat{U}\hat{U}^\dagger=\mathbb{I}$, which gives rise to a unitary transformation such that on a pure state it acts in the following manner:
\begin{equation}
    \begin{split}
        &|\psi\rangle\to|\psi\rangle'= \hat{U}|\psi\rangle\\
        &\hat{\rho}\to\hat{\rho}'= \hat{U}|\psi\rangle\langle\psi|\hat{U}^\dagger.
    \end{split}
\end{equation}
One consequence of what we have just exposed is that 
\begin{equation}
    \left(\hat{\rho}'\right)^2= \hat{U}|\psi\rangle\langle\psi|\hat{U}^\dagger\hat{U}|\psi\rangle\langle\psi|\hat{U}^\dagger=\hat{U}|\psi\rangle\langle\psi|\hat{U}^\dagger=\hat{\rho}'.
\end{equation}
Therefore, this means that if the initial state is pure and we apply to it a unitary transformation, then the state remains unitary. As anticipated, the consequence is that if we start in a pure state, we cannot end up in a mixed state, and therefore the evaporation of a black hole implies the violation of unitarity.

Here, we aim to draw attention to selected noteworthy endeavors that have been undertaken in an effort to address this quandary.
\begin{itemize}
    \item Hawking made an attempt to resolve this problem by proposing a non-unitary super-scattering matrix~\cite{Hawking_1994}. However, this concept was eventually abandoned due to the challenges in constructing a consistent quantum mechanics framework that could accommodate it.
    \item the information trapped inside a black hole is released during the quantum gravity regime at the end of its evaporation. However, there is a challenge if the black hole becomes too small, as there may not be enough energy left to encode all the missing information. Don Page's observation~\cite{PhysRevLett.71.3743} highlights that the decrease in the entanglement entropy of the Hawking radiation, called \emph{purification}, must begin before half of the original black hole entropy remains. Otherwise, there would not be sufficient internal degrees of freedom to purify the emitted radiation by the end of the evaporation process. This critical time, known as the Page-time, marks a turning point where a significant amount of information starts to come out as the black hole shrinks and its entropy decreases. This delayed release of information provides a potential solution to the information paradox, as it allows for a gradual unveiling of the encoded information during the later stages of black hole evaporation.
    \item Instead of a complete event horizon, one can consider scenarios where only trapping horizons form. This can be achieved by regularizing the inner singularity through quantum gravity, leading to the concept of \emph{regular black holes}~\cite{Carballo_Rubio_2018}. In such cases, the singularity is removed from spacetime, and the entire manifold is covered by the causal past of future null infinity. This means there is no longer an event horizon but rather a transient trapped region. The main challenge with regular black holes is their stability, as the presence of an inner horizon can lead to the mass inflation phenomenon, which destabilizes the horizon on short timescales unless the solution is fine-tuned. On the other hand, in the context of black hole bounces, the challenge is to ensure consistency with observations of long-lived black hole-like objects. This can be achieved by having a significant difference in timescales between the bounce for freely-falling observers and static observers at infinity, while still being shorter than the evaporation time. However, the preference for such scenarios is still unclear and depends on computable quantum gravity models.
\end{itemize}

\chapter{Analogue gravity\label{chap:3}}

Analogies may offer fresh and creative perspectives on problems that enable the exchange of concepts between different scientific fields. In analogue gravity, links are created between curved-space quantum field theory and condensed matter physics, with the aim of learning lessons that can be useful to approach a theory of quantum gravity. The most well-known of these analogies is that between light waves in a curved spacetime and sound waves in a flowing fluid. The acoustic equivalent of a "black hole" is a "dumb hole", which can then be created by transonic fluid flow. The analogy can be taken so far, to mathematically prove the existence of phononic Hawking radiation from the acoustic horizon. 

Playing analogue gravity can be thought of as having two primary benefits. Generally speaking, the very fundamental act of connecting ideas from seemingly unrelated systems can inspire original thinking that results in new insights or concepts, whether from one system to the other or vice versa. In reality, this specific illustration offers a practical laboratory model for curved-space quantum field theory in an area where experimentation is practically feasible. 

In the pursuit of various objectives, there exists a wide array of "analogue models" that can prove beneficial. From an experimental standpoint, certain analogue models present intriguing possibilities, while others contribute by shedding light on perplexing theoretical issues. The exchange of information, theoretically speaking, flows in both directions, allowing for the utilization of general relativity-related concepts to comprehend aspects of the analogue models.
By accurately capturing and reflecting a significant number of noteworthy characteristics of general relativity, the analogue models prove their efficacy. It is important to note, however, that analogy should not be confused with identity, and we do not make any claims regarding complete equivalence between the analogue models we consider and general relativity.

From the perspective of the general relativity community, the development of the analogue models can be divided into two periods: the "historical" and the "modern" periods, that are before and after 1981. In fact, the specialty of 1981 is in the publication of the article by Unruh~\cite{PhysRevLett.46.1351}, where he used a fluid-flow analogy to implement an analogue model, and then applied the power of the analogy to tackle basic questions about Hawking radiation from "real" general relativistic black holes. 

In this chapter, firstly, we provide a historical overview of various articles published over time that have contributed to the definition of Analogue Gravity. Subsequently, we introduce the modern analogue models~\cite{analogue} that are most commonly used. In particular, we delve into more detail about the most significant models: from a classical perspective, we discuss a model that allows for the realization of an acoustic metric; from a quantum perspective, we explore a model based on a Bose-Einstein condensate (BEC). Finally, we examine acoustic Hawking radiation. Specifically, we first introduce it and then discuss some pivotal articles that have been instrumental in its study and experimental detection~\cite{PhysRevA.78.021603,stein1,stein,unruh}. In this chapter, we closely follow the living review by Barcelo, Visser and Liberati~\cite{analogue} to trace the story of analogue-gravity models and experiments.

\section{Historical overview of analogue models}

The first article to address analogue models is that of Gordon~\cite{Gordon}, who was interested in using the gravitational field to describe a dielectric medium. The following is now frequently referred to as the Gordon metric
\begin{equation}
    [g_{effective}]_{\mu\nu}= \eta_{\mu\nu}+[1-n^{-2}]V_\mu V_\nu, 
\end{equation}
where $\eta_{\mu\nu}$ is the flat metric, $n$ is the position dependent refractive index, and $V_\mu$ is the quadrivelocity of the medium.

To understand this model basis', we refer to one of the problems from Landau and Lifshitz's book "Classical physics of field"~\cite{landau2}. In the problem discussed in Chapter 10, Paragraph 90, the electromagnetic field is studied in the context of a static gravitational field. The reached bottom line is that
\begin{equation}
    \textbf{D}=\textbf{E}/\sqrt{h}\qquad \textbf{B}= \textbf{H}/\sqrt{h}, 
\end{equation}
with $h=g_{00}$ and $g_{\mu\nu}$ the metric, thus demonstrating that a static gravitational field behaves like an electric and magnetically permeable material with electric and magnetic permeabilities $\mu=\varepsilon=1/\sqrt{h}$. 

Pham Mau Quan~\cite{Pham} adopted this concept and demonstrated how Maxwell's equations in a medium can be expressed in terms of an effective metric, and how the trajectories of electromagnetism beams are geodesics of this new effective metric with null length. In later articles, the dielectric analogy is directly applied to analyze particular physics issues. One such article was written by Balazs~\cite{PhysRev.110.236}, who demonstrated that for electromagnetic waves, the gravitational field of a rotating body acts as an optically active medium, causing the wave's plane of polarization to rotate, even though the effect is only slight. 

The article published by De Felice~\cite{de.Felice} provides a helpful summary of this time. Here, the characteristics of the equivalent medium are discussed. Whenever a general gravitational field is present, it can always be represented as a medium with magnetic permittivity $\mu$ and electric permeability $\varepsilon$ that fulfill $\mu\propto\varepsilon$. De Felice examines the use of geometrical optics in the final paragraphs. and illustrates an equivalent medium for a spherically symmetric gravitational field and a cosmological model, where he discovers the luminosity-red-shift relationship popular in cosmology.

Following that, Anderson and Spiegel~\cite{spiegel} expanded Gordon's measure to include the possibility of a flowing medium. 

In a groundbreaking 1973 article, White~\cite{white} showcased the remarkable utility of a geometric ray-trace theory in analyzing sound propagation through moving inhomogeneous inviscid fluids. By leveraging this theory, White demonstrated how the null geodesics of the metric tensor, derived from the second-order coefficients in the partial differential equation for sound, could be interpreted as the spacetime path histories of sound pulses. This innovative approach paved the way for subsequent investigations in the 1980s, wherein numerous articles explored the propagation of shockwaves in astrophysical contexts using an acoustic analogy.

Of particular significance, Moncrief~\cite{monte} conducted a comprehensive examination of perturbations in stationary, spherical accretion onto a Schwarzschild black hole. Notably, Moncrief's analysis yielded the remarkable conclusion that both unstable normal modes and unstable modes representing standing shocks at the sound horizon were non-existent. This seminal work by Moncrief can be viewed as a precursor to later advancements in the understanding of acoustic geometries and acoustic horizons.

The findings by Moncrief, in a sense, surpass the scope of mainstream acoustic gravity research that followed, as they also encompass a general relativistic Schwarzschild background. Consequently, Moncrief's results hold broader implications and serve as a significant contribution to the field of acoustic gravity.

The most significant event in the "modern" era was the publication in 1981 of Unruh's paper "Experimental black hole evaporation"~\cite{PhysRevLett.46.1351}, which used a fluid-flow analogy to implement an analogue model and explore key issues surrounding Hawking radiation from "real" general relativistic black holes.
Unruh's 1981 paper is considered to be the first observation of the now well-known fact that Hawking radiation is a fundamental curved-space quantum field theory phenomenon, that arises whenever a horizon is present in an effective geometry and has nothing to do with general relativity \emph{per se}. Unruh's 1981 paper was important in this respect, but for many years it went largely unnoticed. In 2016, Steinhauer~\cite{stein} has experimentally demonstrated Unruh's findings and established the fluid/gravity correspondence~\cite{rangamani}.

\section{Analogue gravity models}

A variety of analog models have been investigated over the last few decades to tackle specific phenomena, having in mind different experimental platforms. These have included e.g. water, fluids with an appropriate background velocity, quantum gases and graphene~\cite{analogue,elio4,Volovik:2003fe}.
In the majority of analogous matter systems, kinematics mimics gravity while the microphysics controls dynamical features.
Here, we would like to review selected among the analogue models among those discussed in literature~\cite{analogue}. In general, the models are divided into classical and quantum models. In table~\ref{tab:models} is a more detailed list.

\begin{table}[]
    \centering
    \begin{tabular}{c|c}
        \textbf{Classical models} & \textbf{Quantum models} \\
        \hline
        Classical sound & Bose–Einstein condensates (BECs)\\
        Sound in relativistic hydrodynamics & Helium-based models\\
        Water waves (gravity waves) & Slow light\\
        Classical refractive index & Graphene \\

    \end{tabular}
    \caption{\emph{Short list of classical (left) and quantum (right) models (see text).}}
    \label{tab:models}
\end{table}

In the following sections, we will begin with a concise overview of several selected models, included in those previously mentioned. Subsequently, our attention will be directed towards the most significant and captivating ones, as they hold great value in comprehending the underlying principles behind the construction of analogue models.

\subsubsection{Classical sound}

The basic idea of how these models are created in a non-relativistic moving fluid is the following: sound waves are drawn along by flowing fluids, and if a fluid ever reaches supersonic speed, sound waves will never be able to travel back upstream. In this way, one can create a "\emph{dumb hole}", i.e. a hole where the sound cannot escape from. This sounds very familiar to the concept of black hole in general relativity. To have a solid analogy, we have to translate the above telling in a mathematical language. We can proceed with two approaches:

\begin{itemize}
    \item Geometrical acoustics: it focuses on the creation of an effective metric. It has the advantage that the obtained metric is very simple and has a very general derivation. The disadvantage instead, is that only the spacetime's causal structure can be inferred in the geometrical acoustics limit, and a single useful metric is not obtained.

    \item Physical acoustics: it focuses on how the sound propagates, to build from this an effective metric. The disadvantage is that the derivation is not as general as in the previous case because the analogy is valid only in a restricted regime. The advantage is that the analogy determines a specific effective metric and accommodates a wave equation for the sound waves. 
\end{itemize}

\subsubsection{Sound in relativistic hydrodynamics}

The limit of relativistic ray acoustics allows us to ignore the wave properties of sound. Indeed, here, we are only interested in the "sound cones". We are going to sum up this procedure, a detailed discussion being given in~\cite{mattcarmen}.

We consider a curved manifold with a metric $g_{\mu\nu}$ and a spacetime's point where the fluid's velocity is $V^\mu$ and the speed of sound is $c_s$. We adopt Gaussian coordinates, so that the metric in the chosen point becomes flat and we are in the fluid rest frame. At this point, it is possible to build an analogue metric. In fact, locally in the fluid rest frame, the system is homogeneous. Thus, sound moves here like a particle would do in a flat spacetime, where the light cones are identified by the equation
\begin{equation}
    c^2dt^2-||d\vec{x}||^2=0, 
\end{equation} 
with $c$ is the speed of light.
If we now build an analogy between the speed of light and the speed of sound $c_s$, the sound cones are given by 
\begin{equation}
    c_s^2dt^2-||d\vec{x}||^2=0, 
    \label{eq:minksound}
\end{equation}
implying the existence of a metric $\mathcal{G}_{\mu\nu}\propto\mathrm{diag}(-c_s^2,1,1,1)$. Therefore, the sound propagating in the fluid behaves like a field in a spacetime described by the flat metric $\mathcal{G}_{\mu\nu}$. If we now come back to the first arbitrary coordinates, we have got an analogue metric.

Now, the only "problem" is to find an overall conformal factor so that we can obtain a relativistic wave equation suitable for describing physical acoustics. Indeed, working only on the sound cones it is not possible to find an overall conformal factor.

This approach is necessary in three cases:
\begin{enumerate}
    \item when operating in a backdrop of nontrivial curved general relativity;
    \item any time the fluid is moving at relativistic velocities;
    \item even if the general flow of the fluid is not relativistic, when the internal degrees of freedom of the fluid are relativistic.
\end{enumerate}

\subsubsection{Shallow water waves}

\begin{figure}
    \centering
    \includegraphics[scale = 0.8]{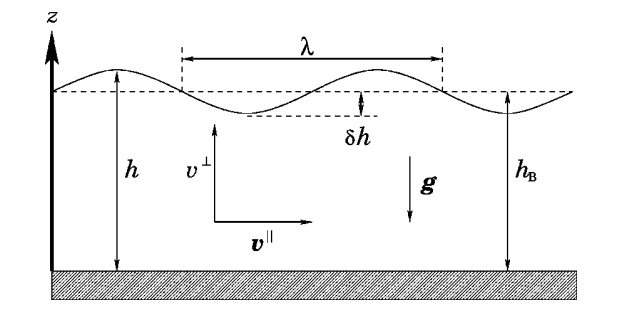}
    \caption{\emph{Classical analogue models. Shallow water waves. A water container with height $h$ and gravitational force $\bm g = -g\bm z$ is schematically shown. The wavelength of the disturbance $\delta h$ is $\lambda$ and it is moving along the vector $\bm v_{||}$. From}~\cite{PhysRevD.66.044019}.}
    \label{fig:basin}
\end{figure}

Under suited circumstances, sound waves in moving fluids are controlled by the same wave equation as a scalar field in a curved spacetime. The simplest system exhibiting an emergent metric is a fluid in a basin~\cite{PhysRevD.66.044019}. In Fig.~\ref{fig:basin} is reported a schematic representation of the system.

If one considers an inviscid and irrotational fluid, $\bm v = \bm \nabla \phi$, one can write the Bernoulli's equation for a shallow basin of height $h$ along the $z$ axis in presence of Earth’s gravitational acceleration $g$: 
\begin{equation}
    \partial_t\phi+\frac12(\bm \nabla \phi)^2= -\frac{p}{\rho} - gz -V_{||}, 
\end{equation}
where $\rho$ is the density of the fluid, $p$ its pressure and $V_{||}$ a potential associated with some external force necessary to generate an horizontal flow in the fluid. Once it is established a flow $\bm v _{||}$ in the direction of the driving force related to $V_{||}$, the perturbation $\delta \phi$ satisfies
 \begin{equation}
     \partial_t \delta\phi+ \bm v _{||} \cdot \bm \nabla \delta \phi = -\frac{\delta p}{\rho}.
 \end{equation}
It can be demonstrated that, if one expands $\delta \phi$ in Taylor series, surface waves with long wavelength $\lambda\gg h$ behave like a field $\delta \phi_0$ which feels an effective metric of the form
\begin{equation}
    ds^2 = \frac{1}{gh}[(g_z h -v_{||}^2)dt^2+2\bm v _{||}\cdot\bm {dx} dt - \bm {dx} \cdot \bm {dx}], 
\end{equation}
with $\sqrt{gh}$ playing the same role as the speed of light. In this manner, we obtain an effective metric
\begin{equation}
    g_{\mu\nu} = \frac{1}{gh}\left(
    \begin{matrix}
        (gh-v_{||}^2) & \bm v_{||}\\
        \bm v_{||} & -\mathbb{I}_{3\times3}
    \end{matrix} \right). 
\end{equation}
The primary benefit of this model is how easily the depth of the basin can be changed in order to change the velocity of the surface waves. The construction of ergoregions should be somewhat simpler than in other models because this velocity can be made very slow. This model is classic, and although this feature makes it easier to actually implement analog horizons, it is a drawback when attempting to identify analog Hawking radiation because the relative temperature will inevitably be very low. 

\subsubsection{Classical refractive index}

This model is based on the analogy between electromagnetism in curved spacetime and electromagnetic waves in a medium. In fact, the macroscopic Maxwell equations inside a dielectric are
\begin{align}
    &\bm \nabla \cdot\bm B = 0,\qquad \bm\nabla\times\bm E + \partial_t \bm B=0,
    \label{eq:vacuum}\\
    &\bm \nabla \cdot\bm D = 0,\qquad \bm\nabla\times\bm H + \partial_t \bm D=0, 
    \label{eq:medium}
\end{align}
where $\bm H = \bm \mu^{-1}\cdot \bm B$ and $\bm D = \bm \epsilon\cdot \bm E$, with $\bm \epsilon$ the $3\times3$ permittivity tensor and $\bm \mu$ the $3\times3$ the permeability tensor of the medium. One can recast Eqs.~\eqref{eq:vacuum},~\eqref{eq:medium} in terms of a tensor $Z^{\mu\alpha\nu\beta}$ as
\begin{equation}
    \partial_\alpha\left(Z^{\mu\alpha\nu\beta}F_{\nu\beta}\right)=0, 
\end{equation}
where $F_{\nu\beta}= A_{[\nu,\beta]}$ is the electromagnetic tensor.
The non-vanishing components of the tensor $Z$ are given by
\begin{align}
    \label{eq:ze1}
    &Z^{0i0j}=-Z^{0ij0}=Z^{i0j0}=-Z^{i00j}=-\frac12\epsilon^{ij}\\
    \label{eq:ze2}
    &Z^{ijkl}=\frac12\varepsilon^{ijm}\varepsilon^{kln}\mu^{-1}_{mn}, 
\end{align}
where $\varepsilon^{ijm}$ is the completely antisymmetric Levi-Civita tensor.

Taken the Lagrangian for electromagnetism in curved spacetime, one can build a tensor of the form
\begin{equation}
    Z^{\mu\nu\alpha\beta}= K\sqrt{-g}[g^{\mu\alpha}g^{\nu\beta}-g^{\mu\beta}g^{\nu\alpha}], 
\end{equation}
to obtain a static emergent gravitational metric $g_{\mu\nu}=\Omega^2(-1\oplus g_{ij})$. It is not difficult to find the components of the tensor $Z$, that are:
\begin{align}
    \label{eq:zg1}
    &Z^{0i0j}=-Z^{0ij0}=Z^{i0j0}=-Z^{i00j}=K\sqrt{-g}g^{ij}\\
    \label{eq:zg2}
    &Z^{ijkl}=K\sqrt{-g}[g^{ik}g^{jl}-g^{il}g^{jk}].
\end{align}
We notice that the non-vanishing components of the tensor $Z$ for electromagnetism in spacetime have the same properties of the components of the tensor $Z$ for electromagnetism in a dielectric. In particular, equating~\eqref{eq:ze1} with~\eqref{eq:zg1} and~\eqref{eq:ze2} with~\eqref{eq:zg2}, one gets a relation that links the metric of a curved manifold to the electric permittivity and the magnetic permeability of a medium:
\begin{equation}
    \varepsilon^{ij}=4K^2\mu^{ij}\qquad
    g^{ij}=\left[\left( \frac{\bm \mu^{1/2}\bm\epsilon\bm\mu^{1/2}}{\mathrm{det}(\bm{\mu\, \epsilon})}\right)^{1/2}\right]^{ij}.
\end{equation}
Thus, a static gravitational field can always be treated like a medium with $\bm\epsilon\propto\bm \mu$.

This approach can be expanded to include a medium in motion with a little more effort, which would result in an extension of the Gordon metric.

\subsubsection{The color-flavor locked superfluid in the limit of vanishing temperature}

The color-flavor locked (CFL) phase is a superfluid. Indeed, by picking a phase, its order parameter spontaneously breaks the baryonic number $U(1)_B$ symmetry. In the regime of sufficiently high densities, it is possible to derive the hydrodynamic equations from the microscopic physics, as shown by Son in~\cite{https://doi.org/10.48550/arxiv.hep-ph/0204199} for superfluid phonons. One needs to take into account the dynamics of phonons travelling in the superfluid backdrop, in order to fully generalize thermal effects in the low temperature limit. It is possible to interpret the phonon field's equation of motion in the superfluid backdrop as a scalar field travelling in a "acoustic" non-flat metric. In this system~\cite{PhysRevD.77.103014}, only very low temperature regimes are considered, because in so doing we can ignore other quasiparticles that can be excited in the CFL phase upon increasing the temperature. 

The Goldstone boson associated to the breaking of the $U(1)_B$ symmetry is the superfluid phonon in the CFL phase, and it can be introduced as the phase of the diquark condensate.
However, the gradient of the condensate phase defines the superfluid velocity as well. It is possible to write the phase of the condensate as a decomposition of two fields, the first describing the hydrodynamical variable $\Bar{\varphi}(x)$, and the second the quantum fluctuations associated to the phonons $\phi(x)$:
\begin{equation}
    \varphi(x)=\Bar{\varphi}(x)+\phi(x). 
\end{equation}
From the low-energy effective action of the system
\begin{equation}
    S[\varphi]=\int d^4x\lag_{eff}[\partial \varphi], 
\end{equation}
it is possible to obtain the effective action for the phonon field
\begin{equation}
    S[\varphi]=S[\Bar{\varphi}]+\frac12\int d^4x\left.\frac{\partial^2\lag_{eff}}{\partial(\partial_\mu\varphi)\partial(\partial_\nu\varphi)}\right|_{\Bar{\varphi}}\partial_\mu\phi\partial_\nu\phi+..
    \label{eq:action}
\end{equation}
Knowing that
\begin{align}
    &P(\mu_q)=\frac{3}{4\pi^2}\mu_q^4&\lag_{eff}=\frac{3}{4\pi^2}[(\partial_0\varphi-\mu_q)^2-(\partial_i\varphi)^2]^2 \\
    &\Bar{\mu}=[(\partial_\alpha\Bar{\varphi}-\mu_q\delta_{0\alpha})(\partial^\alpha\Bar{\varphi}-\mu_q\delta^{0\alpha})]^{1/2}& n_0=\left.\frac{d P}{\mu}\right|_{\mu=\Bar{\mu}}=\frac{3}{\pi^2}\Bar{\mu}^3, 
\end{align}
one gets
\begin{equation}
    f^{\mu\nu}=\left.\frac{\partial^2\lag_{eff}}{\partial(\partial_\mu\varphi)\partial(\partial_\nu\varphi)}\right|_{\Bar{\varphi}}=\frac{n_0}{\Bar{\mu}}\left[\eta^{\mu\nu}+\left(\frac{1}{c_s^2}-1\right)v^\mu v^\nu\right], 
\end{equation}
with $c_s=1/\sqrt{3}$ the speed of sound in CFL quark matter. The action of the linearized fluctuation is represented by the second term in the right-hand-side of Eq.~\eqref{eq:action}, and it can be written as the action of a boson moving in a non-trivial gravity background
\begin{equation}
    S[\phi]=\frac{1}{2}\int d^4x\sqrt{-\mathcal{G}}\mathcal{G}^{\mu\nu}\partial_\mu\phi\partial_\nu\phi \qquad \mathcal{G}^{\mu\nu}=\eta^{\mu\nu}+\left(\frac{1}{c_s^2}-1\right)v^\mu v^\nu. 
    \label{eq:actionfluc}
\end{equation}
Here, $\mathcal{G^{\mu\nu}}$ is the metric tensor inverse and $\mathcal{G}=1/\mathrm{det}\mathcal{G}^{\mu\nu}$. In the action~\eqref{eq:actionfluc}, assuming $\Bar{\mu}$ and $n_0$ to be constants, the phonon field has been rescaled
\begin{equation}
    \phi\to\sqrt{\frac{c_s\Bar{\mu}}{n_0}}\phi\,.
\end{equation}
The normalized field obeys the classical equation of motion: 
\begin{equation}
    \frac{1}{\sqrt{-\mathcal{G}}}\partial_\mu\left(\sqrt{-\mathcal{G}}\mathcal{G}^{\mu\nu}\partial_\nu\phi\right)=0, 
\end{equation}
that has the same structure of the equation of motion for a massless scalar field on curved spacetime with metric $\mathcal{G}_{\mu\nu}$.

\subsubsection{Bose-Einstein condensates \label{sec:BEC}}

The idea of Bose-Einstein condensation of atoms (BEC) was born with Einstein in 1925 while studying the statistical description of light quanta, when he discovered that a gas of non-interacting atoms can undergo a phase transition, as a statistical effect, associated to the condensation of atoms in the lowest energy state. In general, working with condensed matter systems can be especially useful to focus on quantum effect, like Hawking radiation. Indeed, the analog condensed matter models frequently have a microscopic quantum mechanical description that is well known, allowing full control of their physics, in the theory and in experimental platforms. In fact, among the many systems proposed to create black-hole-like configurations, BECs appear to be quite attractive from the experimental point of view. 

This model will be explained in Sec.~\ref{sec:BEC}.

\subsubsection{Helium-based models}

Helium~\cite{elio4,Volovik:2003fe} is paradigmatic in that it involves numerous and diverse macroscopic characteristics that can be traced back to its rich microscopic properties. Here, we're interested in the development of effective shapes in helium and how semiclassical gravity might be tested using these geometries.
\begin{itemize}
    \item $^4$He

    It is a bosonic system~\cite{elio4,Volovik:2003fe} presenting superfluid properties when cooled below $2.17$ K. Since superfluids are inviscid and irrotational, this element behavior can be advantageous for all the models needing these characheristics. The superfluid behavior is connected to the macroscopically large number of atoms condensing in the vacuum state, acquiring all the characteristics of the BEC, but with a significant interactions-driven depletion of the condensate. With a logical extension to quantum field theory, the propagation of sound waves over the background fluid can be described by an effective Lorentzian geometry, or the acoustic metric.

    \item $^3$He

    It is a fermionic system~\cite{elio3,Volovik:2003fe}, becoming superfluid at much lower temperatures than $^4$He, below $2.5$mK. Sperfluidity in $^3$He is due to the formation of Cooper pairs, which undergo Bose-Einstein condensation. In the B-phase, the quasi-particle spectrum exhibits complete gap formation and comprises interacting fermionic and bosonic quantum fields. However, $^3$He-B does not possess the requisite properties of a relativistic quantum field theory necessary for simulating the quantum vacuum. Notably, it lacks Lorentz invariance, and only certain components of the order parameter bear distant resemblance to gravitons. Nonetheless, $^3$He-B can effectively serve as a model system for simulating various phenomena in particle-physics and cosmology. In the A-phase and in equilibrium configuration, instead of the Fermi surface with momenta $|p| = p_F$ equal to the Fermi momentum $p_F$, the fermionic vacuum for $^3$He can be represented by two points $p_{x,y} = 0$ and $p_z = p_F$.
    As a result, the quasiparticles spectrum is the same as a Weyl fermion spectrum over planar spacetime. When the superfluid collective excitations are considered, these are experienced by Weyl quasiparticles as an effective electromagnetic field and as a curved geometry. 
\end{itemize}

\subsection{Acoustic metric \label{sub:acustic}}

It is well known that a sound wave in a static, homogeneous, inviscid fluid propagates with a speed of sound $c_s$ according to the wave equation:
\begin{equation}
    \partial_t^2\phi = c_s^2\bm \nabla ^2\phi.
\end{equation}
This solution can be extended to fluids which are simply barotropic and inviscid and such that the flow is irrotational: the equation of motion for the sound wave is identical to the d’Alembertian equation of motion for a minimally-coupled massless scalar field propagating in a $3+1$-dimensional Lorentzian geometry~\cite{landau2013fluid}. Thus, the propagation of sound is lead by an acoustic metric, which describes a pseudo-Riemannian geometry and depends on density, flow velocity and the medium speed of sound. This behavior exists even if the fluid dynamics is Newtonian, non-relativistic and lives on flat space.

The Euler's equation
\begin{equation}
    \bm{\mathrm{f}} = \rho\frac{d\bm v}{dt} \equiv \rho\{\partial_t\bm v +(\bm v\cdot \bm \nabla)\bm v\}, 
\end{equation}
and the equation of continuity 
\begin{equation}
    \partial_t \rho + \bm \nabla \cdot(\rho\bm v )=0
\end{equation}
are the fundamental equations of a non-relativistic fluid. Here, $\rho$ is the matter density, $\bm v$ the flow velocity, and $\bm{\mathrm{f}}$ the force density. Since the fluid is inviscid, the only forces that play a role are caused by the pressure gradients, thus
\begin{equation}
    \bm{\mathrm{f}} = -\bm \nabla p.
\end{equation}
Knowing that $\bm v \times(\bm \nabla \times \bm v)= \frac12\bm \nabla v^2-(\bm v\cdot\bm \nabla)\bm v$, one gets
\begin{equation}
    \partial_t \bm v = \bm v \times(\bm \nabla\times \bm v) -\frac1\rho\bm\nabla p -\frac12\bm \nabla\left(v^2\right).
\end{equation}
If we now take into account the hypothesis of irrotational and barotropic fluid, we can recast the fluid velocity as 
\begin{equation}
    \bm v = -\bm \nabla \phi , 
\end{equation}
where $\phi$ is the velocity potential. We can also usefully define the entalpy $h(p)$ as a function of $p$ alone:
\begin{equation}
    h(p) = \int_0^p\frac{d\Tilde{p}}{\rho(\Tilde{p})}\, ,
\end{equation}
so that$\bm \nabla h = \frac1\rho\bm \nabla p$.

Thus, the Euler's equation reduces to
\begin{equation}
    -\partial_t \phi + h+\frac12(\bm \nabla \phi)^2=0.
\end{equation}

We are now ready to linearize our equations around a given background\\ $(\rho_0,\, p_0,\, \phi_0)$:
\begin{align}
    &\rho = \rho_0+\epsilon\rho_1 O(\epsilon^2),\\
    &p= p_0 + \epsilon p_1+ O(\epsilon^2),\\
    &\phi = \phi_0+\epsilon \phi_1+ O(\epsilon^2), 
\end{align}
with $\epsilon\ll1$, where the perturbations describe the sound waves, that are long-wavelength fluctuations of dynamical quantities. If we now call $\bm v_0 = \bm \nabla \phi_0$ the background velocity flow, and apply the linearization procedure to the continuity equation, we get
\begin{align}
    &\partial_t\rhoo+\bm \nabla\cdot(\rhoo\bm v_0)=0\\
    &\partial_t\rho_1+\bm \nabla\cdot (\rho_1\bm v_0+\rho_0\bm v_1)=0.
    \label{eq:continuityfluct}
\end{align}
Linearizing the barotropic condition instead, yields:
\begin{equation}
    h(p)=h(p_0+\epsilon p_1+O(\epsilon^2))=h_0+\epsilon\frac{p_1}{\rhoo}+O(\epsilon^2). 
\end{equation}
Applied to Euler's equation, this gives
\begin{align}
    &-\partial_t\phi_0+h_0+\frac12(\bm\phi_0)^2=0\\
    &-\partial_t\phi_1+\frac{p_1}{\rhoo}-\bm v_0\cdot \bm \nabla \phi_1=0.
    \label{eq:eulerfluct}
\end{align}
If we now write $p_1$ in terms of $\rho_0$, $\phi_1$ and $\bm v_0$ from Eq.~\eqref{eq:eulerfluct}, and use the barotropic assumption, we get
\begin{equation}
    \rho_1=\frac{\partial\rho}{\partial p}p_1 =\frac{\partial\rho}{\partial p} \rhoo(\partial_t\phi_1 +\bm v_0\cdot \bm\nabla\phi_1).
\end{equation}
Substituting now this equation in ~\eqref{eq:continuityfluct} we obtain:
\begin{equation}
    -\partial_t\left(\frac{\partial\rho}{\partial p}\rhoo(\partial_t\phi_1+\bm v_0\cdot\bm \nabla\phi_1)\right)+\bm \nabla\cdot\left(\rhoo\bm\nabla\phi_1-\frac{\partial\rho}{\partial p}\rhoo\bm v _0(\partial_t\phi_1+\bm v_0(\partial_t\phi_1+\bm v_0\cdot\bm \nabla\phi_1)\right)=0.
    \label{eq:soundwave}
\end{equation}
Equation~\eqref{eq:soundwave} describes the propagation of the linearized scalar potential $\phi_1$, and completely determines the propagation of an acoustic disturbance. We can thus define the speed of sound, the velocity at which the sound waves propagates:
\begin{equation}
    \frac{1}{c_s^2}\equiv\frac{\partial\rho}{\partial p}.
\end{equation}
Eq.~\eqref{eq:soundwave} can be re-written in the compact form
\begin{equation}
    \partial_\mu(f^{\mu\nu}\partial_\nu\phi_1)=0, 
    \label{eq:waveeqf}
\end{equation}
where $f^{\mu\nu}$ is a $4\times4$ tensor
\begin{equation}
    f^{\mu\nu}(t,\bm x)\equiv\frac{\rhoo}{c_s^2}\left(
    \begin{matrix}
        -1 & - v_0^j\\
        - v_0^i & f^{ij}
    \end{matrix}\right)\, , \qquad \mathrm{where}\ f^{ij}=c_s^2\delta^{ij}-v_0^iv_0^j.
\end{equation}
Notice that we use the notation $\mu=0,1,2,3$, $i,j=1,2,3$.
We now want to progress further, and write the wave equation in terms of a metric $g_{\mu\nu}(t,\bm x)$ which describes a $3+1$ dimensional Lorentzian geometry. Thus, Eq.~\eqref{eq:waveeqf} must be written as
\begin{equation}
    \Box\phi_1 = \frac{1}{\sqrt{-g}}\partial_\mu(\sqrt{-g}g^{\mu\nu}\partial_\nu\phi_1)=0, 
\end{equation}
with $g=\mathrm{det}(g_{\mu\nu})$.
For this purpose, we define 
\begin{equation}
    f^{\mu\nu}= \sqrt{-g}g^{\mu\nu},
\end{equation}
that implies $\mathrm{det}(f^{\mu\nu})=g=-\rhoo^4/c_s^2$. As a result, the acoustic metric is
\begin{equation}
     g^{\mu\nu}(t,\bm x)\equiv\frac{1}{\rhoo c_s}\left(
    \begin{matrix}
        1 \quad&  v_0^j\\
         v_0^i \quad& v_0^iv_0^j-c_s^2\delta^{ij}
    \end{matrix}\right)\qquad g_{\mu\nu}(t,\bm x)\equiv\frac{\rhoo}{c_s}\left(
    \begin{matrix}
        c_s^2-v_0^2 \quad&  -v_{0,j}\\
        - v_{0,i} \quad& -\delta_{ij}
    \end{matrix}\right). 
    \label{eq:metric1}
\end{equation}

The general features of this effective metric are as follows: 
\begin{itemize}
    \item The signature of the acoustic metric in Eq.~\eqref{eq:metric1} is the signature of a Lorentzian metric, i.e. $(-,+,+,+)$.
    \item $g^{\mu\nu}$ has only 2 degrees of freedom: it is completely determined by $\phi_0$, $\rhoo$ and $c_s$, but two among these three parameters are linked by the continuity equation. Thus, while a general $3+1$ dimensional Lorentzian geometry has 6 degrees of freedom, simple acoustic models can only mimic a subclass of all Lorentzian geometries.
    \item Two different metrics play a relevant role for the discussion in the present thesis, the \emph{physical spacetime} and the \emph{acoustic metric}. The \emph{physical spacetime metric} is the Minkowski metric
    \begin{equation}
            \eta_{\mu\nu}=[\mathrm{diag}(-c^2,1,1,1)]_{\mu\nu}, 
    \end{equation}
    and is coupled to the fluid particles. The \emph{acoustic metric} $g_{\mu\nu}$ couples to acoustic perturbations. Indeed, the sound waves do not "see" the physical metric: they propagate as they were lying on a curved manifold described by the effective metric.
    The acoustic geometry derived from the acoustic metric retains important properties from the underlying physical metric, such as the shared topology of $\reale^4$. The acoustic manifold may have certain regions removed due to imposed boundary conditions, but its overall topology remains consistent with the physical metric. 
    \item The fluid velocity field's integral curves are perpendicular to the constant time surfaces (in the Lorentzian metric). In fact, the normalized four-velocity of the fluid is
    \begin{equation}
        u^\mu=\frac{1}{\sqrt{\rhoo c_s}}(1,\bm v_0)\, , \qquad g_{\mu\nu}u^\mu u ^\nu=-1, 
    \end{equation}
    and it is related to the gradient of the natural time parameter by
    \begin{equation}
        \nabla_\mu t=(1,0,0,0)\, , \qquad \nabla^\mu t = -\frac{u^\mu}{\sqrt{\rho c_s}}
    \end{equation}
    \item The quality of "stable causality" is immediately transferred by the acoustic geometry.
    \item The use of concepts like horizons and trapped surface~\cite{viss} are introduced and allows for the simulation of black holes' features. In particular:
    \begin{itemize}
        \item Trapped surface: consider any closed two-surface. If here the fluid goes everywhere inward, and the normal component to the surface is always greater than speed of sound, it is therefore claimed that the surface is outer-trapped: the sound cannot escape, it will be swept inward by the fluid flow, and trapped within the surface. Inner-trapped surfaces are defined by requiring that the fluid flow be outward-pointing throughout with a supersonic normal component.
        The acoustic trapped region is defined as the region containing outer trapped surfaces.
        \item Acoustic apparent horizon: it is the boundary of the trapped region. Thus, it is a closed two-surface for which the normal component of the fluid speed is everywhere equal to the speed of sound.
        \item Acoustic event horizon: the boundary of the region from which null geodesics, i.e. phonons, cannot escape.
    \end{itemize}
    \begin{figure}[ht!]
        \centering
        \includegraphics[scale=0.6]{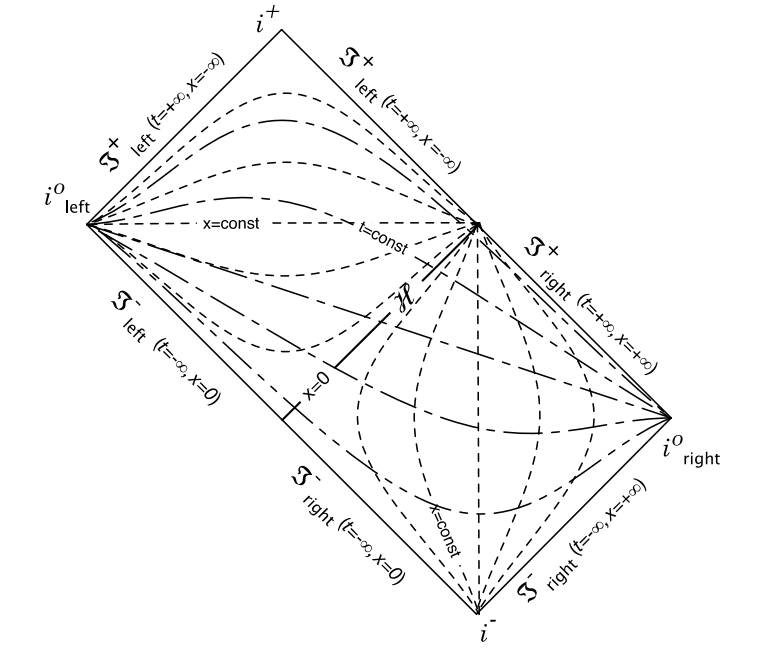}
        \caption{\emph{Acoustic black hole. Carter-Pensorse conformal diagram of an isolated acoustic horizon. From }\cite{analogue1}.}
        \label{fig:pc_bh}
    \end{figure}
    Notice that the notions we have just exposed have the same properties as their astrophysical analogues discussed in the previous chapter. 
    \item The metric of a non-compact $1+1$-manifold can always be conformally mapped onto the metric of a compact geometry. Thus, considering the just found metric~\eqref{eq:metric1} with coordinates $(t,x)$, one has to introduce first suited appropriate null coordinates $(u,v)$, then null Kruskal-like coordinates $(U,V)$, and finally compactify by means of a new coordinate pair $(\mathcal{U},\mathcal{V})$ involving an appropriate function that maps an infinite to a finite range. Fig.~\ref{fig:pc_bh} displays a Carter-Penrose diagram of an isolated acoustic black hole horizon.
\end{itemize}

\subsection{Bose-Einstein condensates}

BECs have more recently taken the lead among all potential analogue systems, for a number of reasons. In general, because they are extremely simple to build and operate in the lab, and specifically because the speed of sound is as low as a few centimeters per second. BECs can provide a particularly fascinating analogue model. In fact, unlike gravitational theories, measurements can be engineered and performed in a variety of lab configurations by using well-known BECs theories. The analogy holds dual significance. Firstly, by conducting tabletop experiments with precise control over the physical system, we can explore the existence of Hawking radiation. Secondly, our profound comprehension of the microscopic quantum theory underlying these systems allows us to address a subtle concern within the theory of Hawking radiation and potentially resolve certain inherent contradictions in its conventional derivation. In this context, we focus on the straightforward scenario of a non-relativistic condensate.

A BEC is the ground state of a second quantized many-body Hamiltonian for a system of $N$ bosons which interact with an external potential $V_{ext}(t)$~\cite{RevModPhys.71.463}. When the number of atoms is large and the atomic interactions suitably weak, near absolute zero almost all of the atoms are in the same single-particle quantum state $\oppsi(x, t)$. In the diluite-gas approximation for neutral atomic systems at ultracold temperatures, the evolution of the mean field $\oppsi$ is described by the Gross-Pitaevskii equation
\begin{equation}
    i\hbar\frac{\partial}{\partial t}\oppsi = \left( -\frac{\hbar^2}{2m}\bm{\nabla}^2+V_{ext}(\bm x)+ \lambda\oppsi^\dagger\oppsi \right)\oppsi, 
    \label{eq:grosspitaevskii}
\end{equation}
where $\lambda$ is the coupling length that can be expressed in terms of the s-wave scattering length $a$
\begin{equation}
    \lambda = \frac{4\pi a\hbar^2}{m}.
\end{equation}
In the mean-field approximation, $\bm{\hat{\Psi}}$ can be expressed as sum of a macroscopic, classically-describable condensate $\langle\oppsi\rangle = \psi$, and of a fluctuation $\opphi$ on top of it:
\begin{equation}
    \oppsi = \psi + \opphi.
\end{equation}
After substituting this relation in ~\eqref{eq:grosspitaevskii}, and considering that $\opphi^\dagger\opphi\opphi\simeq2\langle\opphi^\dagger\opphi\rangle\opphi+\langle\opphi\opphi\rangle\opphi^\dagger$, one gets
\begin{equation}
    \begin{split}
        i\hbar\frac{\partial}{\partial t}\psi(t,\bm x)=&\left(-\frac{\hbar^2}{2m}\bm\nabla ^2 +V_{ext}(\bm x)+ \lambda |\psi(t,\bm x)|^2\right)\psi(t,\bm x) +\\
        &+\lambda\left(2\langle\opphi^\dagger\opphi\rangle\psi(t,\bm x)+\langle\opphi\opphi\rangle\psi^*(t,\bm x)\right)
    \end{split}
\end{equation}
\begin{equation}
    \begin{split}
        i\hbar\frac{\partial}{\partial t}\opphi(t,\bm x)=&\left[ -\frac{\hbar^2}{2m}\bm \nabla ^2 +V_{ext}(\bm x) +2\lambda\left(|\psi(t,\bm x)|^2+\langle\opphi^\dagger\opphi\rangle\right)\right]\opphi(t,\bm x)+ \\
        &+ \lambda \left(\psi^2(t,\bm x)+\langle\opphi\opphi\rangle\opphi^\dagger(t,\bm x) \right) .
    \end{split}
        \label{eq:gpfluc}
\end{equation}
These two equations must be solved together and self-consistently. To connect them with the hydrodynamic equations, we adopt the \emph{Madelung representation} for the wave function of the condensate: 
\begin{equation}
    \psi(t, \bm x) = \sqrt{n_c(t,\bm x)}e^{-i\frac{\theta(t,\bm x)}{\hbar}}
\end{equation}
where we have defined $n_c \equiv |\psi(t,\bm x)|^2$, and the \emph{quantum acoustic representation} for the quantum perturbations of the system
\begin{equation}
    \opphi(t,\bm x)=e^{-i\theta/\hbar}\left(\frac{1}{2\sqrt{n_c}}\hat{n}_1-i\frac{\sqrt{n_c}}{\hbar}\hat{\theta}_1\right), 
    \label{eq:flucbec}
\end{equation}
where $\opn_1$ and $\optheta$ are real quantum fields.
First,  we study the wave function of the condensate.
Given that the fluid is irrotational, we can also define $\bm v = \bm \nabla \theta/m$ as the "velocity field". Now, the Gross-Pitaevskii equation can be written as
\begin{align}
    \label{eq:continuitybec}
    &\frac{\partial}{\partial t}n_c+\bm \nabla\cdot(n_c\bm v)=0\\
    \label{eq:eulerbec}
    &m \frac{\partial}{\partial t}\bm v +\bm \nabla\left(\frac{mv^2}{2}+V_{ext}(t,\bm x)+\lambda n_c-\frac{\hbar^2}{2m}\frac{\nabla^2\sqrt{n_c}}{\sqrt{n_c}}\right)=0.
\end{align}
Eq.~\eqref{eq:continuitybec} has the form of a continuity equation, while Eq.~\eqref{eq:eulerbec} has looks like the Euler's equation. Thus, these two equations are completely equivalent to those of an irrotational and inviscid fluid, apart from the existence of the quantum potential
\begin{equation}
    V_q=-\frac{\hbar^2}{2m}\frac{\nabla^2\sqrt{n_c}}{\sqrt{n_c}}.
\end{equation}
This potential leads in fact to define the quantum stress tensor $\sigma_{ij}$, with the dimensions of a pressure: it is a quantum anisotropic pressure contributing to the Euler's equation. Notice that when the density gradients are small, we can neglect this quantum term: this is called \emph{hydrodynamic approximation}.
After substituting the relation between $\bm v$ and $\theta$, the Euler equation assumes the Hamilton-Jacobi form
\begin{equation}
    m\frac{\partial}{\partial t}\theta+\left(\frac{(\bm \nabla\theta)^2}{2m}+V_{ext}(t,\bm x)+\lambda n_c-\frac{\hbar^2}{2,}\frac{\nabla^2\sqrt{n_c}}{\sqrt{n_c}}\right)=0.
\end{equation}
Let us now consider the quantum perturbation of the system. Linearizing around the classical solution ($n_c\to n_c+\opn_1$, $\phi\to\phi+\opphi_1$), the quantum potential represented by the second-order differential operator $D_2$ is
\begin{equation}
    D_2\opn_1\equiv-\frac12n_c^{-3/2}[\nabla^2(n_c^{1/2})]\opn_1+\frac12n_c^{-1/2}\nabla^2(n_c^{-1/2}\opn_1)\,.
\end{equation}
Substituting the representation in Eq.~\eqref{eq:flucbec} into Eq.~\eqref{eq:gpfluc}, we get:
\begin{align}
    \label{eq:eqn1}
    &\partial_t\opn_1+\frac{1}{m}\bm \nabla\cdot\left(\opn_1\bm\nabla\theta+n_c\bm\nabla\optheta_1\right)=0\\
    \label{eq:eqth1}
    &\partial_t\optheta_1+ \frac1m\bm\nabla\theta\cdot\bm\nabla\optheta_1+\lambda \opn_1-\frac{\hbar^2}{2m}D_2\opn_1=0.
\end{align}
It is crucial to understand that in Eqs.~\eqref{eq:eqn1} and~\eqref{eq:eqth1}, the backreaction, i.e. the impact of quantum matter fields on the structure and dynamics of spacetime as a result of their interactions~\cite{hu_verdaguer_2020}, of the quantum fluctuations on the background solution has been considered negligible.
Substituting $\opn_1$, derived from ~\eqref{eq:eqth1} into Eq.~\eqref{eq:eqn1}, one gets a wave equation for $\optheta_1$. This is a partial differential equation (PDE), second-order in time derivatives and infinite-order in space derivatives. To simplify things, it is convenient to build a symmetric $4\times4$ matrix $f^{\mu\nu}(t,\bm x)$ such that
\begin{align}
    &f^{00}=-\left[ \lambda-\frac{\hbar^2}{2m}D_2 \right]^{-1}\\
    &f^{0j}=-\left[\lambda-\frac{\hbar^2}{2m}D_2\right]^{-1}\frac{\nabla^j\thetao}{m}\\
    &f^{i0}=-\frac{\nabla^i\thetao}{m}\left[\lambda-\frac{\hbar^2}{2m}D_2\right]^{-1}\\
    &f^{ij}=\frac{n_c\delta^{ij}}{m}-\frac{\nabla^i\thetao}{m}\left[\lambda-\frac{\hbar^2}{2m}D_2\right]^{-1}\frac{\nabla^j\thetao}{m}.
\end{align}
Indeed, after introducing $3+1$-dimensional spacetime coordinates $x^\mu\equiv(t,\bm x)$, we obtain the following wave equation for $\optheta_1$
\begin{equation}
    \partial_\mu(f^{\mu\nu}\partial_\nu\optheta_1)=0.
\end{equation}
If we perform a spectral decomposition of the field $\optheta_1$, we can see that for wavelengths greater than $\hbar/\sqrt{2m\lambda n_c}$ the terms resulting from the linearized quantum potential can be neglected, in which case $f^{\mu\nu}$ can be approximated by numbers rather than differential operators. The quantity $\xi=\hbar/\sqrt{2m\lambda n_c}$ is called \emph{healing length}, and since for wavelength grater than $\xi$ we have equations which are equivalent to that of hydrodynamics, for this constraint we are in the hydrodynamic approximation~\cite{hydro,RevModPhys.80.1215}. If we now identify
\begin{equation}
    \sqrt{-g}g^{\mu\nu}= f^{\mu\nu}, 
\end{equation}
the wave equation for $\optheta_1$ becomes
\begin{equation}
    \Box\optheta_1\equiv \frac{1}{\sqrt{-g}}\partial_\mu(\sqrt{-g}g^{\mu\nu}\partial_\nu)\optheta_1=0. 
\end{equation}
This is the equation of motion of a massless, minimally coupled, quantum scalar field over a curved background, with an effective metric
\begin{equation}
    g_{\mu\nu}(t,\bm x)\equiv\frac{n_c}{m c_s(a,n_c)}\left(
    \begin{matrix}
        -c_s^2(a,n_c)+v^2\ & -v_j\\
        -v_i\ & \delta_{ij}
    \end{matrix}
    \right), 
\end{equation}
with 
\begin{equation}
    c_s^2(a,n_c)=\frac{\lambda n_c}{m}
\end{equation}
the phonon speed in the medium.

\section{Hawking radiation}

We have already seen that black holes are not completely black
objects: when including quantum effects, they emit a thermal radiation at the Hawking temperature. Hawking radiation is a quantum-field in curved space effect. Its existence does not originate from Einstein's equation: it is a kinematic effect depending only on the existence of a Lorentzian metric and some sort of horizon. As such, Hawking radiation can be reproduced in a condensed matter system under appropriate constraints (see~\cite{PhysRevLett.46.1351}). In particular, to obtain Hawking radiation in laboratory one has to:
\begin{itemize}
    \item choose an appropriate quantum analogue model with a classical effective background spacetime, and some standard relativistic quantum fields residing there;
    \item tune the effective geometry so to equip the analogue system with an acoustic horizon.
\end{itemize}
In reality, the existence of a locally definable apparent horizon is quite sufficient to obtain the Hawking effect.

\begin{figure}[ht]
    \centering
    \includegraphics[scale=1]{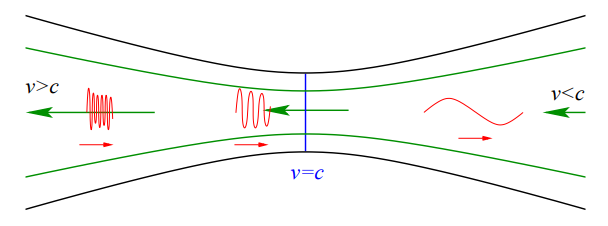}
    \caption{\emph{Analogue horizons and Hawking radiation. Illustration of a de Laval nozzle producing a transonic fluid flow. The blue segment is the analogue of an event horizon, where Hawking radiation originated (red curves). The green lines/arrows denote the fluid flow. The black lines are the walls of the nozzle. From}~\cite{shutz}.}
    \label{fig:hawkingradiation}
\end{figure}
Studying the analogue Hawking radiation is very useful, because it allows us to work on the unresolved problems we have already discussed in Chapter~\ref{chap:2}. Fig.~\ref{fig:hawkingradiation} illustrates a toy model capturing the relevant features that one can expect from the emission of Hawking radiation. As a consequence, considering for example the metric in Eq.~\eqref{eq:metric1}, instead of the Schwarzschild geometry~\cite{PhysRevLett.46.1351}, one gets the Hawking temperature
\begin{equation}
    T_{H}=\frac{\hbar}{2\pi k_B}\left|\opn\cdot\bm\nabla(v_0-c_s)\right|_{v_0=c_s}\, .
\end{equation}
Here, $\opn$ is the unit vector normal to the surface at $\bm x_H$, i.e. the analogous of the event horizon.
The value of this temperature depends on the experimental realization and it can range from a few nano-Kelvin in BECs~\cite{PhysRevA.63.023611} up to fractions of a Kelvin in electromagnetic waveguides~\cite{PhysRevLett.95.031301}. Though yet very hard to be measured, these values are still much larger than the Hawking temperatures of real solar-size black holes.

Since the concept of analogue Hawking radiation was introduced, there has been a significant theoretical effort to look into it within analog black holes. We have seen that the Hawking temperature is proportional to the gradient of the velocity at the acoustic horizon. This value cannot be arbitrary large: it has to be on the same scale of the coherence length, otherwise the hydrodynamic hypothesis breaks. This suggests that the expected power loss from the Hawking emission may be calculated to be on the order of $P\simeq 10-48$ W in a cold system with a low speed of sound like in a BEC~\cite{analogue2}. Thus, due to the condensate's finite temperature, it may be too faint to be seen above the thermal phonon background. Despite this, Carusotto \emph{et al.}~\cite{PhysRevA.78.021603, num} predicted that Hawking radiation could be observed after measuring the density correlation function between emitted and absorbed Hawking particles. In a 2015 preliminary paper~\cite{stein1}, Steinhauer showed that the
density-density correlation function can be used to measure the entanglement between the Hawking and the partner particles; the following year, Steinhauer observed spontaneous Hawking radiation emanating from an analogue black hole in an atomic Bose–Einstein condensate~\cite{stein}.

\subsection{Experimental black hole evaporation}

The possibility of observing black-hole evaporation is one motivation of the present thesis. For this reason, we now dedicate a writeup space to recall in deeper detail what Unruh derived in his 1981 paper~\cite{PhysRevLett.46.1351}. He was indeed motivated by the fact that the experimental investigation of the black hole evaporation would seem to be virtually impossible. He decided to consider a model which has all of the properties of a black hole, as far as the quantum thermal radiation is concerned, but in which all of the basic physics is completely understood. The motion of sound waves in a convergent fluid flow serves as a model for the behavior of a quantum field in a classical gravitational environment.
He considered an inviscid, irrotational and barotropic fluid. Unruh's hypotesis are the same made in Sec.~\ref{sub:acustic}, thus, following the same steps, we obtain for the field $\phi$ Eq.~\eqref{eq:soundwave}.
This is the equation for a massless scalar field in a geometry with metric that, after defining $c^2(\rhoo)=g'(\ln\rhoo)$ as the local velocity of sound (considered by the author as constant for simplicity), is:
\begin{equation}
    ds^2=\frac{\rho_0}{c(\rho_0)}\{\left[c^2(\rhoo)-\bm v_0\cdot \bm v_0\right]dt^2+2dt\bm v_0\cdot d\bm x-d\bm x\cdot d\bm x\}.
\end{equation}
At this point, Unruh wanted to reproduce an acoustic metric which was very similar to the Schwarzschild one. Thus, he considered the background flow to be with a spherical symmetry, stationary and convergent, defined a new time, and assumed that at some value $r=R$ the velocity of the fluid become faster that the speed of sound:
\begin{equation}
    \tau=t\int\frac{v_0^r(r)dr}{c^2-v_0^{r2}(r)}\qquad v_0^r=-c+\alpha(r-R)+O((r-R)^2). 
\end{equation}
As a result, and after dropping the angular part, the new metric became
\begin{equation}
    ds^2\simeq\frac{\rhoo(R)}{c}\left(2c\alpha(r-R)d\tau^2-\frac{dr^2}{2\alpha(r-R)}\right).
\end{equation}
This metric is very similar to the Schwarzschild metric near the horizon. 

At this point, Unruh was interested in quantizing the field $\phit$. The latter can be expanded in terms of the modes $\phit_\omega$
\begin{equation}
    \phit=\sum_\omega(a_\omega\phit_\omega+a_\omega^\dagger\phit_\omega^*)+\mathrm{ingoing\ parts\ of\ }\phit, 
\end{equation}
 with
 \begin{equation}
    \phit_\omega
     \begin{cases}
         \exp\left(i\omega(\tau-\frac{1}{2\alpha}\ln(r-R))\right)+B_\omega\exp\left(i\omega(\tau+\frac{1}{2\alpha}\ln(r-R))\right) \qquad & r\to R\\
         A_\omega\exp\left(i\omega(\tau-\frac{r}{c})\right) & r\to\infty .
     \end{cases}
 \end{equation}
Here, Unruh considered that an observer travelling with the fluid would interpret the field's $\phit$ state as resembling the vacuum state, and thatthe appropriate physical time for this observer is $t$. The state for the fluid near the horizon is given by the part with positive frequency with respect to $t$. If now one defines the co-moving radial coordinate $\Tilde{r}(r,t,t_0)$, one finds that near the horizon the modes are
\begin{equation}
    \phit_\omega\simeq\left(t-t_0+\frac{\Tilde{r}}{c}\right)^{i\frac{\omega}{\alpha}}\times(\mathrm{a\ smooth\ function}), 
\end{equation}
which corresponds to the modes in the Schwarzschild case.
This implies that the sonic black hole emits sound waves with a thermal spectrum characterized by the temperature
\begin{equation}
    T=\left.\frac{\hbar}{2\pi k_B}\frac{\partial v^r}{\partial r}\right|_H,
\end{equation}
where with subscript $H$ we mean that the derivative of $v^r$ with respect to $r$ is valuated at the horizon.

\subsection{The density-density correlation function\label{sub:carusotto}}

As anticipated, in 2008 Roberto Balbinot \emph{et al.}~\cite{PhysRevA.78.021603} proposed a method to spot an acoustic black hole's emission in a moving BEC. Given that their results are relevant to the discussion presented in Chapter~\ref{chap:4}, we now recall the details of their derivation. Notice that in the case of astrophysical black holes, in principle, we could access only to the outside region. As a consequence, for these objects it would be impossible to study the correlation function between the Hawking partners. In the case of acoustic black hole instead, both its external and internal region are accessible to the experiment. A significant long-range correlation between the density fluctuations at different locations inside and outside the acoustic black hole is caused by quantum-optical correlations between the correspondingly (inside and outside) emitted Hawking radiation. 

\begin{figure}
    \centering
    \includegraphics[scale = 0.6]{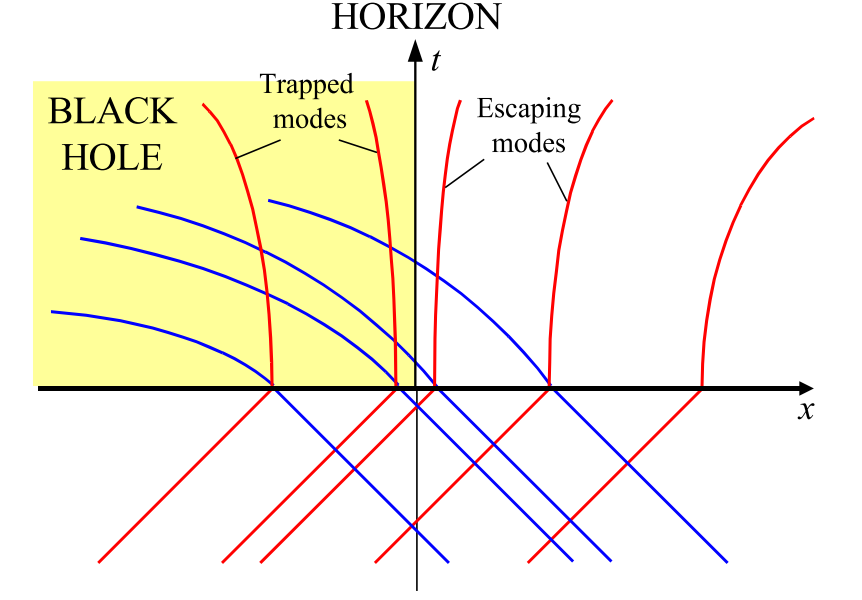}
    \caption{\emph{Hawking radiation. Pictorial representation of the creation process. The blue lines represent the ingoing modes, and the red lines the outgoing modes. Modes propagate in straight lines before the formation of the black hole ($t< 0$). When the horizon forms (in $x = 0$), propagation is greatly warped, trapping the outgoing modes inside the black hole. From}~\cite{PhysRevA.78.021603}.}
    \label{fig:hawkproc}
\end{figure}

At the basis of the~\cite{PhysRevA.78.021603} reasoning, for both astrophysical and acoustic black holes, it was considered that at early times before the horizon creation, the quantum field associated to Hawking radiation is in its vacuum state $|in\rangle$. At later times after the horizon formation, the vacuum state would then be $|out\rangle$. It is then possible to define a basis for the \emph{out modes}, i.e. modes obtained in correspondence of $|out\rangle$ as vacuum state. This is composed by:
\begin{itemize}
    \item ingoing modes, those which propagate downstream in the free-falling frame, and which have a vacuum state $|0_{ing}\rangle$;
    \item outgoing modes, those which propagate upstream, and which have a vacuum state $|0_{outg}\rangle$.
\end{itemize}
The outgoing modes include modes which are allowed to propagate outside the horizon and to reach the asymptotic region, and modes which are trapped in the horizon. Hawking radiation relates to particle production in the outgoing sector only. Limiting the discussion to this latter sector, the late-time decomposition for the $|in\langle$ vacuum in terms of these fleeing and trapped modes takes the form of a two-mode squeezed vacuum state
\begin{equation}
    |in\rangle\propto \exp\left(\frac1\hbar\sum_\omega e^{-\frac{\hbar\omega}{2k_B T_H}}a_\omega^{(esc)\dagger}a_\omega^{(tr)\dagger}\right)|0_{outg}\rangle.
\end{equation}
Here, $a_\omega^{(esc)\dagger}$ and $a_\omega^{(tr)\dagger}$ are the creation operators for outgoing escaping and trapped modes, respectively, and $T_H$ is the Hawking temperature. Thus, the horizon formation causes the Hawking process, which produces correlated pairs of outgoing quanta~\cite{PhysRevD.54.7444}. A scheme of this process is shown in Fig.~\ref{fig:hawkproc}.

Moving now to an analog model based on a flowing atomic BEC, in~\cite{PhysRevA.78.021603} they were next interested in the small fluctuations around a stationary and fully condensed state. Thus, considering the Bose field operator
\begin{equation}
    \bm \oppsi = e^{i\optheta}\sqrt{\opn},
\end{equation}
with $\optheta$ a phase and $\opn$ the number density, the expansion
\begin{align}
    \opn=n+\opn_1\\
    \optheta=\thetat+\optheta_1
\end{align}
is performed around the classical background values. Also, the hydrodynamic limit is considered, with $\lambda>>\xi$ and $\xi$ the healing length. From the linearization of the equation of motion for $\optheta_1$, one obtains that 
\begin{equation}
    \Box\optheta_1=\frac{1}{\sqrt{-g}}\partial_\mu(\sqrt{-g}g^{\mu\nu}\partial_\nu)\optheta_1=0, 
\end{equation}
with $g_{\mu\nu}$ the acoustic metric 
\begin{equation}
    ds^2=g_{\mu\nu}dx^\mu dx^\nu =\frac{n}{mc}[-c^2dt^2+(d\bm x-\bm v dt)\cdot(d\bm x - \bm v dt)]\, , 
    \label{eq:metric2}
\end{equation}
in fact equivalent to a curved-spacetime field
equation for a massless scalar field. In Eq.~\eqref{eq:metric2}, $\bm v =\hbar\bm \nabla\theta/m$ is the local flow velocity and $c=\sqrt{gn/m}$ is the local sound speed.
The one-time density-density correlation function is then defined as
\begin{equation}
    G_2(x,x')=\langle\opn(x)\opn(x')\rangle-\langle\opn(x)\rangle\langle\opn(x')\rangle.
\end{equation}
Exploiting the relation between $\opn_1$ and $\optheta_1$, it can be recast in the form:
\begin{equation}
    G_2(x,x')=\frac{\hbar^2}{g(x,t)g(x',t)}\lim_{t'\to t}\mathcal{D}\langle\optheta_1(x,t)\optheta_1(x',t')\rangle, 
\end{equation}
with $\mathcal{D}=[\partial_t\partial_{t'}+\bm v(\bm x)\cdot\bm\nabla\partial_{t'}+\bm v(\bm x')\cdot(\partial_t\bm\nabla')+\bm (\bm v(\bm x)\cdot\bm \nabla)(\bm v(\bm x')\cdot\bm \nabla')]$. Having in mind a quasi-one-dimensional (1D) experimental geometry, after performing a dimensional reduction to 1D with transverse size $\ell_T$ assumed to be much smaller than the healing length $\xi$, one obtains:
\begin{equation}
    \langle\optheta_1(x,t)\optheta_1(x',t')\rangle\simeq-\frac{1}{4\pi\sqrt{C(x,t)C(x',t')}}\ln(\Delta x^-\Delta x^+)\, .
\end{equation}
Here, $C=n_{1D}\xi$ is the conformal factor of the metric~\eqref{eq:metric2}, with $n_{1D}=n\ell_T^2$ the 1D effective density, and $x^\pm = t\pm\int\frac{dx}{c\mp v}$ are the light cone coordinates.
In this derivation, the Polyakov approximation \footnote{When applying adimensional reduction in the equation of motion for a massless scalar field, the latter is coupled not only to the metric, but also to a "dilation" term, which depends on the transverse area of the system. This term causes back-scattering of the modes. Neglecting the back-scattering term is called Polyakov approximation~\cite{balbinot2006hawking}.} has been applied, the result remaining affected only in terms of a slight quantitative overestimation.

Using this approach to a spatially homogeneous one-dimensional BEC with density $n_{1D}$ and moving at a constant and spatially uniform speed $v$, for $|x-x'|\gg\xi$ one finds
\begin{equation}
    G_2^{1D}(x,x')=-\frac{n_{1D}\xi}{2\pi(x-x')^2}.
\end{equation}

Instead, when applying the same procedure to a one-dimensional BEC
in the presence of the sound-velocity profile
\begin{equation}
    c(x,t)=
    \begin{cases}
        c_r>v \qquad & t<0\\
        \begin{cases}
            c_l<v & x<-x_0\\
            \frac{c_r-c_l}{2x_0}x+\frac{c_r+c_l}{2} & -x_0\le x\le x_0\\
            c_r>v & x>x_0
        \end{cases} & t\ge0, 
    \end{cases}
\end{equation}
\begin{figure}[ht!]
    \centering
    \includegraphics[scale=0.5]{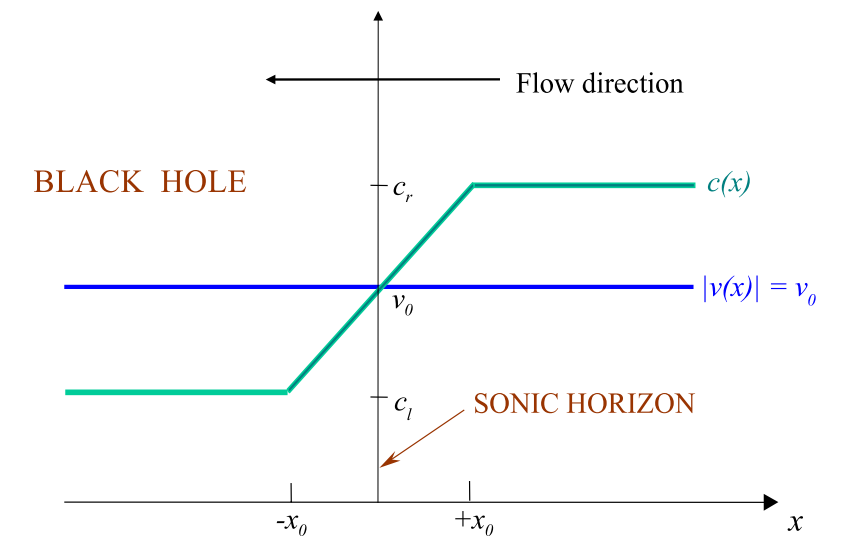}
    \caption{\emph{Detection of Hawking radiation in acoustic black holes. Scheme of the speed of sound profile proposed in}~\cite{PhysRevA.78.021603}. \emph{From}~\cite{PhysRevA.78.021603}.}
    \label{fig:horizoncarusotto}
\end{figure}
such that in $x=0\ v=c$ (see Fig.~\ref{fig:horizoncarusotto}), the correlation function is
\begin{equation}
    \frac{G_2^{1D}(x,x')}{n_{1D}^2}\simeq-\frac{k^2\xi_l\xi_r}{16\pi c_l c_r}\frac{1}{\sqrt{(n_{1D}\xi_r)(n_{1D}\xi_l)}}\frac{c_rc_l}{(c_r-v)(v-c_l)}\frac{1}{\cosh^2{\left[\frac{k}{2}\left(\frac{x}{c_r-v}+\frac{x'}{v-c_l}\right)\right]}}\, ,  
\end{equation}
where $k=\left(\frac{dc}{dx}\right)_H$ is the surface gravity and $x$ and $x'$ are taken such that $x'<-x_0$ and $x>x_0$. The correlation function presents a valley shaped feature centered in $\frac{x'}{v-c_l}=-\frac{x}{c_r-v}$. The interpretation of this valley is that at each time, pairs of Hawking phonons emerge from the horizon region almost at the same time, and propagate in the sub-sonic and super-sonic region with velocities $c_r-v$ and $v-c_l$, respectively. 

This result from Roberto Balbinot \emph{et al.} offered a promising signature to identify and isolate the Hawking emission of phonons from acoustic black holes.

\subsection{From the density-density correlation function to the entanglement of analogue Hawking radiation}

In 2015, Steinhauer~\cite{stein1} has demonstrated that the entanglement of Hawking radiation pairs emitted by an analogue black hole, can be measured by the experimentally accessible density-density correlation function, thus greatly simplifying the measurement. Since the phonons going in the direction of flow are ignored, this approach depends on the fact that the thermal populations of phonons are negligible. The measurement of the phonon populations is more challenging than the measurement of the correlations between the Hawking and companion particles, because of the background quantum fluctuations. The method proposed in~\cite{stein1} is applicable to an analogue black hole with homogeneous upstream and downstream regions. Subsonic (or supersonic) flow in the $x$ direction is present in the upstream (downstream) region. The necessary Bogoliubov excitations travel against the flow in the fluid's local rest frame, and have positive wavenumbers $k$.

After defining the annihilation operator for a Bogoliubov excitation with wavenumber $k$ as $\opb_{k}^{u/d}$, the Peres-Horodecki criterion~\cite{PhysRevD.89.105024} can be used to determine the entanglement
\begin{equation}
    \label{eq:delta}
    \Delta\equiv\langle\opb_{k_{HR}}^{u\dagger} \opb_{k_{HR}}^u\rangle\langle\opb_{k_{P}}^{d\dagger}\opb_{k_{P}}^{d}\rangle-\left|\langle\opb_{k_{HR}}^u\opb_{k_{P}}^{d}\rangle\right|^2, 
\end{equation}
where $d$ and $u$ indicate the downstream and upstream regions, respectively, and $HR$ and $P$ the Hawking radiation and the partner. In Eq.~\eqref{eq:delta}, the first term represents the correlations between the Hawking and partner particles, while the second is the population of the Hawking particles. The correlations between the Hawking and partner particles are strong enough to show that they are entangled when $\Delta$ is negative.
It is convenient to write $\Delta$ in terms of the Fourier transform of the density operator~\cite{hydro} in the Bogoliubov approximation
\begin{equation}
    \rho_k=\sum_p\opa^\dagger_{p+k}\opa_p=\sqrt{N}(u_k+v_k)\left(\opb^\dagger_k+\opb_{-k}\right), 
\end{equation}
where $\opa_p$ is the annihilation operator for a single atom with momentum $\hbar p$, $N$ is the total number of atoms and $u_k$ and $v_k$ are the Bogoliubov amplitudes. The upstream and downstream regions can be treated separately
\begin{align}
    &\rho_k=\sqrt{N^u}(u_{k_{HR}} +v_{k_{HR}} ) \left( \opb^{u\dagger}_{k_{HR}}+\opb_{-k_{HR}}^u\right)\\
    &\rho_k=\sqrt{N^d}(u_{k_{P}} +v_{k_{P}} ) \left( \opb^{d\dagger}_{k_{P}}+\opb_{-k_{P}}^d\right).
\end{align}
Here, in~\cite{stein1},it was defined a generalized version of the usual static structure factor for a homogeneous system, that is:
\begin{equation}
    \begin{split}
        \langle \rho_{k_i}^i\rho_{k_j}^j\rangle=\sqrt{N^iN^j}\left(u_{k_i}+v_{k_i}\right)\left(u_{k_j}+v_{k_j}\right)\Big[\langle&\opb_{k_i}^{i\dagger}\opb_{k_j}^{j\dagger}\rangle+\langle\opb_{k_i}^{i\dagger}\opb_{-k_j}^{j}\rangle+ \\
        +\langle& \opb_{-k_i}^{i}\opb_{k_j}^{j\dagger}\rangle+\langle\opb_{-k_i}^{i}\opb_{-k_j}^{j}\rangle+\delta_{ij}\delta_{-k_ik_j}\Big] , 
    \end{split}
\end{equation}
 where $i$ and $j$ can each be either $u$ or $d$.
 The aim is to compute $\Delta$, thus one has to consider all the combinations for $i,j=u,d$. To do this, Steinhauer assumed a sufficiently low condensate's temperature, to be able to neglect the number of generated excitations travelling with the flow. Under these conditions, any term with negative $k$-value can be set to zero. As a consequence, one obtains that:
\begin{align}
   &\langle\rho_{-k_{HR}}^u\rho_{-k_{P}}^d\rangle=\sqrt{N^uN^d}\left(u_{k_{HR}}+v_{k_{HR}}\right)\left(u_{k_{P}}+v_{k_{P}}\right)\langle\opb_{k_{HR}}^{u}\opb_{k_{P}}^{d}\rangle\\
   &\langle\rho_{k_{HR}}^u\rho_{-k_{P}}^u\rangle=N^u\left(u_{k_{HR}}+v_{k_{HR}}\right)^2\left[\langle\opb_{k_{HR}}^{u\dagger}\opb_{k_{HR}}^{u}\rangle+1\right]\\
   &\langle\rho_{k_{P}}^d\rho_{-k_{P}}^d\rangle=N^d\left(u_{k_{P}}+v_{k_{P}}\right)^2\left[\langle\opb_{k_{P}}^{d\dagger}\opb_{k_{P}}^{d}\rangle+1\right]. \\
\end{align}
\begin{figure}[ht!]
    \centering
    \subfloat[]{\includegraphics[scale = 0.5]{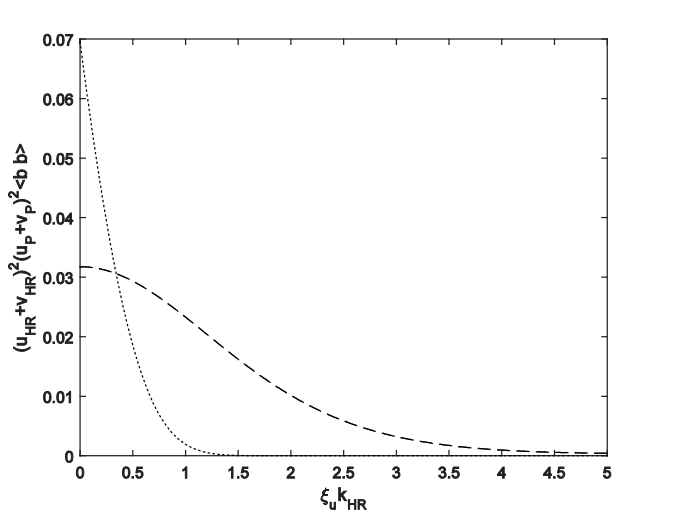}}
    \subfloat[]{\includegraphics[scale = 0.5]{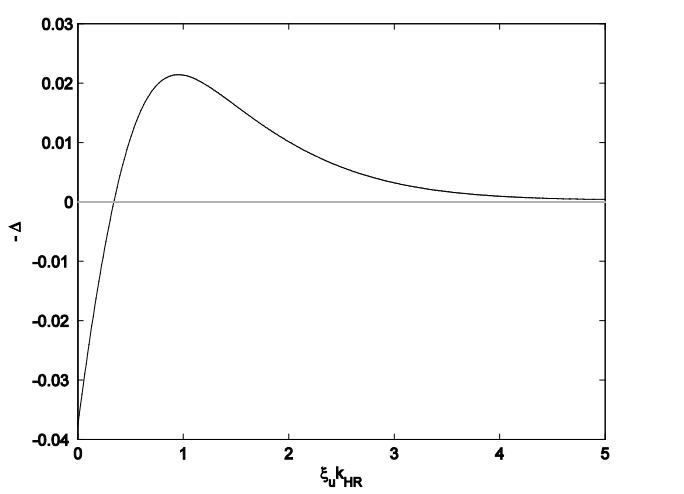}}
    \caption{\emph{Entanglement between the Hawking and partner particles. (a) The two terms in the entanglement parameter $\Delta$ as functions of $\xi_uk_{HR}$, with $\xi_u$ the upstream healing length. The correlations between the Hawking and companion particles are shown by the dotted curve. The population of Hawking particles (i.e. the Planck distribution) is represented by the dashed curve. Entanglement is indicated by the dotted curve exceeding the dashed curve. (b) $-\Delta$ as a function of $\xi_uk_{HR}$. Positive values of $-\Delta$ correspond to entanglement. From}~\cite{stein1}.}
    \label{fig:delta}
\end{figure}
Substituting these expressions in Eq.~\eqref{eq:delta}, one gets the entanglement criterion in terms of $\rho_k$. To connect with the experimental images, it is convenient to pass from the 2$^{\mathrm{nd}}$ quantized form of $\rho_k$ to its explicit expression as a Fourier transform
\begin{equation}
    \langle \rho_{k_i}^i\rho_{k_j}^j\rangle=\int dxdy e^{-ik_ix}e^{-ik_jy}\langle n^i(x)n^j(y)\rangle.
\end{equation}
Therefore, the entanglement function $\Delta$ is computed from
\begin{equation}
    \langle n^u(x)n^d(y)\rangle\qquad\langle n^u(x)n^u(y) \rangle \qquad\langle n^d(x)n^d(y)\rangle.
\end{equation}
Fig.~\ref{fig:delta}(a) displays the two terms which constitute Eq.~\eqref{eq:delta}. Here, one can notice that the correlations between the Hawking and partner particles are weaker but broader than the Planck distribution. As a result, only for large $k$ the Hawking/partner correlations outperform the Planck distribution, indicating that these $k$-values are entangled.
This is evident in Fig.~\ref{fig:delta}(b), which shows $\Delta$ and, therefore, the entanglement. The Planck distribution's high energy tail is observed to have entangled $k$-values, whereas its low $k$ values do not.

\subsection{Observation of analogue quantum Hawking radiation}

\begin{figure}[ht!]
    \centering
    \subfloat[]{\includegraphics[scale = 0.7]{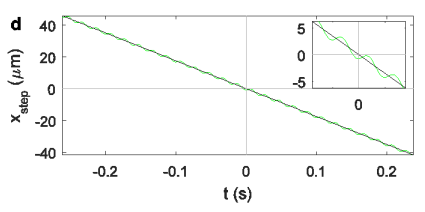}}
    \subfloat[]{\includegraphics[scale = 0.6]{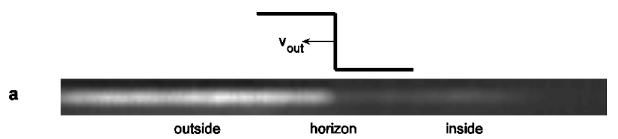}}
    \caption{\emph{Experimental setup for the observation of Hawking radiation in Steinhauer lab~\cite{stein}. (a) The position of the step potential vs. time. The black line represents the step potential used to generate the horizon, the green line the step potential used to make the horizon oscillate. (b) Experimental data representing the realized acoustic hole horizon. In particular, the 1D BEC, which traps the phonons in the right region. From}~\cite{stein}.}
    \label{fig:steppot}
\end{figure}
The preparatory theoretical work illustrated so far has preluded to the first ever observation in 2016 of analogue Hawking radiation in the same Steinhauer lab~\cite{stein}, happened in 2016. In the following, we briefly present the experiment setting and results. A BEC of $^{87}$Rb atoms was radially confined by a narrow laser beam. As the radial trap frequency is higher than the maximum interaction frequency, the behavior is effectively one-dimensional. The horizon was created by means of a very sharp potential step, swept along the condensate at a constant speed, and made possible by short-wavelength laser light and high-resolution optics. The step potential is the black line shown in Fig.~\ref{fig:steppot}(a).
Since the horizon width amounts to a few healing lengths, the horizon is hydrodynamic. A sketch of the system is shown in Fig.~\ref{fig:steppot}(b). Deep in the area outside the left of the step, the condensate is at rest. While one gets closer to the step, the flow speed picks up. The potential drop causes the condensate to flow at supersonic speed to the right of the step. The reference frame is important, in which the step is at rest in the origin. In this configuration, in the outside region, the condensate flows from left to the right at velocity $\nu_{out}$, which is slower than the speed of sound, and thus the phonons are allowed to travel against the flow and escape from the acoustic black hole. In the inside region instead, the condensate flows at velocity $\nu_{in}$, which is faster than speed of sound, so that phonons are trapped in, as it would happen for a black hole. 
A preliminary experiment is carried out to assess the horizon sharpness, by using an oscillating potential in order to explore the analog horizon (see Fig.~\ref{fig:steppot}(a) and~\cite{stein} for further details). The dispersion relation for phonons inside and outside the horizon has been studied thanks to this preliminary activity. In fact, a phonon is predicted to only exhibit linear dispersion at low energies, hence the validity of the experiment is determined by the energy level at which the quadratic correction can be disregarded.
\begin{figure}
    \centering
    \subfloat[]{\includegraphics[scale = 0.6]{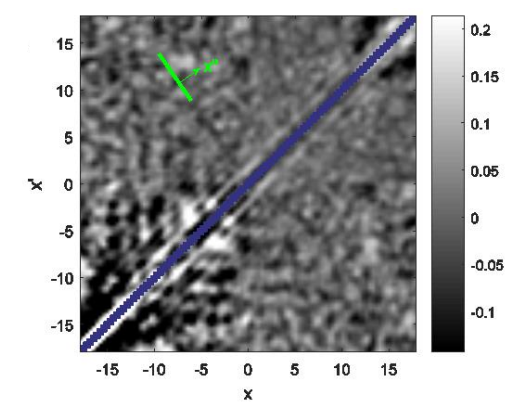}}
    \subfloat[]{\includegraphics[scale = 0.7]{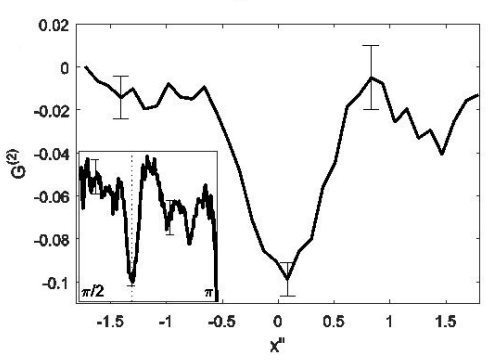}}
    \caption{\emph{Observation of Hawking/partners pair correlations in Steinhauer lab~\cite{stein}. (a) Two-body correlation function. The horizon is located in the origin, so that the dark bands represent the correlations between the Hawking and partner particles. (b) Measured profile of the Hawking-partner correlations. The entanglement of the Hawking pairs is measured after Fourier transforming this curve. The error bars indicate the standard error of the mean. From}~\cite{stein}.}
    \label{fig:correlation}
\end{figure}

The correlation function was then computed for a set of 4600 repetitions of the experiment. In the result, the effects due to imaging shot-noise at high frequencies have been removed, as well as effects of imaging fringes and overall slopes caused by the profile of the acoustic black hole. The data expressing the correlation between upwards and downwards phonons are shown in Figs.~\ref{fig:correlation}. 

\begin{figure}
    \centering
    \includegraphics[scale = 0.8]{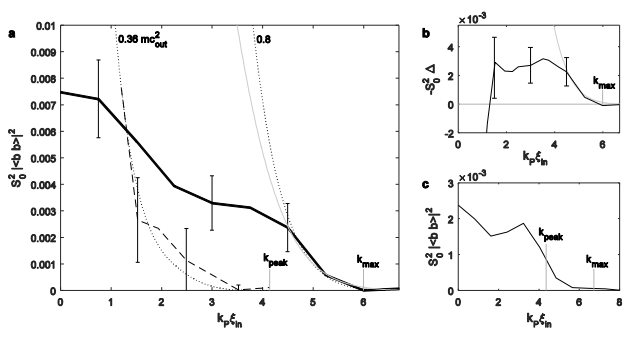}
    \caption{\emph{Observation of entanglement between the Hawking pairs in Steinhauer lab~\cite{stein}. (a) Correlations between the Hawking and partner particles (solid line) and squared population (dashed line). The occurrence of entanglement corresponds to the solid curve exceeding the dashed line. The dotted curves represent the theoretical population squared for the values $k_BT_H=0.36 mc^2_{\mathrm{out}}$ and $k_BT_H=0.8 mc^2_{\mathrm{out}}$ of the Hawking temperature, respectively. The maximally entangled Heisenberg limit corresponding to $k_BT_H=0.36 mc^2_{\mathrm{out}}$ is indicated by the solid gray curve. The error bars indicate the standard error of the mean. (b) Difference between the solid and the dashed lines of (a), for enhanced view of entanglement occurrence: positive values of $-\Delta$ signal entanglement. (c) Numerical simulation corresponding to (a). From}~\cite{stein}.}
    \label{fig:entanglement}
\end{figure}

In the experiment, a technique was also developed to observe the real and virtual phonons. From the static structure factor of the density fluctuations in the spatial region outside the analogue black hole (i.e. where the Hawking radiation is observed in Fig.~\ref{fig:steppot}(a)), the phonon population was obtained (included the Hawking radiation as well as any background phonons). This is shown in Fig.~\ref{fig:entanglement}(a) with the dashed curve line.

As a final achievement, the entanglement between the Hawking pairs was measured. The result is shown in Figs.~\ref{fig:entanglement}.

Triggered by these findings, a number of BECs experiments have been performed in diverse geometries. For more recent platforms, see e.g.~\cite{2005.04027.pdf}.

\chapter{Effective field theory for an analogue model\label{chap:4}}

In the present chapter we present our innovative results about the low-energy excitations of a transonic fluid in a BEC. 
In Sec.~\ref{sec:BEC}, we have already presented an approach to build an analogue model for a Bose-Einstein condensate (BEC) starting from the Gross-Pitaevskii equation. Here, we use an effective field theory (EFT) for a system with a $U(1)$ and Lorentz symmetry breaking. This method presents several benefits. First, it allows us to integrate out the degrees of freedom which we are not interest in. Second, it is possible to apply this procedure to any system which presents a Lorentz and $U(1)$ symmetry breaking. Third, the solutions found in this chapter in the mean field approximation, can be easily extended to the local field approximation, after considering 1-loop correction terms beyond tree-level terms. 

For generality and notation convenience, we discuss a relativistic theory, however we will also illustrate the procedure for recovering the non-relativistic limit.

We apply this method to a system described by the Lagrangian density for a complex scalar field with a quartic interaction. In particular, 
we first study a standard Lagrangian with a $\phi^4$ self-interaction. This is useful because the introduction of this potential causes a $U(1)$ symmetry breaking that generates Nambu-Goldstone bosons. Therefore, we leverage this context to represent a background in which a fluid flows.
Then, we present a general effective Lagrangian~\cite{doi:10.1146/annurev.ns.43.120193.001233, manohar2018introduction,kaplan2005lectures,Dubovsky_2012} which describes the BEC's excitations at the leading order (LO) and at the next to the leading order (NLO). As we stated above, this innovative approach allows us to apply the results to any system in which there is a $U(1)$ and Lorentz symmetry breaking. We compare our results with those obtained by considering different approaches, as, for instance, that proposed by Son and Wingate~\cite{https://doi.org/10.48550/arxiv.hep-ph/0204199, Son2}. In the presence of an acoustic horizon, generated by the BEC's flow, we determine the dispersion law for the phonons, corresponding to the low-energy excitation of the microscopic Lagrangian. We find that, while considering a generic effective Lagrangian, in some specific cases the next-to-leading order phonon dispersion law has a nontrivial minimum that is dependent on the superfluid-flow velocity and the next-to-leading order low-energy constants (LECs). The non-trivial minimum's presence identifies a translations symmetry breaking, which indicates a phase transition, and a rotonic phonon creation. We think that the final phase is supersolid, but this hypothesis needs a demostration that is not object of study of this thesis. Then, we determine which values of the LECs give a nontrivial dispersion law. Finally, we compute the density-density correlation function for a homogeneous BEC for the system described by the microscopic Lagrangian and for the system in which the phononic dispersion law presents a non-monotony behavior. This correlation function is a fundamental quantity to experimentally detect and describe the analogue Hawking radiation. We notice that the propagator's leading term is given by the phonon, contribution which disappears when we take the limit in the normal phase. The obtained density-density correlation function for the system described by the microscopic Lagrangian, is compared with Haldane's result~\cite{PhysRevLett.47.1840}, which it is found to be consistent with. The novel approach that we propose to obtain this correlation function is useful in that it allows us to directly calculate the quantity we are interested in, and to easily calculate also all others correlation functions, though not subjects of this thesis work. 

\section{Brief overview of effective field theories}

In the universe in which we live, interesting physics phenomena occur at all scales. Consequently, it is often convenient to isolate certain phenomena from others, so that we can concentrate on specific ones. Low-energy EFTs are powerful tools to parameterize our lack of knowledge about short-range physics and describe the dynamics of experimentally accessible macroscopic physical systems.

The general concept is that if there exist energy (or length) scales that are significantly large or small compared to those of the physical quantities we are interested in, we can approximate the physics more easily by setting the small scale to zero and the large scale to infinity. The finite impacts of these scales can then be considered as minor deviations from this simplified approximation.

For example, Newtonian mechanics is still taught separately rather than as an upper bound of relativistic mechanics for low velocities. When all velocities are much slower than the speed of light, we can completely disregard relativity. It's not that a fully relativistic treatment of mechanics is inherently flawed; it is just simpler to ignore relativity when it is not necessary.

However, an EFT is not solely a matter of convenience. We now understand that interactions exist at each energy level in our universe. Given a certain energy scale $E$, we can describe our system with a certain accuracy $\varepsilon$ by using an EFT that involves a finite number of parameters. In EFTs only the relevant interactions are included, by an appropriate power counting approximation. This allows us to include only the necessary terms in the EFT Lagrangian to ensure accurate results.

Another reason to employ an EFT is that these theories make symmetries apparent. The chiral Lagrangian~\cite{PhysRevLett.18.188,PhysRev.183.1245,PhysRev.183.1261}, which describes pions as Goldstone bosons, provides a classic example of this concept. The derivative expansion serves as a natural perturbative scheme to characterize the dynamics of Goldstone bosons, for momenta below the symmetry breaking scale. The symmetries dictate the possible allowed interactions.

In general, constructing a Lagrangian in EFTs requires to identify the dynamical degrees of freedom present in our system. In some cases, this task is straightforward, such as for weakly coupled low-energy versions of a theory, where only the light fields need to be retained. However, in many cases, identifying the degrees of freedom in an EFT can be challenging. Additionally, there is no unique or correct way to choose the fields used in an interacting quantum field theory; the same system can be described by two different EFTs. However, these two EFTs must share the same symmetries, thus it is possible to map one EFT to the other. Clearly, the physical observables do not depend on the chosen EFT. 

\section{The standard BEC microscopic Lagrangian\label{sec:micro_lagrangian}}

In this section we introduce a microscopic Lagrangian, that is able to describe and emphasise the features of our system. We are interested in describing a bosonic system at finite density and vanishing temperature in the BEC phase. 

\begin{figure}
    \centering
    \includegraphics[scale=0.6]{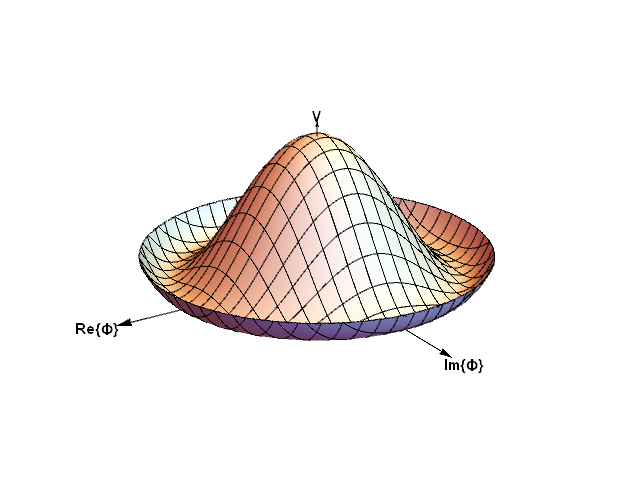}
    \caption{\emph{Potential of the field $\Phi$ displaying the $U(1)$ symmetry breaking. The 
    degenerate minimum corresponds to 
    $|\Phi|=\sqrt{\frac{\mu^2-m^2}{2\lambda}}$; the potential also presents a local maximum in $|\Phi|=0$.}}
    \label{fig:potenziale}
\end{figure}

To take into account the non-vanishing chemical potential, we start with a Lagrangian for a minimally coupled complex scalar field, $\Phi$, written as
\begin{equation}
    \lag=(D_\nu\Phi)^*D^\nu\Phi-m^2\Phi^*\Phi-\frac14F_{\mu\nu}F^{\mu\nu},
\end{equation}
where $m$ is the field's mass, $D_\mu=\partial_\mu-iA_\mu$ is the covariant derivative taking into account that $\Phi$ interacts with an external field $A_\mu$, and $F_{\mu\nu}=\partial_\mu A_\nu-\partial_\nu A_\mu$ is the strength of the field $A_\mu$. We can introduce the chemical potential $\mu$ as 
\begin{equation}
    A_\nu=\mu\delta_{\nu0},
\end{equation}
which we can assume to be constant. This expression clearly shows that the chemical potential explicitly breaks the Lorentz symmetry, leaving rotations and translations as the system's symmetries. Thus the Lagrangian becomes
\begin{equation}
    \lag=\partial_\nu\Phi^*\partial^\nu \Phi+i\mu(\Phi^*\partial_t\Phi-\Phi\partial_t\Phi^*) -(m^2-\mu^2)|\Phi|^2.
    \label{eq:lagphi}
\end{equation}
Note that the chemical potential gives two different contributions:
\begin{itemize}
    \item the second term on the RHS of Eq.~\eqref{eq:lagphi} corresponds to an explicit Lorentz symmetry breaking;
    \item the third term on the RHS of Eq.~\eqref{eq:lagphi} shows an effective mass shift that, as we shall see, in certain condition may lead to the breaking of $U(1)$ symmetry.
\end{itemize}
The latter can be shown by the fact that for $\mu=m$ the effective mass vanishes. In addition, for $\mu>m$ the potential
\begin{equation}
    V=(m^2-\mu^2)|\Phi|^2
\end{equation}
presents a maximum in $|\Phi|=0$. As a consequence, to have a non-zero vacuum expectation value, we have to introduce a repulsive term in the potential. The simplest is the repulsive contact interaction
\begin{equation}
    V=(m^2-\mu^2)|\Phi|^2+\lambda|\Phi|^4.
\end{equation}
In this way, for $\mu>m$ the potential (see Fig.~\ref{fig:potenziale}) has a minimum for 
\begin{equation}
    |\Phi|^2=\frac{\mu^2-m^2}{2\lambda}\qquad V=-\frac{(\mu^2-m^2)^2}{4\lambda}.
\end{equation}

Since the potential depends only on the modulus of $|\Phi|$, it is convenient to use the Madelung representation
\begin{equation}
    \Phi=\frac{1}{\sqrt{2}}\rho e^{i\theta}.
\end{equation}
Upon substituting this expression in the above Lagrangian and taking into account the contact interaction term, we obtain 
\begin{equation}
    \lag=\frac12\partial_\nu\rho\partial^\nu\rho+ \frac12\rho^2\partial_\nu\theta\partial^\nu\theta-\mu\rho^2\partial_t\theta-\frac12(m^2-\mu^2)\rho^2-\frac\lambda4\rho^4.
    \label{eq:microscopic}
\end{equation}
We notice that the field $\rho$ is characterized by both couplings, derivative and not derivative, and it is also massive; the field $\theta$ is massless and is characterized only by derivative couplings. Thus, the field $\theta$ is a cyclic variable and generates a current
\begin{equation}
\label{eq:conserved_current}
    J^\mu=\frac{\delta\lag}{\delta\partial_\mu\theta}=\rho^2\partial^\mu\theta,
\end{equation}
which is conserved, meaning that the system can have non-dissipative currents, i.e. it is a superfluid.

\section{The most general BEC effective Lagrangian \label{sec:effectivelagrangian}}

In the present section, we want to generalize the Lagrangian~\eqref{eq:lagphi} discussed in the the previous one. More in detail, we determine the low-energy Lagrangian of a system with a spontaneously broken global $U(1)$ symmetry. As a consequence, we assume that the system we are working on is described by the Lagrangian density
\begin{equation}
    \lag\equiv\lag(\rho,\partial_\mu\rho,\partial_\mu\theta)\,,
\label{eq:Lag_generic}
\end{equation}
with a cyclic coordinate $\theta$ as in Eq.~\eqref{eq:microscopic}. We want to linearize our equations, to make them tractable. In order to do this, we expand the fields as follows:
\begin{align}
    &\rho=\rhoo+\rhot+..\\
    &\theta=\thetao+\thetat+..\ .
\end{align}
Here, $\rhoo$ and $\thetao$ are the solutions of the classical equations of motion, corresponding to a stationary point of the action. Physically, they are associated to the density and to the currents of the background, respectively. The field $\rhot$ is the Higgs mode, and represents the density fluctuations; the field $\thetat$ is the Nambu-Goldstone boson (NGB) and represents the flowing fluid's excitations, which we will generically call phonons. 

In the following, we assume that the only external source to the system is given by the chemical potential $\mu$, and that the background presents a (super)flow with velocity $\bm v$, given by $\bm v= \bm \nabla \thetao$. The reason is that, since $\theta$ is a cyclic variable, there will be a conserved current as in~\eqref{eq:conserved_current}. We assume that the Langrangian system has non-dissipative terms, and therefore all the operators are hermitian. We also work in \emph{local density approximation}~\cite{article}: the momentum associated to the velocity is much lower than the momentum associated to $\partial_\mu\thetat$ and, therefore, the velocity derivatives can be considered negligible with respect to the $\partial_\mu\thetat$ derivatives.

We expand the action around the stationary point ($\rhoo,\thetao$), and, apart from surface terms, we have:
\begin{equation}
    \lag=\lag(\rhoo, \partial_\mu\rhoo, \partial_\mu \thetao)+\left.\rhot\left(\frac{\delta\lag}{\delta\rho}-\partial_\mu\frac{\delta \lag}{\delta\partial_\mu\rho}\right)\right|_{\rhoo,\thetao}
    +\left.\pmtht\frac{\delta\lag}{\delta\partial_\mu\theta}\right|_{\rhoo,\thetao}+\lag_2+\lag_3+..., 
\end{equation}
where $\lag_2$ and $ \lag_3$ are the quadratic and the cubic Lagrangian, respectively, and the dots $...$ refer to higher order terms. Since $\rhoo$ e $\thetao$ are solutions of the classical equations of motion, it follows $\lag(\rhoo,\partial_\mu\rhoo,\partial_\mu\thetao)$ is the background pressure. In addition, the linear terms in the perturbations vanish, and thus in the we are left with quadratic, cubic, and so on, terms in the fluctuations.
We are interested in developing the LO terms $\lag_2$, $\lag_3$, and $\lag_4$ to have an effective Lagrangian in terms of $\partial_\mu\thetat$, which describes also the field's interactions. In particular, we develop the Lagrangian up to fourth order because the third and the quartic ones are interaction terms.
As we have seen in the previous section, interaction terms play an essential role: in the first place, they allow to have a spontaneous symmetry breaking. In addition, as we shall see, they allow us to achieve a better approximation than the mean field approximation: the local-field approximation. 

In the following, we expose a general method to get all these terms without explicitly specifying the microscopic Lagrangian.

\subsection{Quadratic terms \label{sub:quadratic}}

We build the quadratic term of the general effective Lagrangian in Eq.~\eqref{eq:Lag_generic}. The quadratic terms in the fluctuations are obtained developing the Lagrangian density $\lag(\rho,\partial_\mu\rho,\partial_\mu\theta)$ in terms of the fluctuations $\rhot, \partial_\mu\rhot$ and $\partial_\mu\thetat$:
\begin{equation}
    \begin{split}
        \lag_2 =& \left.\frac{\rhot^2}{2}\frac{\delta^2\lag}{\delta\rho^2}\right|_{\rhoo,\thetao} + \left.\frac{\partial_\mu\rhot\partial_\nu\rhot}{2}\frac{\delta^2\lag}{\delta\partial_\mu \rho\delta\partial_\nu \rho}\right|_{\rhoo,\thetao} + \left.\frac{\partial_\mu\thetat\partial_\nu\thetat}{2}\frac{\delta^2\lag}{\delta\partial_\mu\theta\delta\partial_\nu\theta}\right|_{\rhoo,\thetao} +\left. \rhot\partial_\mu\rhot\frac{\delta^2\lag}{\delta\rho\delta\partial_\mu\rho}\right|_{\rhoo,\thetao} +\\
        &+ \left.\rhot\partial_\mu\thetat\frac{\delta^2\lag}{\delta\rho\delta\partial_\mu\theta}\right|_{\rhoo,\thetao} + \left.\partial_\mu\rhot\partial_\nu\thetat\frac{\delta^2\lag}{\delta\partial_\mu\rho\delta\partial_\nu\theta}\right|_{\rhoo,\thetao}.
    \end{split}
    \label{eq:quadratic}
\end{equation}
The microscopic Lagrangian in Eq.~\eqref{eq:microscopic} has no terms in which the field $\rho$ has derivative couplings. Here, we assume that, even in this effective Lagrangian, the field $\rho$ has no derivative couplings, and therefore the fourth and the sixth terms in Eq.~\eqref{eq:quadratic} vanish. We will discuss the effects of such terms later on.

The first term in Eq.~\eqref{eq:quadratic} is
\begin{equation}
    \left.\frac{\rhot}{2} \frac{\delta^2\lag}{\delta\rho^2}\right|_{\rhoo,\thetao}=-\frac{\mt^2}{2}\rhot^2, 
    \label{eq:mt}
\end{equation}
where we have defined $\mt$ as the effective mass of the field $\rhot$. 

The second term of $\lag_2$ is
\begin{equation}
    \left.\frac{\partial_\mu\rhot\partial_\nu\rhot}{2}\frac{\delta^2\lag}{\delta\partial_\mu\rho\delta\partial_\nu\rho}\right|_{\rhoo,\thetao} = \frac{E^{\mu\nu}}{2}\partial_\mu\rhot\partial_\nu\rhot,
\end{equation}
where we have defined $E^{\mu\nu}$, a tensor that may depend on the background. Since the medium effects are encoded in the minimal coupling and the $\rho$ field has no derivative couplings, $E^{\mu\nu} = \eta^{\mu\nu}$. 

The third term is
\begin{equation}
    \left.\frac{\partial_\mu\thetat\partial_\nu\thetat}{2}\frac{\delta^2\lag}{\delta\partial_\mu\theta\delta\partial_\nu\theta}\right|_{\rhoo,\thetao}=\frac{L^{\mu\nu}}{2}\partial_\mu\thetat\partial_\nu\thetat,
    \label{eq:b}
\end{equation}
where $L^{\mu\nu}$ is a matrix which may depend on the background. With assumptions similar to those used before, we can write $L^{\mu\nu}=B^2\eta^{\mu\nu}$, with $B$ a constant of dimension 1. 

The fifth term is
\begin{equation}
    \left.\rhot\partial_\mu\thetat\frac{\delta^2\lag}{\delta\rho\delta\partial_\mu\theta}\right|_{\rhoo,\thetao}= V^\mu\rhot\partial_\mu\thetat,
    \label{eq:v}
\end{equation}
where $V^\mu$ is a generic vector with energy dimension 2, which breaks the Lorentz symmetries of the Lagrangian. In our system, we have two sources of Lorentz symmetry breaking, the chemical potential $\mu$ and the background fluid velocity $v_\mu$. The velocity field is associated relative to the background, so it follows that $V^\mu$ is a linear combination of the two quantities. The vector is a real quantity: because of our assumption of non-dissipative terms, thus all the operators are hermitian. We note that the hermitian derivative operator is defined as $i\partial_\mu$ and not as $\partial_\mu$. Thus, one could think that $V^\mu$ is pure imaginary. This is not the case. This fact can be show in two ways:
\begin{enumerate}
    \item the $\thetat$ field is a phase, and this means that its hermitian conjugate is $\thetat\to-\thetat$ and thus $\partial_\mu\thetat\to-\partial_\mu\thetat$;
    \item we have introduced the field $\thetat$ as a phase, but the choice is arbitrary: instead of $e^{i\thetat}$, we could consider $e^{i\int dx^\mu A_\mu(x)}$, with $A_\mu$ an hermitian gauge field. Once applied the derivative to this term, instead of $\partial_\mu\thetat$, we get $A_\mu$, which is hermitian.
\end{enumerate}

In the end, the quadratic Lagrangian reads
\begin{equation}
    \lag_2 = \frac{\eta^{\mu\nu}}{2}\partial_\mu\rhot\partial_\nu\rhot-\frac{\mt^2}{2}\rhot^2 + V^\mu\rhot\partial_\mu\thetat+\frac{B^2\eta^{\mu\nu}}{2}\partial_\mu\thetat\partial_\nu\thetat.
    \label{eq:l22}
\end{equation}
The operator $\thetat$ is not written in the canonical form, indeed it does not follow the commutation relations. The latter can be satisfied by introducing a new operator, like $\optheta=B\thetat$. In the following, we find more convenient working with the field $\thetat$.

In order to determine the low-energy Lagrangian of the phonon field, we have to integrate out the field $\rhot$. In this way we obtain a Lagrangian only in terms of $\partial_\mu\thetat$. We leverage on the Nambu- approximation for integrating out the field $\rhot$. At the LO level, this operation consists in substituting in the Lagrangian equation of motion in its density. Thus, exploiting the Euler-Lagrangian equation, the equation of motion for the field $\rhot$ is
\begin{equation}
    (\Box+\mt^2)\rhot=V^\mu\partial_\mu\thetat.
    \label{eq:motorho1}
\end{equation}
We notice that the theory becomes non-local. However, higher derivative terms are suppressed as power of $p^2/\mt^2$, where $p$ is the Nambu-Goldstone boson (NGB) momentum. Indeed, knowing that the field $\partial_\mu\thetat\sim p$, we have that the $\rhot$ field can be expressed as the sum of terms that are odd in the momenta. If we restrict to terms which behave like $\sim p$ and $\sim p^3$, we have that the terms in Eq.~\eqref{eq:l22} give the following contributions
\begin{itemize}
    \item the first term behaves like $\sim \alpha_{11}p^4+\alpha_{12}p^6+..$;
    \item the second term behaves like $\sim \alpha_{21}p^2+\alpha_{22}p^4+\alpha_{23}p^6+..$;
    \item the third term behaves like $\sim \alpha_{31}p^2+\alpha_{32}p^4+\alpha_{33}p^6+..$;
    \item the forth term behaves like $\sim \alpha_{41}p^2$,
\end{itemize}
where $\alpha_{ij}$ are constants.
At the leading order, we consider only the first contribution in the equation of motion for $\rhot$, the term $\sim p$. The terms with powers higher than $p^2$ will be useful afterwards, when we will search for the next-to-leading order Lagrangian. Thus, at the leading order in momenta, the effective Lagrangian is
\begin{equation}
    \lag_2 = \frac12\left( B^2\eta^{\mu\nu}+\frac{V^\mu V^\nu}{\mt^2}\right)\partial_\mu\thetat\partial_\nu\thetat.
    \label{eq:lag2}
\end{equation}
We notice that this equation has a form very similar to that of a Lagrangian for a free massless field $\thetat$ which moves along a curved spacetime described by a metric $g_{\mu\nu}$, such that
\begin{equation}
    \sqrt{-g}g^{\mu\nu}=B^2\eta^{\mu\nu}+\frac{V^\mu V^\nu}{\mt^2}.
\end{equation}
Thus, we have created the searched analogue model. We notice that the curvature is given by the superfluid velocity: if we consider $V^\mu(v^\mu)\sim$const, we obtain an analogue metric which is conformal to the Minkowski metric. 

\wl
Now we specialize this procedure to the Lagrangian in Eq.~\eqref{eq:microscopic}, with the aim of obtaining the quadratic term for the effective Lagrangian that we are searching. We assume that the background field $\theta_0$ has no time dependence, but is space dependent. If we substitute the microscopic Lagrangian in Eq.~\eqref{eq:mt}, we find
\begin{equation}
    \mt^2 = -(\mu^2-(\nabla\thetao)^2)+m^2+3\lambda\rhoo^2.
\end{equation}
If we now define $v^\mu=\gamma(1,\mathbf{v})$, with $\mathbf{v}=\nabla\thetao/\mu$ and $\gamma$ the Lorentz factor, and we use the stationary condition for the $\rhoo$ field, assuming that its space dependence is negligible we get
\begin{equation}
    \mt^2 = 2\lambda\rhoo^2.
\end{equation}
Applying also~\eqref{eq:b} and~\eqref{eq:v} to our Lagrangian, we find
\begin{align}
    &B^2 = \rhoo^2,\label{eq:Bsq} \\
    &V^{\mu} = -2\rhoo\frac{\mu}{\gamma}v^\mu.
\label{eq:Vmu}
\end{align}
Plugging these expression in~\eqref{eq:lag2}, we get
\begin{equation}
    \lag_2 = \frac{\rhoo^2}{2}\left(\eta^{\mu\nu}+\left(\frac{1}{c_s^2}-1\right)v^\mu v^\nu\right)\partial_\mu\thetat\partial_\nu\thetat,
    \label{eq:l2end}
\end{equation}
where we have defined
\begin{equation}
    c_s^2=\frac{\lambda\rhoo^2}{2m^2+3\lambda\rhoo^2}
\end{equation}
as the speed of sound. We notice that $c_s$ vanishes for $\lambda\rhoo^2\to0$, corresponding to the transition point to the normal phase. The largest possible speed of sound of this model is $c_s=\sqrt{1/3}$ when $m\to0$. 

The analogue metric of the system described by the microscopic Lagrangian in Eq.~\eqref{eq:microscopic} is
\begin{equation}
    g^{\mu\nu}=\frac{c_s}{\rhoo^2}\left(\eta^{\mu\nu}+\left(\frac{1}{c_s^2}-1\right)v^\mu v^\nu\right).
    \label{eq:sound}
\end{equation}
With this approach we have just obtained an analogy between a massless field on a curved manifold and a phonon in the BEC. 

Now, we go further in developing the Lagrangian.

\subsection{Cubic terms\label{sub:cubic}}

So far, no interaction terms have been considered.
We want to go on with the development of the Lagrangian to see which kind of interaction terms play a role in our system. The cubic terms in the fluctuations are obtained by the expansion of the Lagrangian density $\lag(\rho,\partial_\mu\rho,\partial_\mu\thetat)$ in terms of the fluctuations $\rhot, \partial_\mu \rhot$ and $\partial_\mu\thetat$ at the third-order:
\begin{equation}
    \begin{split}
        \lag_3=&\left.\frac{\rhot^3}{6}\frac{\delta^3\lag}{\delta\rho^3}\right|_{\rhoo,\thetao}+\left.\frac{\rhot^2\partial_\mu\rhot}{2}\frac{\delta^3\lag}{\delta\rho^2\delta\partial_\mu\rho}\right|_{\rhoo,\thetao}+\left.\frac{\rhot\partial_\mu\rhot\partial_\nu\rhot}{2}\frac{\delta^3\lag}{\delta\rho\delta\partial_\mu\rho\delta\partial_\nu\rho}\right|_{\rhoo,\thetao}+\left.\frac{\rhot^2\pmtht}{2}\frac{\delta^3\lag}{\delta\rho^2\delta\partial_\mu\theta}\right|_{\rhoo,\thetao}+\\
        &+\left.\frac{\rhot\pmtht\pntht}{2}\frac{\delta^3\lag}{\delta\rho\delta\partial_\mu\theta\delta\partial_\nu\theta}\right|_{\rhoo,\thetao}+\left.\rhot\partial_\mu\rhot\pntht\frac{\delta^3\lag}{\delta\rho\delta\partial_\mu\theta\delta\partial_\nu\theta}\right|_{\rhoo,\thetao}+\left.\frac{\partial_\mu\rhot\partial_\nu\rhot\pstht}{2}\frac{\delta^3\lag}{\delta\partial_\mu\rho\delta\partial_\nu\rho\delta\partial_\sigma\theta}\right|_{\rhoo,\thetao}+\\
        &+\left.\frac{\partial_\mu\rhot\partial_\nu \rhot\partial_\sigma\rhot}{6}\frac{\delta^3\lag}{\delta\partial_\mu\rho\delta\partial_\nu\rho\delta\partial_\sigma\rho}\right|_{\rhoo,\thetao}+\left.\frac{\pmtht\pntht\pstht}{6}\frac{\delta^3\lag}{\delta\partial_\mu\theta\delta\partial_\nu\theta\delta\partial_\sigma\theta}\right|_{\rhoo,\thetao}+\\
        &+\left.\frac{\partial_\mu\rhot\pntht\pstht}{2}\frac{\delta^3\lag}{\delta\partial_\mu\rho\delta\partial_\nu\rho\delta\partial_\sigma\theta}\right|_{\rhoo,\thetao}.
    \end{split}
    \label{eq:third}
\end{equation}
As in the previous paragraph~\ref{sub:quadratic}, for a microscopic Lagrangian without derivative-couplings terms for the field $\rho$, we have that these derivative couplings do not appear also in the cubic terms. As a consequence, in Eq.~\eqref{eq:third}, the second, third, sixth, seventh, eighth and tenth terms vanish. In addition, the term proportional to $\partial_\mu\tilde\theta\partial_\nu \tilde\theta\partial_\sigma\tilde\theta$ vanishes because there is no microscopic derivative self-interaction. We shall see that a term like this emerges once the radial field fluctuations have been integrated out.

We can now re-write the first term of Eq.~\eqref{eq:third} as
\begin{equation}
    \left.\frac{\rhot^2}{6}\frac{\delta^3\lag}{\delta\rho^3}\right|_{\rhoo,\thetao}=\frac{N}{3}\rhot^3,
    \label{eq:a}
\end{equation}
where we have defined $N$ as a constant of dimension 1. 

The fourth term is 
\begin{equation}
    \left.\frac{\rhot^2\pmtht}{2}\frac{\delta^3\lag}{\delta\rho^2\delta\partial_\mu\theta}\right|_{\rhoo,\thetao}=\Vt^\mu\rhot^2\pmtht,
    \label{eq:vt}
\end{equation}
where $\Vt^\mu$ is a vector. A generic $\Vt^\mu$ breaks the Lorentz symmetries of the Lagrangian, as in the case with $V^\mu$ (see Eq.~\eqref{eq:v}). Thus, here too one can assume that this vector is the result of a linear combination of the two objects that break the symmetry, i.e. $\Vt_\mu=\Tilde{a}\mu\delta_{\mu 0}+\Tilde{b}v_\mu$, with $\Tilde{a}$ and $\Tilde{b}$ two coefficients. Note that, by dimensional analysis, $\Vt^\mu$ follows to have energy dimension 1. 

The fifth term in Eq.~\eqref{eq:third} is
\begin{equation}
    \left.\frac{\rhot\pmtht\pntht}{2}\frac{\delta^3\lag}{\delta\rho\delta\partial_\mu\theta\delta\partial_\nu\theta}\right|_{\rhoo,\thetao}=K^{\mu\nu}\rhot\pmtht\pntht,
    \label{eq:k}
\end{equation}
where $K^{\mu\nu}$ is a symmetric tensor, and depends on the background. With assumptions similar to those considered in Sec.~\ref{sub:quadratic}, we have $K^{\mu\nu}=K\mink$, with $K$ a constant of dimension 1.

In the end, the cubic Lagrangian reads
\begin{equation}
	\lag_3=\frac{N}{3}\rhot^3+\Vt^\mu\rhot^2\pmtht+K\mink\rhot\pmtht\pntht.
\end{equation}
We want to build an effective Lagrangian which contains both the quadratic and the cubic terms. In doing this, we have to properly take into account the power counting in momenta. Keeping this in mind, we consider $\lagt=\lag_2+\varepsilon\lag_3$, where $\varepsilon$ is a dimensionless, infinitesimal constant. 
As in Sec.~\ref{sub:quadratic}, we want to integrate out the massive field $\rhot$ by the Nambu- approximation. Therefore, we calculate the equation of motion for the field $\rhot$ by means of the Euler-Lagrange equation, and obtain a Klein-Gordon that can be resolved perturbatively. 
Always referring back to Sec.~\ref{sub:quadratic}, we are at the leading order in momenta, thus higher derivative terms are suppressed as powers of $p^2/\mt^2$: we only consider the lowest power in the momentum of the equation of motion for $\rhot$. Therefore, we have the effective Lagrangian at LO for the NGB with the quadratic and the cubic terms:
\begin{equation}
	\begin{split}
		\lagt=&\frac12 \left(B^2\eta^{\mu\nu}+\frac{V^\mu V^\nu}{\mt^2}\right)\partial_\mu \thetat\pntht+\varepsilon\left[\frac{\Vt^\alpha V^\nu V^\mu}{\mt^4}+\frac{K V^\alpha\eta^{\mu\nu}}{\mt^2}+\frac{N}{3}\frac{V^\alpha V^\nu V^\nu}{\mt^6}\right]\partial_\alpha\thetat\pntht\pmtht,
	\end{split}
	\label{eq:lt}
\end{equation}
which is valid for momenta $p\ll\mt$.

Now, we specialize this procedure to the system described by the microscopic Lagrangian in Eq.~\eqref{eq:microscopic}. In particular, we are interested in understanding how phonons interact with each other and with the medium. Thus, we can readily substitute the microscopic Lagrangian (Eq.~\eqref{eq:microscopic}) in Eqs.~\eqref{eq:a},~\eqref{eq:vt} and~\eqref{eq:k}, getting
\begin{align}
	&\Vt^\mu=-\frac{\mu}{\gamma}v^\mu,\\
	&K=\rhoo,\\
	&N=-3\lambda\rhoo.
\end{align}
Plugging these expressions and the relations~\eqref{eq:mt},~\eqref{eq:b} and~\eqref{eq:v} in~\eqref{eq:lt} already found in the previous section, we have that
\begin{equation}
	\lagt=\frac12\rhoo^2\left[\eta^{\mu\nu}+\left(\frac{1}{c_s^2}-1\right)v^\mu  v^\nu\right]\pmtht\pntht-\varepsilon\frac{\rhoo^2}{2}\frac{\gamma}{\mu}\left(\frac{1}{c_s^2}-1\right)\eta^{\mu\nu}v^\alpha\partial_\alpha\thetat\pmtht\pntht,
    \label{eq:ltend}
\end{equation}
where we have used also the definition of $c_s$ in Eq.~\eqref{eq:sound}. Therefore, we have now an effective Lagrangian up to third-order in the phonon field, which shows us how phonons propagate in the superfluid background and how they interact.

To complete our construction at the leading order, in the next paragraph we expand the Lagrangian up to the forth-order.

\subsection{Quartic terms}

We now want to derive the quartic terms. The key reason to include these terms is that they provide quartic interaction terms between phonons. Unless the dispersion law is anomalous (we shall clarify this point shortly below) the cubic interaction is kinematically suppressed, and only the four-phonon interaction is allowed. Once again, we assume no derivative coupling in the microscopic Lagrangian. 

Let us consider the quartic terms in the fluctuations, obtained by the expansion of the Lagrangian density $\lag(\rho,\partial_\mu\rhot,\partial_\mu\thetat)$ in terms of the fluctuations $\rhot, \partial_\mu\rhot$ and $\partial_\mu\thetat$ at the fourth order. These are:
\begin{equation}
    \begin{split}
        \lag_4=&\left.\frac{\rhot^4}{4!}\frac{\delta^4\lag}{\delta\rho^4}\right|_{\rhoo,\thetao}+\left.\frac{\rhot^3\partial_\mu\rhot}{3!}\frac{\delta^4\lag}{\delta\rho^3\delta\partial_\mu\rho}\right|_{\rhoo,\thetao}+\left.\frac{\rhot^3\pmtht}{3!}\frac{\delta^4\lag}{\delta\rho^3\delta\partial_\mu\theta}\right|_{\rhoo,\thetao}+\left.\frac{\rhot^2\partial_\mu\rhot\partial_\nu\rhot}{4}\frac{\delta^4\lag}{\delta\rho^2\delta\partial_\mu\rho\delta\partial_\nu\rho}\right|_{\rhoo,\thetao}+\\
        &+\left.\frac{\partial_\mu\rhot\partial_\nu \rhot\partial_\sigma\rhot\partial_\tau\rhot}{4!}\frac{\delta^4\lag}{\delta\partial_\mu\rho\delta\partial_\nu\rho\delta\partial_\sigma\rho\delta\partial_\tau\rho}\right|_{\rhoo,\thetao}+\left.\frac{\pmtht\pntht\pstht\pttht}{4!}\frac{\delta^4\lag}{\delta\partial_\mu\theta\delta\partial_\nu\theta\delta\partial_\sigma\theta\delta\partial_\tau\theta}\right|_{\rhoo,\thetao}+\\
        &+\left.\frac{\rhot^2\partial_\mu\rhot\pntht}{4}\frac{\delta^4\lag}{\delta\rho^2\delta\partial_\mu\rho\delta\partial_\nu\theta}\right|_{\rhoo,\thetao}+\left.\frac{\partial_\mu\rhot\partial_\nu\rhot\partial_\sigma\rhot\pttht}{3!}\frac{\delta^4\lag}{\delta\partial_\mu\rho\delta\partial_\nu\rho\delta\partial_\sigma\rho\delta\partial_\tau\theta}\right|_{\rhoo,\thetao}+\\
        &+\left.\frac{\rhot\partial_\mu\rhot\partial_\nu\rhot\partial_\sigma\rhot}{3!}\frac{\delta^4\lag}{\delta\rho\delta\partial_\mu\rho\delta\partial_\nu\rho\delta\partial_\sigma\rho}\right|_{\rhoo,\thetao}+\left.\frac{\rhot\pmtht\pntht\pstht}{3!}\frac{\delta^4\lag}{\delta\rho\delta\partial_\mu\theta\delta\partial_\nu\theta\delta\partial_\sigma\theta}\right|_{\rhoo,\thetao}+\\
        &+\left.\frac{\rhot^2\pmtht\pntht}{4}\frac{\delta^4\lag}{\delta\rho^2\delta\partial_\mu\theta\delta\partial_\nu\theta}\right|_{\rhoo,\thetao}+\left.\frac{\partial_\mu\rhot\partial_\nu\rhot\pstht\pttht}{3!}\frac{\delta^4\lag}{\delta\partial_\mu\rho\delta\partial_\nu\rho\delta\partial_\sigma\theta\delta\partial_\tau\theta}\right|_{\rhoo,\thetao}+\\
        &+\left.\frac{\rhot\partial_\mu\rhot\partial_\nu\rhot\pstht}{3!}\frac{\delta^4\lag}{\delta\rho\delta\partial_\mu\rho\delta\partial_\nu\rho\delta\partial_\sigma\theta}\right|_{\rhoo,\thetao}+\left.\frac{\rhot\partial_\mu\rhot\pntht\pstht}{3!}\frac{\delta^4\lag}{\delta\rho\delta\partial_\mu\rho\delta\partial_\nu\theta\delta\partial_\sigma\theta}\right|_{\rhoo,\thetao}+\\
        &+\left.\frac{\partial_\mu\rhot\pntht\pstht\pttht}{3!}\frac{\delta^4\lag}{\delta\partial_\mu\rho\delta\partial_\nu\theta\delta\partial_\sigma\theta\delta\partial_\tau\theta}\right|_{\rhoo,\thetao}.
        \label{eq:forth}
    \end{split}
\end{equation}
Since in the microscopic Lagrangian~\eqref{eq:microscopic} $\rho$ has no derivative couplings, we can assume that this also happens in the effective Lagrangian. Thus, we can assume that all the terms in Eq.~\eqref{eq:forth} that contain this type of couplings do vanish. Thus the second, forth, fifth, seventh, eighth, ninth, twelfth, thirteenth, fourteenth and fifteenth terms vanish. Knowing also that the microscopic Lagrangian has not couplings of the type $\partial_\mu\theta\partial_\nu\theta\partial _\sigma\theta\pstht$, $\rhot^3\pmtht$ and $\rhot\partial_\mu\theta\partial_\nu \theta\partial_\sigma \theta$ and, therefore, also the third, sixth and tenth term vanish as well. Now let us analyze the remaining terms.

The first term in Eq.~\eqref{eq:forth} is
\begin{equation}
    \left.\frac{\rhot^4}{4!}\frac{\delta^4\lag}{\delta\rho^4}\right|_{\rhoo,\thetao}=\frac{M}{4}\rhot^4,
    \label{eq:c}
\end{equation}
where $M$ is a dimensionless constant. 

The eleventh term of $\lag_4$ is
\begin{equation}
    \left.\frac{\rhot^2\pmtht\pntht}{4}\frac{\delta^4\lag}{\delta\rho^2\delta\partial_\mu\theta\delta\partial_\nu\theta}\right|_{\rhoo,\thetao}=\frac{D^{\mu\nu}}{2}\rhot^2\pmtht\pntht,
    \label{eq:d}
\end{equation}
where $D^{\mu\nu}$ is a symmetric tensor that may depend on the background. With assumptions similar to those used in Sec.~\ref{sub:quadratic} and~\ref{sub:cubic}, we have $D^{\mu\nu}=D\mink$, with $D$ a dimensionless constant.

Finally, the quartic Lagrangian reads
\begin{equation}
    \lag_4=\frac{M}{4}\rhot^4+\frac{D}{2}\mink\rhot^2\pmtht\pntht.
\end{equation}
Now, we want to integrate out the $\rhot$ field in order to obtain an effective Lagrangian in terms of the phonons. Thus, we consider the expansion up to fourth order of the Lagrangian. We define the total effective Lagrangian $\lag_T=\lag_2+\varepsilon\lag_3+\varepsilon^2\lag_4$, where $\varepsilon$ is a dimensionless, infinitesimal constant. We get the equation of motion of the field $\rhot$ in a perturbative approach and, after substituting it in $\lag_T$, we get the effective Lagrangian for the NGB:
{\small
\begin{equation}
	\begin{split}
		\lag=\frac12& \left(B^2\eta^{\mu\nu}+\frac{V^\mu V^\nu}{\mt^2}\right)\partial_\mu \thetat\pntht+\\
        +\varepsilon&\left[\frac{\Vt^\alpha V^\nu V^\mu}{\mt^4}+\frac{K V^\alpha\eta^{\mu\nu}}{\mt^2}+\frac{N}{3}\frac{V^\alpha V^\nu V^\nu}{\mt^6}\right]\partial_\alpha\thetat\pntht\pmtht+\\
        +\frac{\varepsilon^2}{2}&\left[\frac{N^2V^\alpha V^\beta V^\mu V^\nu}{\mt^{10}}+\frac{4NV^\alpha V^\beta V^\mu \Vt^\nu }{\mt^8}+\frac{2NKV^\alpha V^\beta\mink}{\mt^6}\right.+\frac{4V^\alpha V^\beta\Vt^\mu\Vt^\nu }{\mt^6}+\\
        &+\frac{4KV^\alpha \Vt^\beta\mink }{\mt^4}+ \frac{K^2\eta^{\alpha\beta}\mink}{\mt^2}+\left.\frac{M}{2}\frac{V^\alpha V^\beta V^\mu V^\nu }{\mt^8}+\frac{DV^\alpha V^\beta\mink }{\mt^4} \right]\partial_\alpha\thetat\partial_\beta\thetat\partial_\mu\thetat\partial_\nu\thetat.
	\end{split}
	\label{eq:l_T}
\end{equation}
}
As a final step, we apply this result to the microscopic Lagrangian in Eq.~\eqref{eq:microscopic}. If we consider the same assumptions made for the quadratic and cubic terms, from Eqs.~\eqref{eq:c} and~\eqref{eq:d}, we get
\begin{align}
	&M=-\lambda,\\
	&D = 1.
\end{align}
Plugging in these expressions along with the relations already derived in the previous sections~\ref{sub:quadratic}-\ref{sub:cubic}, we obtain
\begin{equation}
    \begin{split}
        \lag=&\frac12\rhoo^2\left[\eta^{\mu\nu}+ \left( \frac{1}{c_s^2} -1\right) v^\mu  v^\nu\right]\pmtht\pntht-\varepsilon\frac{\rhoo^2}{2}\frac{\gamma}{\mu}\left(\frac{1}{c_s^2}-1\right)\eta^{\mu\nu}v^\alpha\partial_\alpha\thetat\pmtht\pntht+\\
        &+\frac{\varepsilon^2}{2}\frac{\rhoo^2}{4}\left(\frac{\gamma}{\mu}\right)^2\med\eta^{\mu\nu}\eta^{\alpha\beta}\pmtht\pntht\partial_\alpha\thetat\partial_\beta\thetat.
    \end{split}
    \label{eq:ltotend}
\end{equation}
In conclusion, we have found an effective Lagrangian up to fourth order, composed of a first term describing how phonons propagate in the analogue manifold, while the second and the third describe the phonon effective self-interactions.

\section{An alternative procedure to get the effective Lagrangian}

The solutions found in Sec.~\ref{sec:effectivelagrangian} can be obtained also in a different manner: in 2001, D.T.~Son~\cite{https://doi.org/10.48550/arxiv.hep-ph/0204199} showed that all terms in the expansion of the quantum effective action of the Goldstone field can be found to leading order in derivatives, once the equation of state is given.
Such an approach is very useful because it allows to obtain an effective Lagrangian in a simpler way, with lighter computations. On the other hand and as we will explain more in detail later, this approach is valid only under some conditions. For example, one cannot use this method to obtain a NLO Lagrangian. In addition, the radial fluctuations do not appear, meaning that these are always assumed to be integrated out. In turn, this implies that the density-density correlator cannot be evaluated via this approach.

\subsection{Effective action for a relativistic superfluid \label{sub:son}}

The discussion in~\cite{https://doi.org/10.48550/arxiv.hep-ph/0204199} begins with considering QCD at finite baryon chemical potential $\mu$, described by the Lagrangian 
\begin{equation}
    \lag=\lag_0+\frac\mu3\Bar{q}\gamma^0q, 
    \label{eq:qcdmu}
\end{equation}
where $\lag_0$ is the usual in-vacuum QCD Lagrangian at zero chemical potential, which is Lorentz invariant, and $q$ is the quark field. Color, flavor and spinorial degrees of freedom are suppressed. The last term in Eq.~\eqref{eq:qcdmu} is the only Lorentz-breaking term in $\lag$. In addition, in order to be definite, in the color superconducting phase we focus on the color-flavor-locked (CFL) phase~\cite{Alford:1998mk}. In this case the ground state breaks the $U(1)$ global symmetry generated by the baryon charge
\begin{equation}
    q\to e^{i\alpha/3}q.
\end{equation}
Thus, one can define $\phi(x)$ as an order parameter, that is a normalized parameter that measures the ordering degree of a system~\cite{Landau:480039}. We do not impose any requirement for $\phi(x)$ to be an elementary field, as its specific nature will not have any impact on the subsequent discussion. In the Lagrangian, $\mu$ is an external parameter and $\Gamma$ depends on $\mu, \phi$, where the functional $\Gamma$ is the quantum effective action and is defined as the Legendre transform of the connected Feynman graphs' generator
\begin{equation}
    W[J]=-i\ln Z[J], 
\end{equation}
with $Z$ the partition function. As we did in Sec.~\ref{sec:micro_lagrangian}, it is useful to generalize Eq.~\eqref{eq:qcdmu} to
\begin{equation}
    \lag= \lag_0+\frac13 G_\mu(x)\Bar{q}\gamma^\mu q, 
    \label{eq:qcdgmu}
\end{equation}
and treat $G_\mu$ as an arbitrary background field. In last stage we will set $G_\nu=\mu\delta_{\nu0}$. Since the Lagrangian in Eq.~\eqref{eq:qcdgmu} presents a gauge symmetry
\begin{equation}
    q\to e^{i\alpha/3}q\,,\qquad G_\mu\to G_\mu+\partial_\mu\alpha, 
\end{equation}
then this is respected also by $\Gamma$:
\begin{equation}
    \Gamma[G_\mu,\phi]=\Gamma[G_\mu+\partial_\mu\alpha, \phi e^{iM\alpha}], 
\end{equation}
where $M$ is the baryon charge of the order parameter. 

As final goal, Son derives the quantum effective action for the Goldstone field $\varphi$, i.e. for the phase of $\phi$. In order to do this, one has to minimize the amplitude $|\phi|$, so as to perform an integration over it.
Then, one expands in power series of $\varphi$ and in its spatial derivatives $\partial_i\varphi$. Since the effective action is invariant under transformation
\begin{equation}
    A_\mu\to A_\mu+\partial_\mu\alpha \qquad \varphi\to\varphi+\alpha,
    \label{eq:symm}
\end{equation} 
there are no terms which contain $\varphi$ and all the contributions are given by powers of $(\partial_i^n\varphi)$, with $n\ge1$. In addition, we keep only terms with the smallest number of spatial derivatives. Still leveraging the symmetry in Eq.~\eqref{eq:symm}, we can choose $\alpha$ so that $\partial_\mu\alpha=-G_\mu$. Therefore, the effective action depends only on $D_\mu\varphi$, where we define $D_\mu\varphi\equiv\partial_\mu\varphi-G_\mu$ as the covariant derivative. In particular, since in the Lagrangian ~\eqref{eq:qcdgmu} the Lorentz symmetry is broken only by the term $G_\mu$, the effective action depends only on $G_\mu G^\mu$. 
Minimizing the quantum effective action, and exploiting thermodynamic relations (see e.g. Ref.~\cite{https://doi.org/10.48550/arxiv.hep-ph/0204199}), one finally gets the expression for the effective Lagrangian density of the system:
\begin{equation}
    \lag_{\mathrm{eff}}(-G_\mu)=P\left(\left(G_\mu G^\mu\right)^{1/2}\right),
\end{equation}
where $P$ is the thermodynamic pressure at chemical potential $\mu$.
The full low-energy effective action is completely determined from the equation of state: 
\begin{equation}
    \Gamma[G_\mu,\varphi]=\int d^4xP\left(\sqrt{D_\mu\varphi D^\mu\varphi}\right).
    \label{eq:solson}
\end{equation}
Substituting, in the end, $G_\mu=(\mu,0)$, one can expand the resulting equation over powers of $\partial_\mu\varphi$, and find all vertices in the effective actions.

This procedure is valid not only for the CFL phase: one can use it in any system at high density,in which there is a spontaneously breaking of a global $U(1)$ symmetry.

\subsection{Background pressure}

We now apply the previous procedure to the Lagrangian in Eq.~\eqref{eq:microscopic}. As far as we know, this has never been discussed before. First, we have to find the equation of state for the system described by the microscopic Lagrangian in Eq.~\eqref{eq:microscopic}, that we report again for readability convenience:
\begin{equation}
    \lag=\frac12\partial_\nu\rho\partial^\nu\rho+ \frac12\rho^2\partial_\nu\theta\partial^\nu\theta-\mu\rho^2\partial_t\theta-\frac12(m^2-\mu^2)\rho^2-\frac\lambda4\rho^2\,.
\end{equation}
When this Lagrangian is evaluated on the ground state with $\rhoo,\thetao$, it gives the background pressure $P$. From the same Lagrangian, we can obtain the stationary condition for $\rho$, which is
\begin{equation}
    \rhoo^2=\frac{\mu^2}{\lambda\gamma^2}-\frac{m^2}{\lambda}.
\end{equation}
As a consequence, substituting this relation in $\lag(\rho)$, and considering that the background field $\thetao$ has no time dependence, the equation of state turns out to be
\begin{equation}
    P=\lag(\rhoo)=\frac{\mu^4}{4\lambda\gamma^4}+\frac{m^4}{4\lambda}-\frac{\mu^2m^2}{2\lambda\gamma^2}.
    \label{eq:pressure}
\end{equation}
If we now calculate the energy density $\varepsilon$ and the density $n$ of our system, and we use the two relationships in~\eqref{eq:pressure}, we verify that
\begin{equation}
    P=\frac{\mu}{\gamma}\ n-\varepsilon.
\end{equation}
Now that we have computed the pressure $P$, we can go on with the approach discussed in the previous section~\ref{sub:son} to obtain the low-energy effective Lagrangian. For the sake of readability, we move the computation's details to Appendix~\ref{app:den}.

\subsection{Effective Lagrangian at leading order}

For what we have just seen in Sec.~\ref{sub:son}, the low-energy leading-order Lagrangian of any system with a $U(1)$ broken symmetry can be determined from the knowledge of the pressure of the system~\cite{https://doi.org/10.48550/arxiv.hep-ph/0204199,PhysRevD.81.043002}. We search for an effective Lagrangian up to the fourth order. In the following, we assume that the system presents a background flow $\bm v$. Thus, in this case, the quantities which break the $U(1)$ symmetry are the $ v^\mu$. In this case, we consider the external field:
\begin{equation}
    G_\nu=\frac{\mu}{\gamma}v_\nu\,,
\end{equation}
and for the results obtained in Sec.~\ref{sub:son}, the low-energy leading-order Lagrangian is simply given by
\begin{equation}
    \lag=P\left[\sqrt{D_\mu\thetat D^\mu\thetat}\right].
    \label{eq:lag02}
\end{equation}
Here, $D_\mu\thetat=\pmtht-G_\mu$ and $P$ is a functional having the same algebraic expression of the system pressure. For convenience, we replace $\mu/\gamma\to\mu$. Upon expanding the right-hand-side of~\eqref{eq:lag02} around $\mu$, we have that
\begin{equation}
    \begin{split}
        \lag=P(&\mu)-\frac{\partial P}{\partial\mu}v_\mu\Pmtht+\frac12\left[\mink\frac{1}{\mu}\frac{\partial P}{\partial\mu}+v^\mu v^\nu\left(\frac{\partial^2P}{\partial\mu^2}-\frac{1}{\mu}\frac{\partial P}{\partial\mu}\right)\right]\pntht\pmtht+\\
        +\frac12&\left[\frac{1}{\mu}\left(\frac{1}{\mu}\frac{\partial P}{\partial\mu}-\frac{\partial^2 P}{\partial\mu^2}\right)\mink-\left(\frac{1}{\mu^2}\frac{\partial P}{\partial\mu}-\frac{1}{\mu}\frac{\partial^2 P}{\partial\mu^2}+\frac13\frac{\partial^3P}{\partial\mu^3}\right)v^\mu v^\nu \right]v^\alpha\pmtht\pntht\partial_\alpha\thetat+\\
        +\frac14& \left[ \frac{1}{2\mu^2}\left(\frac{\partial^2P}{\partial\mu^2}-\frac{1}{\mu}\frac{\partial P}{\partial\mu}\right)\eta^{\mu\nu}\eta^{\alpha\beta}-\frac{3}{\mu}\left(\frac{1}{\mu}\frac{\partial^2P}{\partial\mu^2}-\frac{1}{\mu^2}\frac{\partial P}{\partial\mu}-\frac13\frac{\partial^3 P}{\partial\mu^3}\right)\eta^{\mu\nu}v^\alpha v^\beta\right.+\\
        &+\left.\left(-\frac{5}{2\mu^3}\frac{\partial P}{\partial \mu}+\frac{5}{2\mu^2}\frac{\partial^2 P}{\partial\mu^2}-\frac{1}{\mu}\frac{\partial^3 P}{\partial\mu^3}+\frac16\frac{\partial^4 P}{\partial\mu^4}\right)v^\mu v^\nu v^\alpha v^\beta \right] \partial_\mu\thetat\partial_\nu\thetat\partial_\alpha\thetat\partial_\beta\thetat.
        \label{eq:son4}
    \end{split}
\end{equation}
If we now consider the background pressure found in Eq.~\eqref{eq:pressure} from the microscopic Lagrangian~\eqref{eq:microscopic}, $P$ can be schematically written as
\begin{equation}
    P=k_1\mu^4+k_2\mu^2,
\end{equation}
where $k_1$ and $k_2$ are coefficients. Therefore, after labelling $P^{(n)}=\frac{\partial^n P}{\partial \mu^n}$, we get
\begin{equation}
    \begin{split}
        &\frac{1}{\mu}P^{(1)}=\frac{n}{\mu}\qquad \qquad P^{(2)}-\frac{1}{\mu}P^{(1)}=\frac n\mu\left(\frac{1}{c_s^2}-1\right)\\
        \frac{5}{2\mu^3}P^{(1)}-\frac{5}{2\mu^2}&P^{(2)}+\frac{1}{\mu}P^{(3)}+\frac16P^{(4)}=\frac{1}{\mu}P^{(2)}-\frac{1}{\mu^2}P^{(1)}-\frac13P^{(3)}=0,
    \end{split}
    \label{eq:pressureson}
\end{equation}
where we have exploited the thermodynamic relations
\begin{equation*}
    \begin{split}
        n=P^{(1)}\qquad c_s^2=\frac{n}{\mu P^{(2)}}.
    \end{split}
\end{equation*} 

Thus, substituting Eqs.~\eqref{eq:pressureson} in ~\eqref{eq:son4}, we get the effective Lagrangian at leading order in momenta up to fourth order in the fields:
\begin{equation}
    \begin{split}
        \lag_T=&\frac{n}{2\mu}\left(\mink+\left(\frac{1}{c_s^2}-1\right)v^\mu v^\nu\right)\pntht\pmtht-\frac{n}{2\mu^2}\left(\frac{1}{c_s^2}-1\right)\mink v^\alpha\partial_\alpha\thetat\pmtht\pntht+\\
        &+\frac{n}{8\mu^3}\med\eta^{\mu\nu}\eta^{\alpha\beta}\partial_\mu\thetat\partial_\nu\thetat\partial_\alpha\thetat\partial_\beta\thetat .
    \end{split}
\end{equation}
As we expected, this solution is the same as that in Eq.~\eqref{eq:ltotend}, i.e. it is the same as the effective Lagrangian that we found with the Lagrangian approach.

Let us now analyze the Lagrangian that we have just obtained. We notice, in the first place, that all the terms arising from the effects of the medium are characterized by the combination 
\begin{equation}
    n\med. 
    \label{eq:mediuml}
\end{equation}
Thus, these terms emerge from the coupling between the phonons and the condensate.
It is possible to study two limiting cases:
\begin{itemize}
    \item $c_s^2=0$: the density $n$ vanishes, so this is the transition point to the normal phase: the system is not any longer a condensate; therefore, our theory and approach are not any longer valid, as we are working with an effective theory which covers only a certain energy scale;
    \item $c_s^2=1$: the expression in Eq.~\eqref{eq:mediuml} vanishes, meaning that the system's excitations are not any longer coupled to the condensate (there is a decoupling); in addition, in this case the equation of state reads $P\propto n^2$, which is that of a degenerate, ideal Fermi gas in 2 spatial dimensions. Indeed, the pressure of a $d$-dimensional system with a completely degenerate Fermi gas is
    \begin{equation}
        P=\frac2d\varepsilon\propto n^{\frac{d+2}{d}},
    \end{equation}
    where $\varepsilon$ is the energy density of the system. We notice that the pressure $P\propto n^2$ only for $d=2$. This happens because, in general, the more rigid is the matter of a certain system, more rapidly its speed of sound $c_s$ increases; the smaller are the interactions, the bigger is the frequencies scale.
\end{itemize}

\section{The effective Lagrangian at next-to-leading order}

We are now interested in searching for an effective Lagrangian at next-to-leading order in momenta, to study the propagator's $p^4$-corrections. To achieve this goal, one can follow two paths: the method proposed by Son and Wingate in~\cite{Son2}, which will be exposed in the next section, or the Lagrangian approach, that we have already used for the effective Lagrangian at leading order in momenta. We illustrate below both routes. 

\subsection{Son and Wingate's approach\label{sub:sonewin}}

In~\cite{Son2}, Son and Wingate proposed a method that allows to determine the next-to-leading order Lagrangian for a fermionic system. This procedure can be extended to a boson system as well.
To get the full picture, let's take a step back and address again the LO Lagrangian. 

As in~\cite{Son2}, let us consider a relativistic field theory of one free complex scalar field $\Psi$ in an external four-dimensional metric $g_{\mu\nu}$. For consistency with what we have written so far, we adopt the signature $(+,-,-,-)$. In both the relativistic and non-relativistic limit, the field can be written in terms of its modulus and phase as
\begin{equation}
    \Psi=|\Psi|e^{-i\Theta}\qquad \Psi=\frac{\psi}{\sqrt{2mc}}e^{-imc^2t} \quad\mathrm{where}\quad \psi=|\psi|e^{-i(\mu t-\varphi)},
\end{equation}
where $\mu$ is the chemical potential, $m$ the mass field, $c$ the speed of light, and $t$ the time. The non-relativistic phase of $\Psi$ is defined as $\mu t-\varphi$, i.e. as a sum of the value $\mu t$ assumed in the background plus its fluctuations $\varphi$. The field $\Theta$ is linked to the Goldstone boson $\varphi$ as 
\begin{equation}
    \Theta=(mc^2+\mu)t-\varphi,
\end{equation}
since $\Theta$ is the phase of the field, that gets its expectation value in the ground state $|\Psi|e^{-i\Theta}$. Thus, in flat spacetime and at the lowest order in $\Theta$, the Lagrangian must be a function of
\begin{equation}
    \chi=\frac12\eta^{\mu\nu}\partial_\mu\Theta\partial_\nu\Theta =\frac12\left(\mu_0^2-2\mu_0\partial_0\varphi+\partial_\nu\varphi\partial^\nu\varphi\right),
\end{equation}
where we have defined $\mu_0=mc^2+\mu$. In this way, the leading order Lagrangian is
\begin{equation}
    \lag_{LO}=P[\sqrt{2\chi}],
    \label{eq:solSon2}
\end{equation}
and, in so doing we find the same result reported in Eq.~\eqref{eq:solson}. Now, the goal is to obtain the next-to-leading order Lagrangian in the derivative expansion. In the relativistic theory, the next-to-leading terms contain two more derivatives acting on the field $\Theta$, and we have a form of the type $Q_iF(\chi)$. Here, $F$ is an arbitrary function of $\chi$, and $Q_i$ are Lorentz scalar operators which are of order $p^4$. In flat spacetime, the only possible operators are the following:
\begin{align}
    &Q_1=\left(\partial_\mu\partial_\nu, \Theta\right)\left(\partial_\lambda\partial_\rho\Theta\right)\partial^\mu\Theta\partial^\nu\Theta\partial^\lambda\Theta\partial^\rho\Theta=\partial_\mu\chi\partial_\nu\chi\partial^\mu\Theta\partial^\nu\Theta,\\
    &Q_2=\left(\partial_\mu\partial_\nu\Theta\right) \left(\Box\Theta\right)\partial^\mu\Theta\partial^\nu\Theta=-\partial_\mu\chi\partial^\mu\Theta\Box\Theta,\\
    &Q_3=\left(\partial_\mu\partial_\nu\Theta\right) \left(\partial^\mu\partial_\rho\Theta\right)\partial^\rho\Theta\partial^\nu\Theta=\partial_\mu\chi\partial^\mu\chi,\\
    &Q_4=\left(\Box\Theta\right)^2,\\
    &Q_5=\left(\partial_\mu\Theta\partial_\nu\Theta\right) \left(\partial^\mu\Theta\partial^\nu\Theta\right).
\end{align}
Then, the field equation can be cast as
\begin{equation}
    \partial_\nu\chi\partial^\mu\Theta=-\frac{\lag'}{\lag''}\Box\Theta.
\end{equation}
At $O(p^4)$ order, one can express $Q_1$ and $Q_2$ through $Q_4$
\begin{equation}
    Q_1=\left(\frac{\lag'}{\lag''}\right)^2Q_4+O(p^4)\qquad Q_2=\frac{\lag'}{\lag''}Q_4+O(p^4).
\end{equation}
Therefore, these terms are not independent. In addition, since the following relation is satisfied
{\small
\begin{equation}
    \partial_\mu\partial_\nu\Theta\partial^\mu\partial^\nu\Theta F(\chi)=\left(\Box\Theta\right)^2\left(F-\frac{\lag'}{\lag''}F'\right)+\partial_\mu\chi\partial^\mu\chi F'(\chi)-\partial_\mu\left[\left(\partial^\mu\chi+\partial^\mu\Theta\Box\Theta\right)F(\chi)\right],
\end{equation}
}
one of the terms written above can be expressed as a linear combination of the others.

In conclusion, the next-to-leading order Lagrangian is given by
\begin{equation}
    \lag_{NLO}=\partial^\mu\chi\partial_\mu\chi F_1(\chi)+\left(\Box\Theta\right)^2F_2(\chi).
\end{equation}
To obtain the functions $F_1$ and $F_2$, one should use the scale and the conformal invariance (see e.g. Ref.~\cite{Son2}). For our aims, it is sufficient to expand the Lagrangian up to terms which are quadratic in $\varphi$~\cite{PhysRevD.81.043002}. We redefine  $F_1$ and $F_2$ so that the new functions
\begin{equation}
    f_1(\chi)=\mu_0^2F_1(\chi), \qquad f_2(\chi)=F_2(\chi)
\end{equation}
are dimensionless. We also assume that they are analytical and can be expanded in a Taylor series around $\mu_0^2$.

Now, we apply this procedure to the system described by the microscopic Lagrangian in Eq.~\eqref{eq:microscopic}. At leading order and after the necessary rescaling to have a normalized kinetic term, the effective Lagrangian obtained by applying Eq.~\eqref{eq:solSon2} is:
\begin{equation}
    \lag_{LO} = \frac12(v_\mu\partial^\mu\thetat)^2+\frac{c_s^2}{2}((\partial_\mu\thetat)^2-(v_\mu\partial^\mu\thetat)^2)+...\ .
\end{equation}
At NLO, one has instead
\begin{equation}
    \lag_{NLO} = \frac{f_1(\chi)}{\mu^2}\partial_\mu\chi\partial^\mu\chi+f_2(\chi)(\Box\Theta)^2,
\end{equation}
where in our model
\begin{align}
    \Theta = \mu v^\alpha x_\alpha -\thetat\,,\qquad
    \chi = \frac12\eta_{\mu\nu}\partial^\mu\Theta\partial^\nu\Theta.
\end{align}
Our aim is to study the effects caused by the modified phononic dispersion law, which contains also the NLO terms. Thus, we can build the Lagrangian considering up to quadratic terms in $\thetat$. We proceed with the same reasonings used before and, after an appropriate rescaling, we have
\begin{equation}
    \lag_{LO}+\lag_{NLO} = \frac12(v_\mu\partial^\mu\thetat)^2+\frac{c_s^2}{2}((\partial_\mu\thetat)^2-(v_\mu\partial^\mu\thetat)^2)+m_1(v^\mu\partial_\nu\partial_\mu\thetat)^2+m_2(\partial_\mu\partial^\mu\thetat)^2+...,
    \label{eq:corr1}
\end{equation}
where we have defined
\begin{equation}
    m_1 = \frac{f_1(\mu^2)}{\frac{\partial^2P}{\partial\mu^2}}\qquad m_2 = \frac{f_2(\mu^2)}{\frac{\partial^2P}{\partial\mu^2}}.
\end{equation}
The functions $f_{1,2}$ can now be expressed in powers of $\chi/\mu^2$ and with coefficients depending on the microscopic model used. In the following section, we will find them for the particular Lagrangian in Eq.~\eqref{eq:microscopic}.

\subsection{Microscopic Lagrangian approach}

We want to calculate the next-to-leading order Lagrangian starting from Eq.~\eqref{eq:microscopic} and using the method discussed in the previous section~\ref{sub:sonewin}. To get the LO Lagrangian density, we had considered only the leading power in the momenta: in order to obtain the NLO Lagrangian density, we consider also the next-to-leading power in the momenta. Therefore, when we treated the leading order Lagrangian, we approximated $\frac{1}{\Box+\mt^2}\simeq\frac{1}{\mt^2}+O(p^2)$. Since we now want $p^4$-corrections, we use the approximation at the next order
\begin{equation}
    \frac{1}{\Box+\mt^2}\simeq\frac{1}{\mt^2}\left(1-\frac{\Box}{\mt^2}\right)+O(p^4).
    \label{eq:app}
\end{equation}
Once we substitute the equation of motion~\eqref{eq:motorho1} in the Lagrangian~\eqref{eq:l22}, and use the approximation shown in Eq.~\eqref{eq:app}, one gets
\begin{equation}
    \lag_2 = \frac12\partial_\mu\thetat\left(\frac{V^\mu V^\nu}{\mt^2}\left(1-\frac{\Box}{\mt^2}\right)+B^2\eta^{\mu\nu}\right)\partial_\nu\thetat.
\end{equation}
We apply this result to the Lagrangian ~\eqref{eq:microscopic}, to get the effective Lagrangian up to the next-to-leading order. We still consider that the field $\theta_0$ has no temporal dependence, only spatial, and that the spatial dependence of $\rho_0$ is negligible. Leveraging on the expressions~\eqref{eq:mt},~\eqref{eq:b} and~\eqref{eq:v}, and rescaling the Lagrangian to have a normalized kinetic term, we find that
\begin{equation}
    \begin{split}
        \lag_{2,R} =& \frac12(v_\mu\partial^\mu\thetat)^2+\frac{c_s^2}{2}((\partial_\mu\thetat)^2-(v_\mu\partial^\mu\thetat)^2)-\frac{c_s^2}{8}\left(\frac{\gamma}{\mu}\med\right)^2v^\mu v^\nu\partial_\mu\thetat\Box\partial_\nu\thetat.
    \end{split}
    \label{eq:corr2}
\end{equation}
Apparently, it seems that comparing Eq.~\eqref{eq:corr2} and~\eqref{eq:corr1}, the coefficients $m_1$ and $m_2$ vanish, while Eq.~\eqref{eq:corr2} contains one more term different from those in Eq.~\eqref{eq:corr1}. This is not true. Indeed, as written before:
\begin{equation}
    Q_2 = \left(\frac{\lag'}{\lag''}\right)Q_4 + O(p^4) \quad \mathrm{where}\quad
    \begin{cases}
        Q_2 = v^\mu v^\nu\partial_\mu\partial_\nu\thetat\Box\thetat\\
        Q_4 = (\Box\thetat)^2.
    \end{cases}
\end{equation}
Therefore, at the end, we find
\begin{align}
    &m_2 = \frac{c_s^2}{8}\left(\frac{\gamma}{\mu}\right)^4\med^2\frac{\lag'}{\lag''}\\
    &f_2(\mu^2) = \frac{\rhoo^2}{8}\med^2\left(\frac{\gamma}{\mu}\right)^4\frac{\lag'}{\lag''}.
\end{align}
The effective Lagrangian that we have just found can be used to derive the dispersion law for the phonons, i.e. for the acoustic Hawking radiation, with $p^3$-corrections terms.

\section{Dispersion law for the excitation energy}

We now want to compute the dispersion law for the phonons belonging to the system described by the microscopic Lagrangian~\eqref{eq:microscopic}.
Consider the Lagrangian ~\eqref{eq:corr2}. It is convenient to switch to momentum space, thus we Fourier transform the $\thetat$ field:
\begin{equation}
    \thetat(x)=\int \frac{d^4 p}{(2\pi)^4}e^{-ip\cdot x}\thetat(p).
\end{equation}
Since we use the local-density approximation, the velocity varies at a much slower rate than the $\thetat$ field: the derivatives terms of $v^\mu$ can be neglected and the dominant contribution is given by how $\thetat$ varies. In this approximation, the Lagrangian density becomes 
\begin{equation}
    \begin{split}
        \lag=-&\frac{\rho_0^2}{2c_s^2}\int \frac{d^4p}{(2\pi)^4}\frac{d^4 q}{(2\pi)^4}e^{-ip\cdot x}e^{-iq\cdot x}\times\\
        &\times\thetat(q)\left[(1-c_s^2)v^\mu v^\nu p_\mu q_\nu+c_s^2p_\mu q^\mu-\frac{c_s^2}{\mt^2}\left(\frac{1}{c_s^2}-1\right)v^\mu v^\nu p_\mu q_\nu p_\alpha q^\alpha\right]\thetat(p).
    \end{split}
\end{equation}
Considering the action $S=\int d^4x\lag$ and integrating with respect to $x$ and $q$, we get
\begin{equation}
    S=\frac12\int \frac{d^4p}{(2\pi)^4}\thetat(-p)D^{-1}(p)\thetat(p),
\end{equation}
where we have defined the inverse propagator
\begin{equation}
    D^{-1}(p)=\frac{\rho_0^2}{c_s^2}\left[(1-c_s^2)v^\mu v^\nu p_\mu p_\nu+c_s^2p_\mu p^\mu+\frac{c_s^2}{\mt^2}\left(\frac{1}{c_s^2}-1\right)v^\mu v^\nu p_\mu p_\nu p_\alpha p^\alpha\right].
\end{equation}
The dispersion law is given by the propagator's poles, thus by the zeros of $D^{-1}(p)$
\begin{equation}
    D^{-1}(p)=\frac{\rho_0^2}{c_s^2}\left[(1-c_s^2)v^\mu v^\nu p_\mu p_\nu+c_s^2p_\mu p^\mu+\frac{c_s^2}{\mt^2}\left(\frac{1}{c_s^2}-1\right)v^\mu v^\nu p_\mu p_\nu p_\alpha p^\alpha\right]=0.
    \label{eq:dispersionlaw1}
\end{equation}
We consider $v^\mu=\gamma(1,\bm v)$, $p^\mu=(\omega,\bm p)$ and $p_{\parallel}=\bm p\cdot \bm v/|\bm v|$. Therefore, Eq.~\eqref{eq:dispersionlaw1} becomes
\begin{equation}
    \underbrace{(1-c_s^2)\gamma^2(\omega-p_{\parallel}v)^2}_{A}+\underbrace{c_s^2(\omega^2-p^2)}_{B}+\underbrace{\frac{c_s^2}{\mt^2}\left(\frac{1}{c_s^2}-1\right)\gamma^2(\omega-p_{\parallel})^2(\omega^2-p^2)}_{C}=0,
    \label{eq:dispersionlaw2}
\end{equation}
where we have defined $p=|\bm p|,\ v=|\bm v|$. Let us first consider only the $O(p^2)$ contributions and, therefore, we momentarily neglect the term $C$ in Eq.~\eqref{eq:dispersionlaw2}. We want to study our system in the non-relativistic limit with $c_s, v\ll c$ and $\gamma\simeq1$. Therefore, we perform the power counting for the velocity considering only the lowest nontrivial contributions, which are $O(v^2)$. In the following, we assume $v$ and $c_s$ to be of the same order. We also make an \textit {a priori} reasonable ansatz: we assume the momentum to be of the same order of the temporal component of the quadrivelocity, $p\sim v^0$, and the energy to be of the same order of the spatial velocity $\omega\sim v$. In this context, the leading terms of Eq.~\eqref{eq:dispersionlaw2} are
\begin{itemize}
    \item $A\sim(\omega-p_{\parallel}v)^2$,
    \item $B\sim-c_s^2p^2$.
\end{itemize}
Therefore the ${\cal O} (p^2)$ dispersion law in the non-relativistic limit is
\begin{equation}
    \omega=p_{\parallel}v\pm c_s\sqrt{p_{\parallel}^2+p_{\perp}^2},
\end{equation}
where $p_{\perp}^2=p^2-p_{\parallel}$. From this solution we understand \textit{a posteriori} that the ansatz of considering $\omega\sim v$ was correct.

We now include the term $C$ of Eq.~\eqref{eq:dispersionlaw2}. We still want to study our system in the non-relativistic limit, so that we still perform the power counting of the velocity considering only the lowest terms. The leading terms of $C$ are
\begin{equation*}
    C\sim-\frac{1}{\mt^2}(\omega-p_\parallel v)^2p^2.
\end{equation*}
The dispersion law with the corrections $p^4$ in the non-relativistic limit is 
\begin{equation}
    \omega=p_\parallel v\pm c_s p\left(1+\frac{p^2}{2\mt^2}\right).
    \label{eq:dispersionlaw3}
\end{equation}
To conclude, we notice that the energy embodies a term behaving like $\bm p \cdot \bm v$: this is a term due to a boost transformation with velocity $\bm v$. 

\section{General high-frequency dispersion law\label{sec:displaw}}

We now consider a generic 1+1-dimensional system with a dispersion law, in the fluid rest frame, of the type:
\begin{equation}
    E_{\pm} = \pm\left(c_s|\bm p|+\frac{\mathcal{A}}{\mathcal{M}^2}|\bm p|^3\right),
\end{equation}
with $\mathcal{A}$ a constant and $\mathcal{M}$ a cutoff scale. Depending on the value of $\mathcal{A}$, we have two different cases, that we treat separately.

\subsection{Monotonic dispersion law}

In this case, the dispersion law in the laboratory frame is
\begin{equation}
    E_{\pm,\mathrm{lab}} = \mathbf{v}\cdot\mathbf{p}\pm\left(c_s|\bm p|+\frac{\mathcal{A}}{\mathcal{M}^2}|\bm p|^3\right). 
\end{equation}
We can notice that this is similar to the dispersion law obtained in the previous case (see Eq.~\eqref{eq:dispersionlaw3}). 
We consider $\mathbf{v} = -v\hat{x}$ and $\mathbf{p} = \pm p\hat{x}$. The momentum can be both antiparallel or parallel to the velocity and, depending on its direction, we can identify two cases corresponding to the plus and minus signs. We first study the case "+". At the horizon, $v=c_s$, the phonon has energy and group velocity given by
\begin{align*}
    E_{+} = \frac{\mathcal{A}}{\mathcal{M}^2}p^3 \quad & \quad v_{g}^{+}=\frac{3\mathcal{A}}{\mathcal{M}^2}p^2,\\
    E_{-} = -2c_sp-\frac{\mathcal{A}}{\mathcal{M}^2}p^3 \quad& \quad v_{g}^{-}=-2c_s-\frac{3\mathcal{A}}{\mathcal{M}^2}p^2,
\end{align*}
respectively. To be physical, the modes must be positive, so that we neglect the solution $E_-$. 
Asymptotically, where $v\ll c_s$ far from the horizon, the modes are characterized by
\begin{align*}
    E_{+} = c_s p +\frac{\mathcal{A}}{\mathcal{M}^2}p^3\quad & \quad
    v_{g}^{+}=c_s + \frac{3\mathcal{A}}{\mathcal{M}^2}p^2,\\
    E_{-} = -c_s p -\frac{\mathcal{A}}{\mathcal{M}^2}p^3\quad & \quad
    v_{g}^{-}=-c_s - \frac{3\mathcal{A}}{\mathcal{M}^2}p^2.
\end{align*}
Also in this case $E_-$ is negative, thus, it is not a physical solution and we neglect it. 

Let us now study now the case "-". At the horizon, the phonon's energy and group velocity are given, respectively, by:
\begin{align*}
    E_{+} = 2c_sp+\frac{\mathcal{A}}{\mathcal{M}^2}p^3  \quad & \quad v_{g}^{+}=-2c_s-\frac{3\mathcal{A}}{\mathcal{M}^2}p^2,\\
    E_{-} = -\frac{\mathcal{A}}{\mathcal{M}^2}p^3 \quad & \quad v_{g}^{-}=\frac{3\mathcal{A}}{\mathcal{M}^2}p^2.
\end{align*}
The solution $E_-<0$ and thus is not physical and we neglect it.
Asymptotically, the modes are described by:
\begin{align*}
    E_{+} = c_s p +\frac{\mathcal{A}}{\mathcal{M}^2}p^3\quad & \quad
    v_{g}^{+}=-c_s - \frac{3\mathcal{A}}{\mathcal{M}^2}p^2,\\
    E_{-} = -c_s p -\frac{\mathcal{A}}{\mathcal{M}^2}p^3\quad & \quad
    v_{g}^{-}=c_s + \frac{3\mathcal{A}}{\mathcal{M}^2}p^2.
\end{align*}
Here too $E_-$ is negative and we neglect it.
The modes are defined as \emph{emitted} or \emph{absorbed}, depending on the sign of their group velocity: if the group velocity is positive, the phonon is emitted; if the group velocity is negative, the phonon is absorbed. Knowing this, in the following we define the emitted and absorbed mode with their corresponding group velocity. At the horizon they are described by:
\begin{align}
    E_{em} = \frac{\mathcal{A}}{\mathcal{M}^2}p^3 \quad & \quad v_{g,em}=\frac{3\mathcal{A}}{\mathcal{M}^2}p^2,\\
    E_{ab} = 2c_sp+\frac{\mathcal{A}}{\mathcal{M}^2}p^3  \quad & \quad v_{g,ab}=-2c_s-\frac{3\mathcal{A}}{\mathcal{M}^2}p^2,\\
\end{align}
while asymptotically by:
\begin{align}
    E_{em} = c_s p +\frac{\mathcal{A}}{\mathcal{M}^2}p^3\quad & \quad
    v_{g,em}=c_s + \frac{3\mathcal{A}}{\mathcal{M}^2}p^2,\\
    E_{ab} = c_s p +\frac{\mathcal{A}}{\mathcal{M}^2}p^3\quad & \quad
    v_{g,ab}=-c_s - \frac{3\mathcal{A}}{\mathcal{M}^2}p^2.
\end{align}
\begin{figure}[ht!]
    \centering
    \subfloat[]{\includegraphics[scale=0.55]{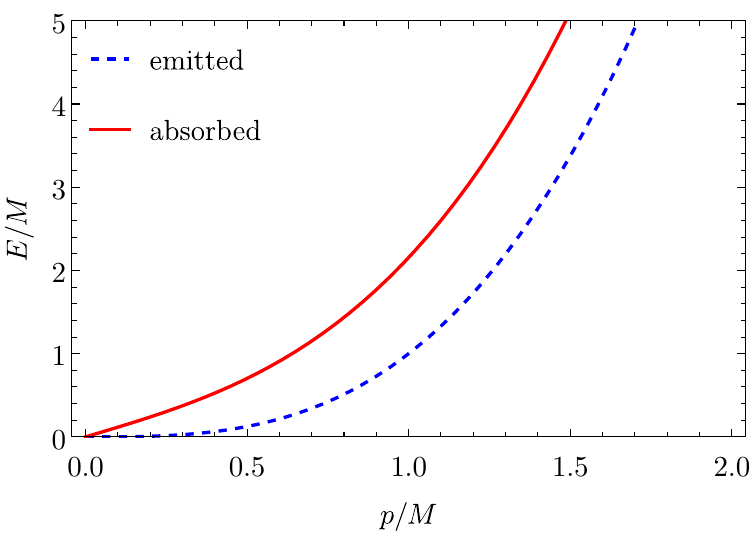}}\quad
    \subfloat[]{\includegraphics[scale=0.55]{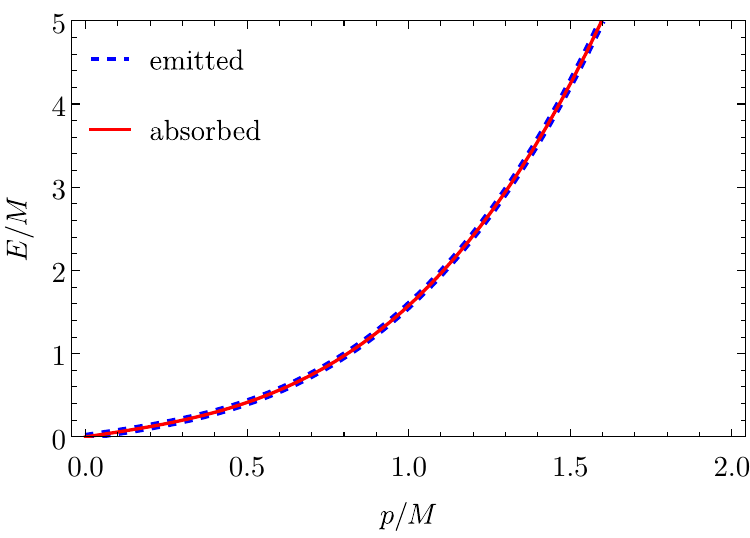}}
    \caption{\emph{Dispersion laws for $\mathcal{A}$ positive, see text. The red solid lines indicate the absorbed modes, the blue dashed lines indicate the emitted modes. (a) Dispersion law for phonons at the horizon. (b) Dispersion law for phonons far from the horizon. In both (a) and (b), for the parameters' value, we consider $\mathcal{A}=1$, and for $c_s$ we take the conformal limit, i.e. $c_s=1/\sqrt{3}$. The two dispersion laws are different. Indeed, far from the horizon (b), the emitted and absorbed phonon modes overlap, meaning that the two modes have the same dispersion law, while at the horizon (a), the emitted and the absorbed modes are separated, and have two different dispersion laws.}}
    \label{fig:modesapositive}
\end{figure}
Figs.~\ref{fig:modesapositive} show the behavior of the dispersion law for the analogue Hawking radiation at the horizon and asymptotically, i.e. far from it.

\subsection{Non-monotonic dispersion law\label{sub:aneg}}

In this case, the coefficient $\mathcal{A}<0$. Therefore, for simplicity, we write it as $\mathcal{A}=-|\mathcal{A}|$. The dispersion law in the laboratory frame is
\begin{equation}
    E_{\pm,\mathrm{lab}} = \mathbf{v}\cdot\mathbf{p}\pm\left(c_s|\bm p|-\frac{|\mathcal{A}|}{\mathcal{M}^2}|\bm p|^3\right). 
\end{equation} 
Like with $\mathcal{A}>0$, we consider $\mathbf{v} = -v\hat{x}$ and $\mathbf{p} = \pm p\hat{x}$. In this case too, we analyze the two cases corresponding to the two possible signs for momentum $\mathbf{p}$. 

We first study the case "+". At the horizon, the phonon's energy and group velocity are given, respectively, by
\begin{align*}
    E_{+,H} = -\frac{|\mathcal{A}|}{\mathcal{M}^2}p^3 \quad & \quad v_{g,H}^{+}=-\frac{3|\mathcal{A}|}{\mathcal{M}^2}p^2,\\
    E_{-,H} = -2c_sp+\frac{|\mathcal{A}|}{\mathcal{M}^2}p^3  \quad& \quad v_{g,H}^{-}=-2c_s+\frac{3|\mathcal{A}|}{\mathcal{M}^2}p^2.
\end{align*}
Once again, we neglect the unphysical modes with negative energy.Thus, since $E_{+,H}$ is negative, we neglect it. Instead, the solution $E_{-,H}$ is physical only for $p^2>2c_s\mathcal{M}^2/|\mathcal{A}|$.
Asymptotically, where $v\ll c_s$ far from the horizon, the modes are characterized by
\begin{align*}
    E_{+,A} = c_s p -\frac{|\mathcal{A}|}{\mathcal{M}^2}p^3\quad & \quad
    v_{g,A}^{+}=c_s - \frac{3|\mathcal{A}|}{\mathcal{M}^2}p^2,\\
    E_{-,A} = -c_s p +\frac{|\mathcal{A}|}{\mathcal{M}^2}p^3\quad & \quad
    v_{g,A}^{-}=-c_s + \frac{3|\mathcal{A}|}{\mathcal{M}^2}p^2.
\end{align*}
We see by inspection that $E_{+,A}$ and $E_{-,A}$ are physical for $p^2<c_s\mathcal{M}^2/|\mathcal{A}|$ and $p^2>c_s\mathcal{M}^2/|\mathcal{A}|$, respectively.

Let us now study the case "-". At the horizon, the phonon's energy and group velocity are given, respectively, by
\begin{align*}
    E_{+,H} = 2c_sp-\frac{|\mathcal{A}|}{\mathcal{M}^2}p^3  \quad & \quad v_{g,H}^{+}=-2c_s+\frac{3|\mathcal{A}|}{\mathcal{M}^2}p^2,\\
    E_{-,H} = \frac{|\mathcal{A}|}{\mathcal{M}^2}p^3 \quad & \quad v_{g,H}^{-}=-\frac{3|\mathcal{A}|}{\mathcal{M}^2}p^2.
\end{align*}
The mode $E_{-,H}$ is physical $\forall p$, while the mode $E_{+,H}$ is physical only for $p^2<2c_s\mathcal{M}^2/|\mathcal{A}|$.
Asymptotically, the modes are 
\begin{align*}
    E_{+,A} = c_s p -\frac{|\mathcal{A}|}{\mathcal{M}^2}p^3\quad & \quad
    v_{g,A}^{+}=-c_s + \frac{3|\mathcal{A}|}{\mathcal{M}^2}p^2,\\
    E_{-,A} = -c_s p +\frac{|\mathcal{A}|}{\mathcal{M}^2}p^3\quad & \quad
    v_{g,A}^{-}=c_s - \frac{3|\mathcal{A}|}{\mathcal{M}^2}p^2.
\end{align*}
Here, the modes $E_{+,A}$ and $E_{-,A}$ are physical for $p^2<c_s\mathcal{M}^2/|\mathcal{A}|$ and $p^2<c_s\mathcal{M}^2/|\mathcal{A}|$, respectively.

\begin{figure}
    \centering
    \subfloat[]{\includegraphics[scale=0.55]{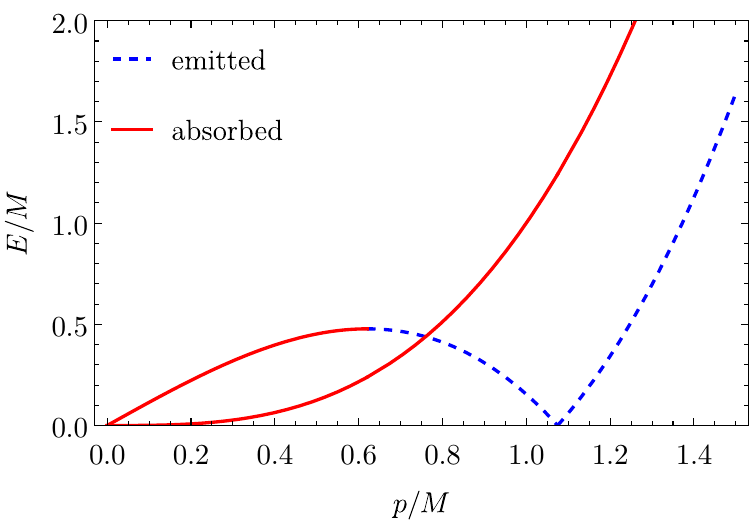}}\quad
    \subfloat[]{\includegraphics[scale=0.55]{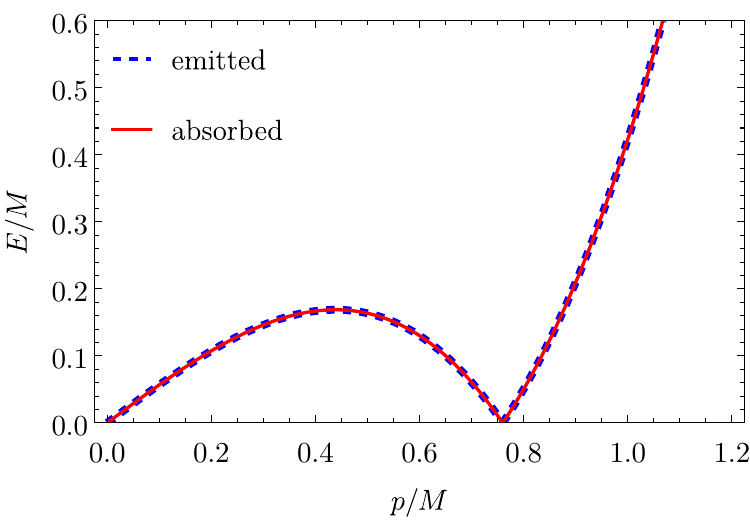}}\\
    \caption{\emph{Dispersion law for $\mathcal{A}$ negative, see text. The red solid lines indicate the absorbed modes, the blue dashed lines indicate the emitted modes. (a) Dispersion law for phonons at the horizon. (b) Dispersion law far from the horizon. In both (a) and (b), for the parameters' value, we consider $|\mathcal{A}|=1$, and for $c_s$ we take the conformal limit, i.e. $c_s=1/\sqrt{3}$. The two dispersion laws are different. Indeed, far from the horizon there are emitted and absorbed phonons for each value of the momentum, while at the horizon, there are emitted phonons only for a certain trigger in the momentum. For momentum lower that this trigger, there are only absorbed phonons.}}
    \label{fig:modesanegative}
\end{figure}

Like with $\mathcal{A}>0$, we classify the modes as emitted or absorbed depending on the sign of their group velocity. Thus, the emitted mode at the horizon is described by
\begin{equation*}
    E_{em}=
    \begin{cases}
        -2c_s p+\frac{|\mathcal{A}|}{\mathcal{M}^2}p^3 & p^2>2c_s\frac{\mathcal{M}^2}{|\mathcal{A}|}\\
        2c_s p-\frac{|\mathcal{A}|}{\mathcal{M}^2}p^3 & \frac23c_s\frac{\mathcal{M}^2}{|\mathcal{A}|}<p^2<2c_s\frac{\mathcal{M}^2}{|\mathcal{A}|},
    \end{cases}        
\end{equation*}
with a group velocity
\begin{equation*}
    v_{g,em}=-2 c_s+3\frac{|\mathcal{A}|}{\mathcal{M}^2}>0 \quad \mathrm{for}\quad p^2>2c_s\frac{\mathcal{M}^2}{|\mathcal{A}|}.
\end{equation*}
The absorbed mode at horizon instead, and its corresponding group velocity are described by:
\begin{equation*}
\begin{cases}
    &E_{ab}=\frac{|\mathcal{A}|}{\mathcal{M}^2}p^3\\
    &v_{g,ab}= -3 \frac{|\mathcal{A}|}{\mathcal{M}^2}p^2
    \end{cases}\qquad \forall p>0.
\end{equation*}
We notice that at the horizon for $p^2<\frac23c_s\frac{\mathcal{M}^2}{|\mathcal{A}|}$, one other solution exists for the absorbed mode, given by
\begin{equation*}
    E_{ab}=2c_sp-\frac{|\mathcal{A}|}{\mathcal{M}^2}p^3\qquad v_{g,ab}=2c_s-3\frac{|\mathcal{A}|}{\mathcal{M}^2}.
\end{equation*}
Asymptotically, far from the horizon, the emitted and the absorbed modes are described by
\begin{equation*}
    E_{em} =E_{ab} = 
    \begin{cases}
        c_sp-\frac{|\mathcal{A}|}{\mathcal{M}^2}p^3 & p^2<c_s\frac{\mathcal{M}^2}{|\mathcal{A}|}\\
        -c_sp+\frac{|\mathcal{A}|}{\mathcal{M}^2}p^3 & p^2\ge c_s\frac{\mathcal{M}^2}{|\mathcal{A}|},
    \end{cases}
    \label{eq:modelec}
\end{equation*}
and the group velocity for the emitted and the absorbed modes are
\begin{equation*}
    v_{g,em}= 
    \begin{cases}
        c_s-3\frac{|\mathcal{A}|}{\mathcal{M}^2}p^2 & 0\le p^2\le \frac13 c_s\frac{\mathcal{M}^2}{|\mathcal{A}|}\\
    -c_s+3\frac{|\mathcal{A}|}{\mathcal{M}^2}p^2 & \frac13 c_s\frac{\mathcal{M}^2}{|\mathcal{A}|}<p^2   
    \end{cases}
\end{equation*}
and 
\begin{equation*}    
    v_{g,ab}=
    \begin{cases}
        -c_s+3\frac{|\mathcal{A}|}{\mathcal{M}^2}p^2 & 0\le p^2\le \frac13 c_s\frac{\mathcal{M}^2}{|\mathcal{A}|}\\
    c_s-3\frac{|\mathcal{A}|}{\mathcal{M}^2}p^2 & \frac13 c_s\frac{\mathcal{M}^2}{|\mathcal{A}|}<p^2  
    \end{cases},
\end{equation*}
respectively.
Thus, the dispersion laws are non-monotonic~\cite{Corley_1996}.
The dispersion laws are summarized in Figs.~\ref{fig:modesanegative}. In particular, we can notice in from Fig.~\ref{fig:modesanegative}(b) that the energy behaves like the dispersion law for a Bose liquid~\cite{Nozires2018TheoryOQ}. The fact that there is a momentum $\Bar{k}$ for which the energy vanishes means that, for this value, the system can produce massless particles, i.e. particles with arbitrarily small energy. In addition, the cancellation of the energy for $\Bar{k}$ implies the introduction of a characteristic length, given by the inverse of $\Bar{k}$, so that the system is not any longer invariant under translations: a phase transition occurs~\cite{leutwyler1996phonons,SCHAKEL_1996}. The fact that the translation symmetry is broken with a characteristic length $\sim 1/\Bar{k}$, while being in a superfluid states suggests that the system might have undergone a transition to a supersolid-like phase~\cite{PhysRevA.107.033301,PhysRevResearch.2.043318,during2011theory,PhysRevLett.25.1543}.

This result is extremely interesting for its potential implications. This scenario opens up intriguing possibilities, where the behavior of the horizon can be connected to signatures of quantum phase transitions. Moreover, this study can have an experimental application in systems akin to dipolar BECs (BECs), where phonons present a non-monotonic dispersion law. Subsequently, we could investigate whether the obtained system exhibits supersolid properties. To achieve this, we would need to expand the scope of our study from (1+1)D to at least (2+1)D. By doing so, we can explore the emergence of superfluid phonon excitations and shear waves~\cite{during2011theory,PhysRevLett.25.1543,PhysRevA.105.023316,Tanzi_2021}, which are recognized as crucial indicators of a system's supersolid character.

Furthermore, this extended investigation could shed light on additional aspects related to supersolidity and offer insights into the underlying phenomena. For example, we could examine the collective behavior of the system, study its response to external perturbations, and investigate the topological properties associated with the supersolid phase. Such an in-depth analysis has the potential to deepen our understanding of supersolidity and provide valuable contributions to the field of condensed matter physics.

Thus, given a generic effective Lagrangian for a complex, massive, scalar field with non-minimal couplings, we would like to understand under which conditions the $\mathcal{A}<0$ case could be realized. This constraints must be searched by considering a general EFT without specifying in which microscopic system we are working. We elaborate on this question in the next section.

\section{Constraints to obtain the non-monotonic dispersion-law \label{sec:lec}}

We now want to find the conditions under which the low-energy Lagrangian allows for a non-monotonic phonon dispersion law. To do this, we proceed to build an effective Lagrangian at next-to-leading order with only tree-level terms, without specifying the microscopic Lagrangian.
Thus, we consider that our system is described by a generic Lagrangian depending on the field $\rho,\ \partial_\mu\rho,\ \partial_\mu\theta$, and expand this Lagrangian around the stationary point $(\rhoo,\partial_\mu\rhoo,\partial_\mu\thetao)$, as we made in Sec.~\ref{sec:effectivelagrangian}. Since we are interested in deriving the phonon dispersion law, we focus on the quadratic terms in the fluctuations, which are
\begin{equation}
     \begin{split}
        \lag_2 =& \left.\frac{\rhot^2}{2}\frac{\delta^2\lag}{\delta\rho^2}\right|_{\rhoo,\thetao} + \left.\frac{\partial_\mu\rhot\partial_\nu\rhot}{2}\frac{\delta^2\lag}{\delta\partial_\mu \rho\delta\partial_\nu \rho}\right|_{\rhoo,\thetao} + \left.\frac{\partial_\mu\thetat\partial_\nu\thetat}{2}\frac{\delta^2\lag}{\delta\partial_\mu\theta\delta\partial_\nu\theta}\right|_{\rhoo,\thetao} +\\
        &+\left. \rhot\partial_\mu\rhot\frac{\delta^2\lag}{\delta\rho\delta\partial_\mu\rho}\right|_{\rhoo,\thetao} + \left.\rhot\partial_\mu\thetat\frac{\delta^2\lag}{\delta\rho\delta\partial_\mu\theta}\right|_{\rhoo,\thetao} + \left.\partial_\mu\rhot\partial_\nu\thetat\frac{\delta^2\lag}{\delta\partial_\mu\rho\delta\partial_\nu\theta}\right|_{\rhoo,\thetao}.
        \label{eq:develope}
    \end{split}
\end{equation}
In Sec.~\ref{sec:effectivelagrangian} we had already defined 
\begin{equation*}
    \begin{split}
         &\left.\frac{\rhot^2}{2}\frac{\delta^2\lag}{\delta\rho^2}\right|_{\rhoo,\thetao}=\frac{\mt^2}{2}\rhot^2\\
         &\left.\frac{\partial_\mu\rhot\partial_\nu\rhot}{2}\frac{\delta^2\lag}{\delta\partial_\mu \rho\delta\partial_\nu \rho}\right|_{\rhoo,\thetao}=\frac12\eta^{\mu\nu}\partial_\mu\rhot\partial_\nu\rhot\\
         &\left.\frac{\partial_\mu\thetat\partial_\nu\thetat}{2}\frac{\delta^2\lag}{\delta\partial_\mu\theta\delta\partial_\nu\theta}\right|_{\rhoo,\thetao}=\frac{B^2}{2}\eta^{\mu\nu}\partial_\mu\thetat\partial_\nu\thetat\\
         &\left.\rhot\partial_\mu\thetat\frac{\delta^2\lag}{\delta\rho\delta\partial_\mu\theta}\right|_{\rhoo,\thetao}=V^\mu\rhot\partial_\mu\thetat    ,   
    \end{split}
\end{equation*}
where $\mt$ is the effective mass of $\rhot$, $B$ a constant with energy dimension 1, and $V^\mu$ a generic vector breaking the $U(1)$ symmetry, and thus depending on the quantities which cause the breaking. We need to define the remaining terms. The fourth term in the expansion ~\eqref{eq:develope}, that can be written as
\begin{equation}
    \left. \rhot\partial_\mu\rhot\frac{\delta^2\lag}{\delta\rho\delta\partial_\mu\rho}\right|_{\rhoo,\thetao}=\Tilde{A}^\mu\rhot\partial_\mu\rhot, 
\end{equation}
where $\Tilde{A}^\mu$ is a vector which depends on the symmetry breaking parameters. Assuming that there is no additional symmetry breaking mechanism, it follows that $\Tilde{A}^\mu\propto V^\mu$, i.e. $\Tilde{A}^\mu=A V^\mu$ with $A$ having dimensions $-1$. In the local-density approximation this term can be cast as a total derivative $\rhot\partial_\mu\rhot=\partial_\mu\rhot^2/2$, and therefore it does not give any contribution to the equation of motion.

The last term in Eq.~\eqref{eq:develope} can be written as
\begin{equation}
    \left.\partial_\mu\rhot\partial_\nu\thetat\frac{\delta^2\lag}{\delta\partial_\mu\rho\delta\partial_\nu\theta}\right|_{\rhoo,\thetao}=C^{\mu\nu}\partial_\mu\rhot\partial_\nu\thetat, 
\end{equation}
where $C^{\mu\nu}$ is a tensor depending on both the background and the symmetry-breaking quantities. Since the medium effects are encoded in the minimal coupling, we assume that this tensor can be expressed as a linear combination of $\eta^{\mu\nu}$ and $V^\mu V^\nu$ 
\begin{equation}
    C^{\mu\nu}=C_1\eta^{\mu\nu}+C_2V^\mu V^\nu, 
\end{equation}
where $C_1$ and $C_2$ are two constants with dimensions $1$ and $-3$, respectively. Since the effective Lagrangian must be invariant under $\rho\to-\rho$, we expect that these two constants depend on some odd power of $\rhoo$. In addition, since we have imposed that in our Lagrangian there are no dissipative terms, the Lagrangian must only contain hermitian operators, and therefore each term must be invariant under $\theta\to-\theta$ transformation. Therefore, the constants $C_1$ and $C_2$ must depend on the sign of the velocity $\bm v=\bm \nabla \thetao$.

To make our equations easier to read, in the following we use the tensor $C^{\mu\nu}$ instead of the constants $C_1$ and $C_2$.
Thus, our effective Lagrangian at tree-level is
\begin{equation}
    \begin{split}
        \lag_2=&\frac12\partial_\mu\rhot\partial^\mu\rhot-\frac{\mt^2}{2}\rhot^2+\frac{B^2}{2}\partial_\mu\thetat\partial^\mu\thetat+V^\mu\rhot\partial_\mu\thetat+AV^\mu\rhot\partial_\mu\rhot+\\
        &+C^{\mu\nu}\partial_\mu\rhot\partial_\nu\thetat.
    \end{split}
    \label{eq:develope2}
\end{equation}
Since we want to find the dispersion law for the phonons, we need an effective Lagrangian for the excitations, so that we need to integrate out the field $\rhot$. To do this, we search for its equation of motion, which is
\begin{equation}
    \rhot=\frac{1}{\Box+\mt^2}\left(V^\mu-C^{\mu\nu} \partial_\nu\right)\partial_\mu\thetat.
    \label{eq:new_rhot_eq}
\end{equation}
We are interested in the $p^4$-corrections, thus we expand the operator $\frac{1}{\Box+\mt^2}\simeq\frac{1}{\mt^2}\left(1-\frac{\Box}{\mt^2}\right)$ and substitute these results in Eq.~\eqref{eq:develope2}. Knowing that the field $\partial_\mu\thetat\sim p$, we see from Eq.~\eqref{eq:new_rhot_eq} that $\rhot$ has contributions up to the third-order in the momenta: a term which behaves like $\sim p$, one like $\sim p^2$, and another like $p^3$. Therefore, with reference to Eq.~\eqref{eq:develope2} we have that 
\begin{itemize}
    \item the first term behaves like $\sim \beta_{11}p^4+\beta_{12}p^5+..$ ;
    \item the second term behaves like $\sim \beta_{21}p^2+\beta_{22}p^3+\beta_{23}p^4+\beta_{24}p^4+\beta_{25}p^5+..$ ;
    \item the third term behaves like $\sim \beta_{31}p^2$ ;
    \item the forth term behaves like $\sim \beta_{41}p^2+\beta_{42}p^3+\beta_{43}p^4+\beta_{44}p^5+..$ ;
    \item the fifth term behaves like $\sim \beta_{51}p^3+\beta_{52}p^4+\beta_{53}p^4+\beta_{54}p^5+..$ ;
    \item the sixth term behaves like $\sim \beta_{61}p^3+\beta_{62}p^4+\beta_{63}p^5+..$ .
\end{itemize}
Here, $\beta_{ij}$ are constants.
Keeping terms up to $p^4$, we have that the effective tree-level Lagrangian for phonons is
{\small
\begin{equation}
    \begin{split}
        \lag_2=\frac{1}{2}\partial_\mu\thetat&\left[\frac{V^\mu V^\nu}{\mt^2}+B^2\eta^{\mu\nu}+\frac{2}{\mt^2}\left(C^{\mu\nu}V^\alpha\partial_\alpha+\frac{AV^\alpha V^\mu V^\nu}{\mt^2}\partial_\alpha\right)\right]\partial_\nu\thetat+\\
        +\frac{1}{2\mt^2}&\Big[\frac{V^\mu V^\nu}{\mt^2}\partial^\alpha\thetat\partial_\mu\thetat\partial_\alpha\partial_\nu\thetat-C^{\mu\nu}C^{\alpha\beta}\partial_\mu\partial_\nu\thetat\partial_\alpha\partial_\beta\thetat-2C^{\mu\nu}C^{\alpha\beta}\partial_\mu\thetat\partial_\nu\partial_\alpha\partial_\beta\thetat+\\
        &-2AC^{\mu\nu}V^\alpha V^\beta \partial_\alpha\thetat\partial_\beta\partial_\mu\partial_\nu\thetat-2AC^{\mu\nu}V^\alpha V^\beta\partial_\mu\partial_\nu\thetat\partial_\alpha\partial_\beta\thetat\Big].
    \end{split}
\end{equation}
}
We notice that setting $A=C_1=C_2=0$ ($C_1=C_2=0$ implies $C^{\mu\nu}=0\ \forall\ \mu,\nu=0,1,2,3$), and we obtain the same Lagrangian that we found in Sec.~\ref{sub:quadratic}.
After Fourier transforming 
\begin{equation}
    \thetat(x)=\int \frac{d^4p}{(2\pi)^4}e^{-ip\cdot x}\thetat(p), 
\end{equation}
we get 
\begin{equation}
    \begin{split}
        \lag_2=\int \frac{d^4p}{(2\pi)^4}\frac{d^4q}{(2\pi)^4}&e^{-ip\cdot x}e^{-iq\cdot x}\thetat(q)\Big[-\frac{V^\mu V^\nu}{2\mt^2} p_\mu q_\nu-\frac{B^2}{2}p_\mu q^\mu+\\
        &+\frac{iC^{\mu\nu} V^\alpha}{\mt^2} q_\mu p_\nu p_\alpha+\frac{iA V^\mu V^\nu V^\alpha}{\mt^4} q_\mu p_\nu p_\alpha+\frac{V^\mu V^\nu}{2\mt^4}q^\alpha p_\alpha q_\nu p_\mu+\\
        &-\frac{C^{\mu\nu}C^{\alpha\beta}}{2\mt^2}q_\mu q_\nu p_\alpha p_\beta-\frac{C^{\mu\nu}C^{\alpha\beta}}{\mt^2}q_\mu p_\nu p_\alpha p_\beta+\\
        &-\frac{AC^{\mu\nu}V^\alpha V^\beta}{\mt^2}q_\mu p_\nu p_\alpha p_\beta-\frac{AC^{\mu\nu}V^\alpha V^\beta}{\mt^2}q_\mu q_\nu p_\alpha p_\beta\Big]\thetat(p).
    \end{split}
\end{equation}
The action $S=\int d^4x\lag_2$ can be written as
\begin{equation}
    \begin{split}
        S=\int d^4x\int \frac{d^4p}{(2\pi)^4}&\frac{d^4q}{(2\pi)^4}e^{-i(p+q)\cdot x}\thetat(q)\Big[-\frac{V^\mu V^\nu}{2\mt^2} p_\mu q_\nu-\frac{B^2}{2}p_\mu q^\mu+\\
        &+\frac{iC^{\mu\nu} V^\alpha}{\mt^2} q_\mu p_\nu p_\alpha+\frac{iA V^\mu V^\nu V^\alpha}{\mt^4} q_\mu p_\nu p_\alpha+\frac{V^\mu V^\nu}{2\mt^4}q^\alpha p_\alpha q_\nu p_\mu+\\
        &-\frac{C^{\mu\nu}C^{\alpha\beta}}{2\mt^2}q_\mu q_\nu p_\alpha p_\beta-\frac{C^{\mu\nu}C^{\alpha\beta}}{\mt^2}q_\mu p_\nu p_\alpha p_\beta+\\
        &-\frac{AC^{\mu\nu}V^\alpha V^\beta}{\mt^2}q_\mu p_\nu p_\alpha p_\beta-\frac{AC^{\mu\nu}V^\alpha V^\beta}{\mt^2}q_\mu q_\nu p_\alpha p_\beta\Big]\thetat(p).
    \end{split}
\end{equation}
In performing these steps, we have neglected the fact that the field $\rhoo$ and the velocity depend on position $x$: we are in local- density approximation, thus the variation of these quantities is negligible with respect to variations in $\thetat$. Therefore, treating the velocity and $\rhoo$ as constants is a legitimate choice. Integrating with respect to $x$, noticing that the dominant term depending from the position is $e^{-i(p+q)\cdot x}$, yields $\delta^{(4)}(p+q)$. Thus, integrating also with respect to $q$ we get
\begin{equation}
    S=\frac12\int\frac{d^4p}{(2\pi)^4}\thetat(-p)D^{-1}(p)\thetat(p), 
\end{equation}
where the inverse propagator is
\begin{equation}
    \begin{split}
        D^{-1}=&\frac{V^\mu V^\nu}{2\mt^2} p_\mu p_\nu+\frac{B^2}{2}p_\mu p^\mu-\frac{iC^{\mu\nu} V^\alpha}{\mt^2} p_\mu p_\nu p_\alpha-\frac{iA V^\mu V^\nu V^\alpha}{\mt^4} p_\mu p_\nu p_\alpha+\\
        &+\frac{V^\mu V^\nu}{2\mt^4}p^\alpha p_\alpha p_\nu p_\mu+\frac{C^{\mu\nu}C^{\alpha\beta}}{2\mt^2}p_\mu p_\nu p_\alpha p_\beta.
    \end{split}
\end{equation}
The cubic terms do not give any contribution to $D^{-1}(p)$. Indeed, since for each term the choice of which momenta, $p$ or $q$, to assign to the fields $\thetat$ is arbitrary, when we exchange them the inverse of the propagator must be the same. We now consider $S_1$ and $S_2$ defined as
\begin{equation}
    \begin{split}
        &S_1=\int d^4x\int\frac{d^4q}{(2\pi)^4}\frac{d^4p}{(2\pi)^4}e^{-i(p+q)\cdot x}\thetat(q)\frac{i}{\mt^2}\Big[C^{\mu\nu}V^\alpha q_\mu p_\nu p_\alpha+\\
        &\qquad \qquad+\frac{AV^\mu V^\nu V^\alpha}{\mt^2}q_\mu p_\nu p_\alpha\Big]\thetat(p)\\
        &S_2=\int d^4x\int\frac{d^4q}{(2\pi)^4}\frac{d^4p}{(2\pi)^4}e^{-i(p+q)\cdot x}\thetat(q)\frac{i}{\mt^2}\Big[C^{\mu\nu}V^\alpha p_\mu q_\nu q_\alpha+\frac{AV^\mu V^\nu V^\alpha}{\mt^2}p_\mu q_\nu q_\alpha\Big]\thetat(p). 
    \end{split}
\end{equation}
With the same reasoning as before, we now integrate with respect to $x$ and $q$ and get
\begin{equation}
    \begin{split}
        &S_1=\int\frac{d^4p}{(2\pi)^4}\thetat(-p)\frac{i}{\mt^2}\Big[-C^{\mu\nu}V^\alpha p_\mu p_\nu p_\alpha-\frac{AV^\mu V^\nu V^\alpha}{\mt^2}p_\mu p_\nu p_\alpha\Big]\thetat(p)\\
        &S_2=\int\frac{d^4p}{(2\pi)^4}\thetat(-p)\frac{i}{\mt^2}\Big[C^{\mu\nu}V^\alpha p_\mu p_\nu p_\alpha+\frac{AV^\mu V^\nu V^\alpha}{\mt^2}p_\mu p_\nu p_\alpha\Big]\thetat(p), 
    \end{split}
\end{equation}
so that $S_1=-S_2$. But we know that $S_1=S_2$ and thus this means $S_1=S_2=0$.

The dispersion law is again given by the poles of the propagator for $\thetat$, thus by the zeroes of $D^{-1}(p)$. We want to stay in the non-relativistic limit. To this purpose, it is convenient to work with dimensionless parameters. Given that $[\rhoo]=1$, we can write the parameters as
\begin{align}
    &[V^\mu]=2 \ \Rightarrow\ V^\mu=\rhoo^2\opV^\mu\,,\quad [\opV^\mu]=0\\
    &[B]=1 \ \Rightarrow\ B=\rhoo\opB\,,\quad [\opB]=0\\
    &[C_1]=1 \ \Rightarrow\ C_1=\rhoo\opC_1\,,\quad [\opC_1]=0\\
    &[C_2]=-3 \ \Rightarrow\ C_2=\frac{\opC_2}{\rhoo^{3}}\,,\quad [\opC_2]=0\,
\end{align}
where $\opV^\mu,\ \opB,\ \opC_1$ and $\opC_2$ are real. The equation that needs to be solved in order to get the dispersion law is given by
\begin{equation}
    \begin{split}
        &\frac{\rhoo^2}{2\mt^2}\opV^\mu\opV^\nu p_\mu p_\nu+\frac
        {\opB^2}{2}p_\mu p^\mu +\frac{\rhoo^2}{2\mt^4}\opV^\mu\opV^\nu p_\mu p_\nu p_\alpha p^\alpha+\frac{\opC_1^2}{2\mt^2}(p_\mu p^\mu)^2+\\
        &+\frac{\opC_2^2}{2\mt^2}\opV^\mu\opV^\nu\opV^\alpha\opV^\beta p_\mu p_\nu p_\alpha p_\beta +\frac{\opC_1 \opC_2}{\mt^2} \opV^\mu \opV^\nu p_\mu p_\nu p_\alpha p^\alpha=0.
        \label{eq:dispersionlaw4}
    \end{split}
\end{equation}
We know that $\opV^\mu$ is proportional to $v^\mu$ times some dimensionless constant.

We now write the velocity and the momenta as $v^\mu=\gamma(1,\bm v)$ and $p^\mu=(\omega,\bm p)$, with $|\bm v|=v$ and $|\bm p|=p$.
We are in the non-relativistic limit thus, given $c_s$ the system speed of sound, this means that $c_s,v\ll 1$ and $\gamma\simeq1$. 
If we consider the case in which $\opC_1=\opC_2=0$ with the microscopic Lagrangian, we notice that
\begin{equation}
    \frac{\opV^\mu\opV^\nu}{\mt^2}\sim \frac{v^\mu v^\nu}{\rhoo^2c_s^2}\qquad \frac{\mt^2}{2\rhoo^2}=\lambda=\frac{2}{\rhoo^2}\left(\frac{\mu}{\gamma}\right)\frac{c_s^2}{1-c_s^2}\sim2c_s^2.
\end{equation}
After replacing all these observations in Eq.~\eqref{eq:dispersionlaw4}, we get
\begin{equation}
    \begin{split}
        &\underbrace{\frac{1}{2}(\omega-p_\parallel v)^2}_{A}+\underbrace{\frac{\opB^2c_s^2}{2}(\omega^2-p^2)}_{B}+\underbrace{\frac{1}{2\mt^2}(\omega-p_\parallel v)^2(\omega^2-p^2)}_{C}+\\
        &\underbrace{\frac{\opC_1^2c_s^2}{2\mt^2}(\omega^2-p^2)^2}_{D}+\underbrace{\frac{8\opC_2^2c_s^2}{\mt^2}(\omega-p_\parallel)^4}_{E}+\underbrace{\frac{4\opC_1\opC_2c_s^2}{\mt^2}(\omega-p_\parallel v)^2(\omega^2-p^2)}_{F}=0.
    \end{split}
    \label{eq:dispersionlaw5}
\end{equation}
Here, we have defined $p_\parallel v=\bm p\cdot \bm v$. In the first place, it is convenient to consider only the term $O(p^2)$ and thus we momentarily neglect the terms $C, D, E$ and $F$ in~\eqref{eq:dispersionlaw5}. Being in the non-relativistic limit, we proceed with a power counting of the velocity and consider only the lowest powers which, as we will see, is $O(v^2)$. As previously performed, we assume $c_s$ to be the same order of $v$, $p\sim v^0$, and $\omega\sim v$. We then find that
\begin{itemize}
    \item $A\sim\frac12(\omega-p_\parallel v)^2$ ;
    \item $B\sim-\frac12\opB^2c_s^2p^2$
\end{itemize}
and

\begin{itemize}
    \item $C\sim-\frac{1}{2\mt^2}(\omega-p_\parallel v)^2p^2$ ;
    \item $D\sim\frac{\opC_1^2}{2\mt^2}c_s^2p^4$ ;
    \item $E\sim F\sim 0$ .
\end{itemize}
Thus, Eq.~\eqref{eq:dispersionlaw5} simplifies to 
\begin{equation}
    (\omega-p_\parallel v)^2-\opB^2c_s^2p^2-\frac{p^2}{\mt^2}(\omega-p_\parallel v)^2+\frac{\opC_1^2}{\mt^2}c_s^2p^4=0
\end{equation}
and its solution is 
\begin{equation}
    \omega=p_\parallel v\pm \opB c_s p\left[1+\left(1-\frac{\opC_1^2}{\opB^2}\right)\frac{p^2}{2\mt^2}\right].
\end{equation}
This solution implies that for $|\opC_1|>|\opB|$ the coefficient $\mathcal{A}$ defined in the previous section~\ref{sec:displaw} is negative and therefore the dispersion law follows the behavior shown in Fig.~\ref{fig:modesanegative}.

\section{Density-density correlation function \label{sec:density}}

In this section, we finally expose a new method to obtain the correlation functions of a superfluid system given its Lagrangian density. These functions give us information about the analogue Hawking radiation. In particular, we study the density-density correlation function, though the approach can be extended to any other correlation functions. We use the field theory notation~\cite{elio3, landau2013statistical} and work with the effective theory at tree-level. The result obtained is compared with that reported in~\cite{PhysRevLett.47.1840}. Although we work in a $1+1$-dimensional system, extension to higher dimensions is straightforward.

We shall discuss two cases: the system described by the microscopic Lagrangian in Eq.~\eqref{eq:microscopic}, and a system described by the effective Lagrangian with generic LECs (see Eq.~\eqref{eq:develope2}).
The starting point is the generating functional $Z[J]$, which, given an action $S[\Psi]$ for a scalar field $\Psi$, is defined as
\begin{equation}
    Z[J]=\int \left[\mathcal{D}\Psi\right]e^{iS[\Psi]+i\int d^4x J(x)\Psi(x)}.
\end{equation}
Its functional derivatives with respect to the external current $J$ give the correlations:
\begin{equation}
    \langle \Psi(x_1)..\Psi(x_n)\rangle=\frac{(-i)^n}{Z[0]}\left.\frac{\delta^n Z[J]}{\delta J(x_1)..\delta J(x_n)}\right|_{J=0}.
\end{equation}
Let us consider a generic quadratic Lagrangian depending on the fluctuations $\rhot$ and $\thetat$.  
If now define the vector $\Psi=(\rhot,\thetat)$, the corresponding partition function is given by
\begin{equation}
    Z[J]=\int \mathcal{D}\Psi\exp\left(i\int d^2x\int d^2y \frac12\Psi^t(x) G^{-1}(x,y)\Psi(y)-i\int d^2x J^t(x)\Psi(x)\right),
    \label{eq:partition}
\end{equation}
where $G^{-1}(x,y)$ is the $2\times2$ matrix inverse propagator, while the quantity $J^t(x)=(J_1,J_2)$ is the vector of the external currents. The first component, $J_1(x)$, is the current coupled to the field $\rhot$, while the second current $J_2(x)$ is coupled to the field $\thetat$.
Since Eq.~\eqref{eq:partition} is Gaussian, we can integrate out $\Psi$ to obtain
\begin{equation}
    Z[J]\propto e^{i\int d^2x \int d^2y\frac12 J^t(x)G(x,y)J(y)}.
\end{equation}
Therefore, the correlation functions are given by the functional derivatives with respect the currents' component, that are:
\begin{align}
    \label{eq:nn}
    &\langle \rhot(x)\rhot(y)\rangle=-\left.\frac{1}{Z[0]}\frac{\delta^2Z[J]}{\delta J_1(x)\delta J_1(y)}\right|_{J=0}=-iG_{11}(x,y),\\
    &\langle \rhot(x)\thetat(y)\rangle=\left.-\frac{1}{Z[0]}\frac{\delta^2Z[J]}{\delta J_1(x)\delta J_2(y)}\right|_{J=0}=-iG_{12}(x,y),\\
    &\langle \thetat(x)\rhot(y)\rangle=\left.-\frac{1}{Z[0]}\frac{\delta^2Z[J]}{\delta J_2(x)\delta J_1(y)}\right|_{J=0}=-iG_{21}(x,y),\\
    &\langle \thetat(x)\thetat(y)\rangle=\left.-\frac{1}{Z[0]}\frac{\delta^2Z[J]}{\delta J_2(x)\delta J_2(y)}\right|_{J=0}=-iG_{22}(x,y).
\end{align}
In order to get the density-density correlation function $\langle\rhot(x)\rhot(y)\rangle$, we are interested in $G_{11}(x,y)$. 

\subsection{Microscopic model}

In the following, for definiteness, we consider the system described by the microscopic Lagrangian~\eqref{eq:microscopic}. The matrix which identifies the inverse of the propagator for the fields, $G^{-1}(x,y)$ is 
\begin{equation}
    G^{-1}(x,y)=\left(
    \begin{matrix}
        -(\Box+\mt^2) & V^\mu \partial_\mu\\
        -V^\mu\partial_\mu & -B^2\Box
    \end{matrix}\right)\delta^{(2)}(x-y),
\end{equation}
where all the derivatives are made with respect to the variable $x$.
As stated before, the density-density correlation function is given by Eq.~\eqref{eq:nn}. Thus, we are interested in obtaining the propagator $G_{11}(x,y)$.
In momentum space we have that
\begin{equation}
    \tilde{G}_{11}(p)=\frac{B^2p_\mu p^\mu}{B^2(p_\mu p^\mu)^2-\mt^2B^2p_\mu p^\mu - V^\mu V^\nu p_\mu p_\nu}\,,
    \label{eq:tildeG11}
\end{equation}
then we substitute the relations~\eqref{eq:Bsq} and~\eqref{eq:Vmu} and we consider the power expansion in $p/\mt$. For simplicity, we consider the fluid rest frame, $V^\mu=-2\rhoo\mu(1,0)$. Leveraging also the relation
\begin{equation*}
    \left(\frac{1}{c_s^2}-1\right)=\frac{2}{\lambda\rhoo^2}\left(\frac{\mu}{\gamma}\right)^2,
\end{equation*}
one gets the following expression for the density-density correlation function
\begin{equation}
    G_{11}(x)=\int\frac{d\omega}{2\pi}\int_I\frac{dp}{2\pi}e^{-i\omega x_0+ipx}\frac{\omega^2-p^2}{(\omega^2-p^2)^2-\mt^2(\omega^2-p^2)-\mt^2\left(\frac{1}{c_s^2}-1\right)\omega^2},
    \label{eq:G11_1}
\end{equation}
where the integration with respect to the spatial momenta is made over the interval $I=[-k_2,-k_1]\cup[k_1,k_2]$. At the end of the calculation, we will take the limits $k_2\to+\infty$ and $k_1\to0$. The calculation of the integrals in Eq.~\eqref{eq:G11_1} is made by considering that $p\ll\mt$. The equal time density-density correlation function is obtained from Eq.~\eqref{eq:G11_1} taking the $x_0\to 0$ limit
\begin{equation}
    \langle \rhot(x)\rhot(0)\rangle= \frac{c_s(1-c_s^2)}{2\pi \mt^2 x^2}.
\end{equation}
In the non-relativistic limit $c_s\ll1$, this becomes
\begin{equation}
    \langle \rhot(x)\rhot(0)\rangle= \frac{c_s}{2\pi \mt^2 x^2}.
    \label{eq:rhotrhot}
\end{equation}
This solution is identical to that obtained in~\cite{PhysRevLett.47.1840} with a completely different procedure. We want to emphasize that $\omega\sim v$ and $p\sim v^0$, so that the leading term in the propagator is $c_s^2p^2$. This means that the density-density correlation function is dominated by the Nambu-Goldstone boson pole.

We also note that when considering the normal phase, i.e. with no condensate, $\mt/c_s=2m$ and $\mt\to0$, the system does not any longer feel the presence of the Goldstone boson, and the density-density correlation function takes the form of a Bessel's function. Indeed, in the $p\ll2m$ limit one gets
\begin{equation}
    G_{11}=-i\int_I \frac{dp}{2\pi}e^{ipx}\frac{1-\frac{p^2}{2m^2}}{4\sqrt{m^2+p^2}}.
\end{equation}
Therefore, the solution in Eq.~\eqref{eq:rhotrhot} only holds for a homogeneous BEC. 

We close this section by noticing that this derivation can be extended to a space-dependent velocity profile in the local density approximation, including the case of a dumb hole. In this way, one could for example calculate the correlation function between two density fluctuations which propagate in the subsonic or supersonic region of an analogue black hole, which is interesting to explore the information paradox. This possibility is not addressed in the present thesis.

\subsection{General effective model}

In the following we revert back to the non-monotonic dispersion law~\eqref{eq:modelec}. For definiteness, we consider the system described by the effective Lagrangian discussed in Sec.~\ref{sec:lec}:
\begin{equation}
    \lag=-\frac12\rhot(\Box+\mt^2)\rhot+\rhoo^2\frac{\opB^2}{2}\partial_\mu\thetat\partial^\mu\thetat+\rhoo^2 \opV^\mu\rhot\partial_\mu\thetat+\rhoo\opC_1\partial_\mu\thetat\partial^\mu\rhot+\rhoo\opC_2\opV^\mu\opV^\nu\partial_\mu\thetat\partial_\nu\rhot,
\end{equation}
where we have already written the LECs in terms of dimensionless parameters and of $\rhoo$, defined such that $[\rhoo]=1$:
\begin{align}
    &[V^\mu]=2 \ \Rightarrow\ V^\mu=\rhoo^2\opV^\mu\,,\quad [\opV^\mu]=0\\
    &[B]=1 \ \Rightarrow\ B=\rhoo\opB\,,\quad [\opB]=0\\
    &[C_1]=1 \ \Rightarrow\ C_1=\rhoo\opC_1\,,\quad [\opC_1]=0\\
    &[C_2]=-3 \ \Rightarrow\ C_2=\frac{\opC_2}{\rhoo^{3}}\,,\quad [\opC_2]=0.
\end{align}
In addition, we remark that the constants $\opC_1,\ \opC_2\in\real$ change their sign with the velocity's sign. To get a dispersion relation as that shown in Fig.~\ref{fig:modesanegative}(b), as found in Sec.~\ref{sec:lec}, in the non-relativistic limit the inequality
\begin{equation}
    \opC_1^2 > \opB^2
\end{equation}
must be satisfied.
We have also neglected the interaction term $\propto\rhot\partial_\mu\rhot$, since it is a total derivative  and does not contribute to the equation of motion of the fields.
The matrix which identifies the inverse of the propagator for the fields, $G^{-1}(x,y)$ is 
\begin{equation}
    G^{-1}(x,y)=\left(
    \begin{matrix}
        -(\Box+\mt^2) & \rhoo^2\opV^\mu \partial_\mu-\rhoo\opC^{\mu\nu}\partial_\mu\partial_\nu\\
        -\rhoo^2\opV^\mu\partial_\mu-\rhoo\opC^{\mu\nu}\partial_\mu\partial_\nu & -B^2\Box
    \end{matrix}\right)\delta^{(2)}(x-y),
\end{equation}
where all the derivatives are made with respect to the variable $x$, and where the operator $\opC$ is defined as
\begin{equation}
    \opC^{\mu\nu}=\opC_1\eta^{\mu\nu}+ \opC_2\opV^\mu\opV^\nu.
\end{equation}
Even in this case, the density-density correlation function is given by Eq.~\eqref{eq:nn}. Thus, we are interested in obtaining the propagator $G_{11}(x,y)$.
In momentum space, we get
\begin{equation}
    \tilde{G}_{11}(p)=\frac{\opB^2p_\mu p^\mu}{\opB^2(p_\mu p^\mu)^2-\mt^2\opB^2p_\mu p^\mu - \rhoo^2\opV^\mu \opV^\nu p_\mu p_\nu-\opC^{\mu\nu}\opC^{\alpha\beta}p_\mu p_\nu p_\alpha p_\beta}.
    \label{eq:tildeG11lec}
\end{equation}
We write the vector $\opV^\mu$ in terms of $v^\mu$ and {in terms} of a dimensionless parameter $k$ such that, in the rest frame where we are, $\opV^\mu=k(1,0)$. 

The expression for the density-density correlation function is given by
\begin{equation}
    G_{11}(x)=\int\frac{d\omega}{2\pi}\int_I\frac{dp}{2\pi}e^{-i\omega x_0+ipx}f(\omega,p),
    \label{eq:G11_1lec}
\end{equation}
where the integration with respect to the spatial momenta is made over the interval $I=[-k_2,-k_1]\cup[k_1,k_2]$. At the end of the calculation we take the limit $k_1\to0$. This time we will not take the limit $k_2\to+\infty$: indeed, we will see that the function, which is integrated in the spatial momenta, has a denominator which vanishes for a finite value of the momentum. This behavior is due to the fact that $\opC_1^2>\opB^2$, therefore we will take the limit for which $k_2$ tends to this finite value of the momenta for which the function which is integrated diverges. In this way we obtain the dominant contribution to the integral.

\begin{figure}
    \centering
    \includegraphics[scale=0.8]{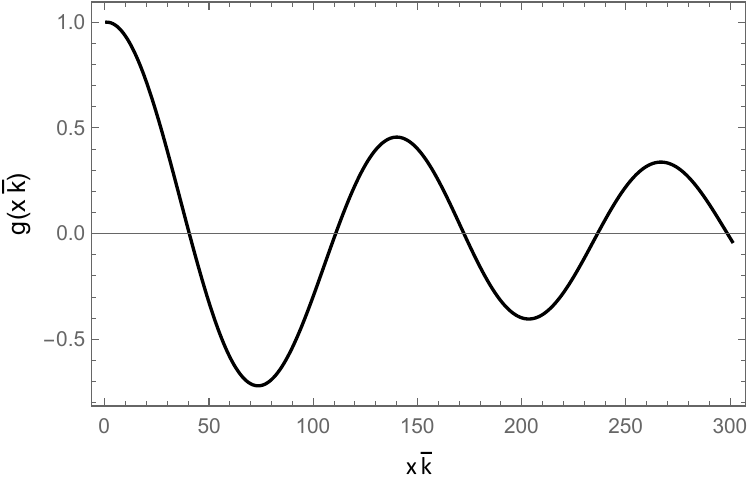}
    \caption{\emph{Numerical integration of $g(x\Bar{k})=\langle\tilde\rho(x)\tilde\rho(0)\rangle/\mathcal{K}$. Apart from the proportionality factor $\mathcal{K}$, given by Eq. \eqref{eq:g11_lecp2}, this is the density-density correlation function.}}
    \label{fig:numerical}
\end{figure}

The function $f(\omega,p)$ is
\begin{equation}
    f(\omega,p)=\frac{\opB^2(\omega^2-p^2)}{F\omega^4-(\opB^2\mt^2+\rhoo^2k^2+Qp^2)\omega^2+\opB^2\mt^2p^2+\opB^2p^2-\opC_1^2p^4},
\end{equation}
with
\begin{equation*}
    F=\opB^2-(\opC_1+\opC_2k^2)^2,\qquad Q=2(\opB^2-\opC_1^2-\opC_1\opC_2k^2).
\end{equation*}
The calculation of the integrals in Eq.~\eqref{eq:G11_1lec} is carried out by considering the limit $p\ll\mt$. After taking only the real part, we obtain:
\begin{equation}
    \langle\rhot(x)\rhot(x)\rangle=\mathcal{K}\int_{0}^{1}dp \frac{p\cos(py)}{\sqrt{1-p^2}}, 
    \label{eq:g11_lecp1}
\end{equation}
with
\begin{equation}
    \mathcal{K}=\frac{\Bar{k}^2}{\pi\opB\mt\sqrt{\opB^2\mt^2+\rhoo^2k^2}}.
    \label{eq:g11_lecp2}
\end{equation}
We have identified
\begin{equation}
    \Bar{k}=\frac{\mt}{\sqrt{\frac{\opC_1^2}{\opB^2}-1}}
\end{equation}
as the momentum value for which the denominator vanishes. We have taken the limit $k_2\to\Bar{k}$, and then we have rescaled $p\to p/\Bar{k}$ and $y=x\Bar{k}$. The solution is given by numerical integration, and is shown in Fig.~\ref{fig:numerical}.

Even in this case, we notice that the leading term in the propagator is given by $c_s|p|$. This means that also in a generic effective model, if there is a condensate, the system is dominated by the Goldstone mode poles.

This effective model satisfying the relation $\opC_1^2>\opB^2$, could represent a condensate in which a phase transition occurs as a result of a translation symmetry breaking. Within our system, a characteristic length is identified, indicating that we could be dealing with a supersolid, as mentioned earlier in Sec.~\ref{sub:aneg}. Indeed, the density-density correlation function's behavior is the same as that expected in a supersolid. While the behavior of the density-density correlation function confirms our hypothesis, these indications are not sufficient to confidently assert the presence of supersolidity. This can be possibly demonstrated by extending the system to at least one additional spatial dimension and verifying the presence of shear waves, for example after looking at the so-called scissor mode~\cite{PhysRevA.105.023316,Tanzi_2021}. 
Also interesting is the possibility to calculate and analyse, yet in local-density approximation for simplicity, the correlation functions for a BEC with a non-monotonic behavior of the energy dispersion law, where a velocity profile is imprinted, so to describe an acoustic black hole.

\chapter{Conclusions and perspectives\label{chap:5}}

In this final chapter, we provide an overview of the main findings of the present thesis work and discuss its implications and possible future research directions.

A significant achievement of this thesis is the introduction of an innovative approach to construct an analogue model utilizing a Bose-Einstein condensate (BEC) by means of an effective field theory (EFT) applied to a microscopic Lagrangian. The BEC is described by a Lagrangian for a massive, complex scalar field with a $U(1)$ and Lorentz symmetry breaking. By employing the Madelung representation $\phi={\rho}\exp(i\theta)/{\sqrt{2}}$, we recast this Lagrangian in terms of the fields $\rho$ and $\theta$. The cyclic nature of the field $\theta$ leads to a conserved current, indicating that the fluid behaves as a superfluid.

We derive the most general effective Lagrangian depending on $\rho$, $\partial_\mu\rho$, and $\partial_\mu\theta$, under the assumption that the $\theta$ field has only derivative couplings. Linearizing the general Lagrangian by separating the fields in background classical contributions and fluctuations, we obtain an effective Lagrangian for the Nambu-Goldstone boson associated to the spontaneous $U(1)$ breaking, up to the fourth order in interaction terms. This procedure allows us to integrate out irrelevant degrees of freedom and work with a low-energy effective theory. Furthermore, this method is applicable to any high density systems with $U(1)$ and Lorentz symmetry breaking, facilitating the inclusion of 1-loop corrections beyond the mean-field approximation and the incorporation of the local field approximation.

The developed method yields consistent results with other approaches in the literature. For instance, we obtain the same low-energy Lagrangian as Greiter, Wilczek, and Witten~\cite{doi:10.1142/S0217984989001400}, and we compare with the next-to-leading order Lagrangian corrections by Son and Wingate~\cite{Son2}. With respect to the mentioned works, the main advantage of our approach is that it allows us to include the fluctuations of the $\rho$ field in a straightforward way. 

Building upon these benchmarks, we present the main original results of this thesis. We calculate the next-to-leading order Lagrangian in terms of the microscopic Lagrangian's parameters. This leads to a dispersion law for phonon pairs in the presence of an acoustic horizon generated by the BEC's flow. Depending on the superfluid flow velocity and the next-to-leading order low-energy constants (LECs), the phonon dispersion law exhibits a nontrivial minimum, indicating a translation-symmetry breaking associated to the appearence of gapless rotonic excitations. This suggests the existence of a supersolid phase transition, which requires further investigation beyond the scope of this thesis. Additionally, we determine the constraints on the LECs necessary for the existence of this minimum and investigate the features of systems that exhibit this type of dispersion law.

We design an original procedure to calculate the density-density correlation function, a fundamental quantity for detecting and describing the analogue Hawking radiation. By recasting the Lagrangian in terms of a two component field $\psi^t=(\rho.\theta)$, representing the density and phase fluctuations, we obtain a $2\times2$ matrix representation of the inverse of the fields' propagators. The correlation functions are then given by the fields' propagators. We calculate the density-density correlation function for a homogeneous BEC in (1+1)D using both the explicit form of a Lagrangian as well as a general effective Lagrangians written in terms of the LECs for a system with $U(1)$ symmetry breaking. From these solutions, we notice that in the broken phase the dominant contribution to the density-density correlation functioncomes from the phonon pole; When the system is in the normal phase, the correlation function does not feel anymore the phonon-pole dominance and the correlation function takes the form of a Bessel function, as expected.
Our method is especially transparent because it allows to directly compute all the correlation functions in a systematic manner, besides the density-density correlation function. 
In a different method developed by Haldane~\cite{PhysRevLett.47.1840} instead, the density-density correlator is indirectly determined in a homogeneous BEC from the phonon-phonon correlation function. Needless to say, our result is consistent with that in~\cite{PhysRevLett.47.1840}.

\wl
In the following, we briefly discuss aspects that remain the subject of ongoing investigation, outlining the potential for new avenues of theoretical advancements.

The tools developed in this thesis work have wide-ranging applications. We provide a few illustrative examples below. One application is the calculation of the density-density correlation function in a non-homogeneous BEC with a velocity profile. To achieve this, we plan to utilize established techniques such as the local-density approximation, which provides a good starting point. In particular, leveraging on this approach, we can calculate the density-density correlation function for a dumb-hole when both density's excitations are in the subsonic region or in the supersonic region. Since in local-density approximation we consider the BEC as locally homogeneous, this method is instead not appropriate to calculate the density-density correlation function between two particles which propagate in two different regions of the acoustic black hole, i.e. with one propagating in the subsonic and the other in the supersonic regions. In order to access such a subsonic-supersonic correlation function, it is necessary to extend the results of the present thesis to implement the current-density approximation, which is better suited to capture time-dependent phenomena. In doing this, we plan to obtain the density-density correlation function between the absorbed and the emitted analogue Hawking particles, as made in~\cite{PhysRevA.78.021603}, but with an approach, that keeps being very transparent and direct.

In addition, our method can be applied to calculate the correlation function between condensate particles and phonons. We can build an effective Lagrangian which depends on the field representing the background density, besides the fields representing system's excitations. In doing this, we can follow the same procedure exposed in Sec.~\ref{sec:density}, which allows to get in a very easy and direct manner all the needed correlation functions. This analysis bears the potential to offer a novel understanding of the information paradox, as proposed in~\cite{Tricella:2020rzl}, where we anticipate that this correlation function plays a crucial role in encoding information while preserving unitarity in the system.

Furthermore, the same methodologies can be employed to investigate systems where the Hawking particles exhibit non-monotonic behavior of the energy dispersion, as demonstratedby the non-trivial minimum we discovered for specific LECs (see Sec.~\ref{sub:aneg}). This scenario opens up intriguing possibilities, where the behavior of the horizon can be connected to signatures of quantum phase transitions. In fact, this study can also have an experimental application in systems akin to dipolar BECs, in which the low-energy excitations exhibit a non-monotonic dispersion law similar to that we have employed in Sec.~\ref{sub:aneg} and characterized in Sec.~\ref{sec:lec}. Then, we could investigate whether the obtained system is a supersolid. In order to do this, we should first extend the study from (1+1)D to at least (2+1)D, to see whether the system displays superfluid phonon excitations and shear waves~\cite{during2011theory,PhysRevLett.25.1543,PhysRevA.105.023316,Tanzi_2021}, one of the signatures of supersolid character.

In summary, the tools developed in the present thesis have broad applicability. The method for calculating the density-density correlation function can be extended to non-homogeneous BECs, enabling a more comprehensive analysis. Additionally, investigating the correlation between condensed particles and phonons can deepen our understanding of the information paradox and the preservation of unitarity. Finally, the application of our methodologies to systems with non-monotonic dispersion relations, as exemplified by the non-trivial minimum found for specific LECs, holds promise for uncovering intriguing connections between the behavior of horizons and quantum phase transitions.

\appendix

\chapter{Density and energy density \label{app:den}}

We want to calculate the energy density and the density of a system described by the microscopic Lagrangian
\begin{equation}
    \lag=\frac12\partial_\nu\rho\partial^\nu\rho+ \frac12\rho^2\partial_\nu\theta\partial^\nu\theta-\mu\rho^2\partial_t\theta-\frac12(m^2-\mu^2)\rho^2-\frac\lambda4\rho^2.
    \label{eq:microscopicapp}
\end{equation}
Given a generic Lagrangia $\lag[\phi]$ for a field $\phi$, the energy momentum tensor (EMT) is defined as
\begin{equation}
    T^{\mu\nu}=\frac{\delta \lag[\phi]}{\delta\partial_\mu\phi}\partial^\nu\phi-\eta^{\mu\nu}\lag.
\end{equation}
Substituting in this equation the Lagrangian in Eq.~\eqref{eq:microscopicapp}, one gets
\begin{equation}
    T^{\mu\nu}=\left(\frac{\mu^4}{\lambda\gamma^4}-\frac{m^2\mu^2}{\lambda\gamma^2}\right)v^\mu v^\nu-\mink\left(\frac{\mu^4}{4\lambda\gamma^4}+\frac{m^4}{4\lambda}-\frac{\mu^2m^2}{2\lambda\gamma^2}\right).
\end{equation}
We observe that this is the Energy-Momentum tensor (EMT) of a perfect fluid with energy density and pressure, respectively,
\begin{equation}
    \varepsilon=\frac{3\mu^4}{4\lambda\gamma^4}-\frac{m^4}{4\lambda}-\frac{m^2\mu^2}{2\lambda\gamma^2}\qquad  P=\lag(\rhoo)=\frac{\mu^4}{4\lambda\gamma^4}+\frac{m^4}{4\lambda}-\frac{\mu^2m^2}{2\lambda\gamma^2}.
\end{equation}
Thus, we have found the energy density.
If we consider the current generated by $\theta$, it is defined as
\begin{equation}
    J^\mu=\left( \frac{\mu^3}{\lambda\gamma^3}-\frac{m^2\mu}{\lambda\gamma} \right)v^\mu
\end{equation}
and it is conserved, as $\theta$ is a cyclic coordinate, and thus, $J^\mu$ is a Noether current. Knowing that $J^\mu=nv^\mu$, we get the density of the system
\begin{equation}
    n=\frac{\mu^3}{\lambda\gamma^3}-\frac{m^2\mu}{\lambda\gamma}
\end{equation}

\bibliographystyle{ieeetr}
\bibliography{mbib.bib}

\end{document}